\documentclass[twocolumn,aps,prc,tightenlines,floats,floatfix,eqsecnum,preprintnumbers,nofootinbib]{revtex4}%%

\def\sp{\kern +3pt}
\def\sm{\kern -3pt}
\def\spQ{\kern +6pt}
\def\bea{\begin{eqnarray}}
\def\eea{\end{eqnarray}}

\def\sfrac#1#2{{\textstyle \frac{#1}{#2}}}

\newcommand{\ket}[1]{|#1\rangle}

\def\be{\begin{equation}}
\def\ee{\end{equation}}
\def\ba{\begin{eqnarray}}
\def\ea{\end{eqnarray}}

\usepackage{graphics}
\usepackage{graphicx}
\usepackage{epsf}
\usepackage{amsmath}
\usepackage{amssymb}
\usepackage[usenames]{color}

\begin{document}

\phantom{0}
\vspace{-0.2in}
\hspace{5.5in}

% include preprint number option
%\preprint{{\bf Version 11}}
\preprint{ }

\vspace{-1in}%\parbox{1.5in}{ \vspace{-9.6in}}  % moves the preprint box down

\title
{\bf
Axial form factors of the octet baryons in a covariant quark model}
\author{G.~Ramalho$^1$ and K.~Tsushima$^{1,2}$}
\vspace{-0.1in}

\affiliation{
$^1$International Institute of Physics, Federal 
University of Rio Grande do Norte, 
Campus Lagoa Nova - Anel Vi\'ario da UFRN, 
Lagoa Nova, Natal-RN, 59070-405, Brazil
\vspace{-0.15in}
}
\affiliation{
$^2$Laborat\'orio de F{\'{i}}sica Te\'orica e Computacional, 
LFTC, Universidade Cruzeiro do Sul, 
S\~ao Paulo 01506-000, Brazil
}

\vspace{0.2in}
\date{\today}

\phantom{0}

\begin{abstract}
We study the weak interaction axial form factors of the octet baryons,   
within the covariant spectator quark model, focusing on 
the dependence of four-momentum transfer squared, $Q^2$.
In our model the axial form factors 
$G_A(Q^2)$ (axial-vector form factor) and $G_P(Q^2)$ 
(induced pseudoscalar form factor), 
are calculated based on the constituent quark axial  
form factors and the octet baryon wave functions.
The quark axial current is parametrized 
by the two constituent quark form factors, 
the axial-vector form factor $g_A^q(Q^2)$,  
and the induced pseudoscalar form factor $g_P^q(Q^2)$. 
The baryon wave functions are composed of a dominant $S$-state 
and a $P$-state mixture for the relative angular momentum 
of the quarks.
First, we study in detail the nucleon  case.
We assume that the quark axial-vector form factor $g_A^q(Q^2)$ 
has the same function form as that of the quark 
electromagnetic isovector form factor.
The remaining parameters of the model,
the $P$-state mixture and the $Q^2$-dependence of  
$g_P^q(Q^2)$,  
are determined by a fit to the nucleon 
axial form factor data obtained by lattice QCD simulations 
with large pion masses.
In this lattice QCD regime the meson cloud effects are small,  
and the physics associated with the valence quarks 
can be better calibrated.
Once the valence quark model is calibrated,  
we extend the model to the physical regime,  
and use the low $Q^2$ experimental data to estimate 
the meson cloud contributions for $G_A(Q^2)$ and $G_P(Q^2)$. 
Using the calibrated quark axial form factors,   
and the generalization of  
the nucleon wave function for the other octet baryon members,   
we make predictions for all the possible weak interaction axial  
form factors $G_A(Q^2)$ and $G_P(Q^2)$ of the octet baryons.
The results are compared with the corresponding 
experimental data for $G_A(0)$, 
and with the estimates of baryon-meson models 
based on $SU(6)$ symmetry. 
\end{abstract}

%\phantom{0}
%\vspace{7.0in}
%\vspace{-6in}
\vspace*{0.9in}  % sets how far the title is below the preprint box
\maketitle

\section{Introduction}

The electromagnetic and the weak structure
of the hadrons can nowadays be accessed  
by electroweak probes, 
and characterized in terms of the corresponding structure form factors.
There is presently a considerable   
information about the vector electroweak  
form factors, including the electromagnetic form factors  
of several baryons and mesons,
where these form factors characterize the spatial
distribution of the charge and magnetism~\cite{NSTAR}. 
As for the weak interaction axial form factors, 
from now on mentioned simply as axial form factors, 
the information is much more scarce.
Only for the nucleon there are some 
data available for finite $Q^2$,
where $Q^2= -q^2$ and $q$ is the four-momentum transfer 
of the corresponding weak-axial transition. 
See Refs.~\cite{Bernard02,Beise05,Gorringe05,Schindler07a}
for a review of the nucleon axial form factors 
and Ref.~\cite{Gaillard84}
for the octet baryon axial form factors.

A better knowledge of the axial form factors of a baryon is very important,  
because it provides complementary information 
to the electromagnetic structure, and also because
involves both the  strong and weak interactions.
The form factors associated 
with the weak interaction axial current 
for the transitions 
$B' \to B \, \ell \, \bar \nu_{\ell}$, with $B,B'$ being spin 1/2 baryons, 
$\ell = e,\mu,\tau$ and $\bar \nu_{\ell}$ is a antineutrino, 
can be decomposed into 
the axial-vector $G_A$ and induced pseudoscalar $G_P$ 
form factors~\cite{Gaillard84,Bernard08}. 
In the limit $Q^2=0$ the octet baryon form factors $G_A$  
can be related with the 
polarized deep inelastic scattering data, 
and used to estimate the spin fraction 
of the baryon carried by the quarks (valence and 
sea)~\cite{Beise05,Myhrer08,Shanahan13,HERMES,NucleonDIS,BCano06,Isgur88}.

The nucleon  axial-vector form factor 
can be accessed by (quasi)elastic 
(anti)neutrino scattering and by charged 
pion electroproduction experiments.
The value for $Q^2=0$ is determined 
accurately by the neutron 
$\beta$ decay~\cite{Bernard02,Beise05,Schindler07a}.
The induced pseudoscalar form factor 
can be determined at very low $Q^2$ 
by pion production experiments and 
muon capture by a proton. 
In general the accuracy is worse  compared 
with the electromagnetic form factors, 
and limited to the region $Q^2<1$ GeV$^2$~\cite{Bernard02,Gorringe05}.
A review of  experimental 
data can be found in Refs.~\cite{Beise05,Schindler07a,Bernard02}.
To improve our knowledge of the weak interaction axial structure 
of nucleon, more precise data for $G_A$ are necessary 
for $Q^2 < 1$ GeV$^2$ as well as higher $Q^2$.
In progress are several experiments 
for quasi-elastic (anti)neutrino scattering 
with proton targets 
(MINER$\nu$\!A~\cite{MINERVA})
and nucleus targets
(T2K~\cite{T2K} and ArgoNeuT~\cite{ArgoNeuT}).
Models for neutrino/antineutrino scattering 
based on baryon-meson coupled-channels 
can be found in  Refs.~\cite{Nakamura15,Alam15,Sato03}.

The $G_P$ data are very scarce, since they cannot be 
obtained by neutrino or antineutrino scattering~\cite{Bilenky13b}.
The available data were obtained 
by pion electroproduction and also
by interaction with muons~\cite{Bernard02,Gorringe05}.
The relevant data can be found in Refs.~\cite{Bardin81,Choi93,Andreev07}.
As for the axial-vector form factors of the octet baryons,  
the available information is limited to 
the values of $G_A(0)$ for a few 
allowed transitions~\cite{Gaillard84,PDG2014}.

The axial form factors of the octet baryons,
including the nucleon,
have been studied using 
constituent and  chiral 
quark models~\cite{Glashow79a,Donoghue87,NRQM,Schlumpf93a,Hellstern97,Glozman01,Boffi02,Merten02,Franklin02,Diaz04,Choi10b,BCano03,BCano06,Li14},
based on the Dyson-Schwinger 
equations~\cite{Eichmann12,Chang13,Qin14,Yamanaka14},
models with meson cloud dressing 
\cite{Thomas84,Kubodera88,Kubodera89,Hogaasen88,Khosonthongkee04,Pasquini07,Silva05,Ledwig08,Lorce08,Adamuscin08,Liu14,Dahiya14,Yang15},
large-$N_C$ and 
chiral perturbation theory~\cite{Bernard94,Jenkins91,Dai96,Savage96,Schindler07a,FMendieta98,Schindler07b,Procura07,Serrano14,Ledwig15,Tiburzi15,Swart63}
and QCD sum rules~\cite{Henley92,Wang06,Erkol11a}.
Recently, lattice QCD simulations for the nucleon  
became available for $Q^2=0$ 
\cite{Sasaki03,Edwards06,Yamazaki08,Bhattacharya14,Horsley14,Green15,ARehim15,Bali15},
for finite $Q^2$~\cite{Liu94,Sasaki08,Yamazaki09,Bratt10,Alexandrou11a,Alexandrou13,Alexandrou13b}, 
and also for the octet baryons~\cite{Lin09,Sasaki09,Erkol10a,Alexandrou14a}. 
These studies are very important 
to understand the role of the valence quarks and  
of the meson cloud dressing.
The role of the meson cloud dressing
in the deep inelastic scattering,
namely in the nucleon parton distribution functions,
was studied in Refs.~\cite{Sullivan72,Thomas85b,Henley90,Kumano}. 
Of interest is also models 
based on the $SU(3)$-flavor symmetry of the  
baryon-meson reactions, 
like the heavy baryon $SU(3)$ chiral perturbation 
theory, and others, which hereafter will simply be referred to as 
$SU(3)$ baryon-meson models~\cite{Bernard08,Gaillard84,Swart63,Jenkins91}.
Furthermore, modifications in the nuclear medium 
are also studied in Refs.~\cite{TsushimaNeuA,Bilenky13,Butkevich14,Amaro15,Bhattacharya15,Musolf94}.

In the present work we study 
the axial form factors of the nucleon 
and octet baryons using the covariant spectator quark model.
The model has successfully been applied in studies of  
the electromagnetic structure of nucleon~\cite{Nucleon,Nucleon2,FixedAxis}, 
several nucleon resonances~\cite{NSTAR,NDelta,Roper,NucleonR,Lattice,LatticeD}
and other baryons~\cite{Omega,OctetFF,DecupletDecays,Omega2,LambdaSigma,Delta,Medium,OctetMM,NucleonR2}.
The covariant spectator quark model is 
based on the assumption that the constituent quarks
have their own internal structure, 
which can be parametrized by individual 
quark (electromagnetic) form factors.

In this work we extend the formalism 
of the covariant spectator quark model 
for the weak interaction axial structure of baryons, by introducing 
the axial-vector $g_A^q$ and induced pseudoscalar $g_P^q$ form factors
at the quark level.\footnote{
We include the upper index $q$ 
to emphasize that the functions are related with 
quarks, and also to avoid the confusion 
with the well established 
notation at the baryon level, $g_A= G_A(0)$
and $g_P= \frac{m_\mu}{2M}G_P(-0.88 m_\mu^2)$,
where $m_\mu$ is the muon mass.}
Based on our construction of the quark 
axial current, we calculate 
the contribution of the valence quarks 
for the {\it macroscopic octet baryon} form factors $G_A$ and $G_P$.
The quark axial-vector form factor $g_A^q$
can be defined {\it naturally} based on its  
isovector character, but $g_P^q$
has to be calibrated through a vector meson dominance form
by the lattice QCD data for the nucleon.

Once the model is calibrated by the lattice QCD data
for the nucleon, we extrapolate the model 
from the lattice QCD regime to the physical regime,
which allows to estimate the magnitude 
of the meson cloud contribution for the 
nucleon axial form factors.

In the present model the 
wave functions of the nucleon 
and the octet baryons are defined 
as in previous works~\cite{Nucleon,OctetFF} 
using an $S$-state structure, 
but we include additionally a $P$-state 
component with an admixture coefficient $n_P$.
The motivation to include the higher angular momentum 
states and the $P$-state is in particular 
comes from 
the Cloudy Bag Model (CBM) and  
nonrelativistic quark models~\cite{Thomas84,Chang13,NRQM,Glashow79a,BCano03}.
The magnitude of the  $P$-state component 
will be fixed by the comparison 
with the lattice QCD data for the nucleon 
with large pion masses, 
where the meson cloud contamination is very small.

After the calibration of the model by the nucleon data
(lattice and physical),
we extend the model parametrization 
to the octet baryons,
and make predictions for the 
valence and valence plus meson cloud contributions 
for the form factors $G_A$ and $G_P$.
The results for the octet baryons 
are compared with the lattice QCD results  
as well as the $SU(3)$ baryon-meson models.

To summarize, in this work we derive 
a successful parametrization 
for the nucleon axial form factors,
valid both in the physical regime and lattice QCD regime. 
Finally, we make predictions 
for the octet baryon axial form factors.

This article is organized as follows. 
In Sec.~\ref{secAxialFF} we introduce definitions 
of axial current and axial form factors,
both for the nucleon and the octet baryons.
In Sec.~\ref{secMethodology} 
we explain briefly the method used
to calibrate the quark axial current and the $P$-state mixture 
in the nucleon wave function, 
based on the available data for the nucleon.
The formalism of the covariant spectator quark model,
including the definition of the quark 
axial current and the octet baryon wave functions,   
are presented in Sec.~\ref{secCSQM}.
In Secs.~\ref{secValence} and~\ref{secValenceOctet}
we present expressions obtained for the 
valence quark components for the axial form factors
(bare or core form factors).
In Sec.~\ref{secMesonCloud} 
we explain how the effects of the meson cloud 
component can be taken into account 
for the physical regime.
Predictions of the  
$SU(3)$ baryon-meson model are 
discussed in Sec.~\ref{secSU6}.
Results for the nucleon and 
octet baryon axial form factors are presented 
respectively in Secs.~\ref{secResults-Nucleon}
and~\ref{secResults-Octet}.
Finally, summary and conclusions 
are presented in Sec.~\ref{secConclusions}.

\section{Axial-vector and induced pseudoscalar form factors}
\label{secAxialFF}

We define below
the axial form factors of nucleon 
and their extensions for the other octet baryon members. 

\subsection{Nucleon}

The  weak-axial transition between the nucleon states   
with an initial momentum $P_-$ 
with a final momentum $P_+$,
where $q=P_+ - P_-$, 
can be defined by 
the weak-axial current as~\cite{Bernard02,Beise05,Gaillard84,Schindler07a,Gorringe05}
\ba
& &(J_5^\mu)_a = \nonumber \\
& &
\bar u(P_+)
\left[
G_A (Q^2) \gamma^\mu + G_P(Q^2) \frac{q^\mu}{2M}
\right] \gamma_5 u(P_-) \,
 \frac{\tau_a}{2}, \nonumber \\
\label{eqAxialFFN}
\ea
where $M$ is the nucleon mass, 
$\tau_a$ ($a=1,2,3$) are the isospin operators
(Pauli matrices),
$u(P_\pm)$ the nucleon Dirac spinors,
and $G_A$ and $G_P$ are respectively the axial-vector and 
induced pseudoscalar form factors.
The "axial-tensor" 
form factor is ignored since it is
associated with the second-class current and consistent
with zero within experimental 
uncertainty~\cite{Beise05,Schindler07a,Gorringe05,Gaillard84}.
The current $(J_5^\mu)_a$ can be projected 
on the nucleon initial and final isospin states,
using the isospin matrices, 
responsible for the $SU(2)$-flavor (isospin) symmetry
[$SU_F(2)$  space], 
acting on the isospin states of the nucleon.

The discussion of the form factors defined by 
Eq.~(\ref{eqAxialFFN}) becomes simpler,
when  we use a spherical representation ($a=0,\pm$).
Then we have neutral transitions when $a=0$ ($n \to n$ and $p \to p$)
and the transitions $n \leftrightarrow p$ for $\tau_{\pm}$.
The weak neutral current ($a=0$),
($\Delta I=0$) corresponds to the
$Z$-boson emission or absorption.
The charged currents ($a=\pm$) are
associated with the
$\Delta I = \pm 1$ transitions 
mediated by the $W$-bosons emission or absorption 
for the $p \leftrightarrow n$ transitions.
In this work we simply assume 
that the function $G_A(Q^2)$ is defined by the
isovector transition form factor,
that corresponds to the 
transition between the $u$ and $d$ quarks
at the tree level.

Note that, experimentally the situation
is more complex, and in practice there are corrections
to the pure $W$- and $Z$-exchanges~\cite{Beise05,Musolf94}.
From the theoretical point of view 
the important issues are, whether or not we can
ignore the effects of the $s$-quark and
sea quarks for the axial form factors of the nucleon 
$G_A$ and $G_P$~\cite{Beise05,Bernard02}.

Since the nucleon axial current
is related with the quarks $u$ and $d$, 
we can perform a flavor decomposition 
defining the isovector combination
of the form factors~\cite{Alexandrou13,Erkol10a,ARehim15}
\ba
G_A^{u - d} \equiv G_A^u  - G_A^d.
\ea
The isoscalar combination can also be defined as
\ba
G_A^{u + d} \equiv G_A^u  + G_A^d.
\ea
Note however, that the isoscalar axial-vector
form factors cannot be obtained by a simple
current operator, but they can be calculated
in lattice QCD simulations using generalized
form factors~\cite{Alexandrou13,Bratt10}.

If we assume charge symmetry using 
$G_A^d = - \sfrac{1}{2} G_A^u$  
according to the relation
between the two quark charges, we obtain
$G_A^{u-d} = \frac{3}{2}G_A^u$
and $G_A^{u + d}= \frac{1}{2} G_A^u$.
Equivalently $G_A^{u-d} = 3 G_A^{u + d}$.

At $Q^2=0$ the values of $G_A^u(0)= \Delta \Sigma^u $
and  $G_A^d(0)= \Delta \Sigma^d$, 
are related with the intrinsic spin
of the valence quark $q$ in the proton, $\Delta \Sigma^q$.
These quantities are very important for the estimate
of the valence quark spin content of the
proton~\cite{Alexandrou13,HERMES,NucleonDIS,Myhrer08,BCano06,Silva05}.

It is important to mention that 
in the lattice QCD simulations, 
contrarily to the isovector axial form factors, 
the isoscalar axial-vector form factor 
has contributions from 
the so-called disconnected diagrams,
which are in general neglected in the simulations~\cite{ARehim15,Alexandrou13b}.
The first calculation indicated 
that the disconnected diagrams 
can contribute about 10\% for $G_A$
and 20\% for $G_P$~\cite{Alexandrou13b}.
In the present work we can ignore these  
corrections, since  
our axial current is identified with the isovector 
axial form factors.

The estimates of $\Delta \Sigma^u$ and $\Delta \Sigma^d$ 
based on deep inelastic scattering data,
combined with the $SU(3)$ symmetry are consistent 
with the charge symmetry 
($\Delta \Sigma^u/2 + \Delta \Sigma^d = -0.006 \pm 0.015 $)~\cite{HERMES}.
The estimates from lattice QCD available 
at the moment are in agreement with the 
estimates based on deep inelastic scattering for $\Delta \Sigma^u$, 
but underestimates $\Delta \Sigma^d$, 
even when the disconnected contributions 
are taken into account~\cite{ARehim15,Alexandrou13b}.
As a consequence, the results from 
lattice QCD violate charge symmetry 
($\Delta \Sigma^u/2 + \Delta \Sigma^d = 
0.097\pm 0,012$)~\cite{Alexandrou13b}.
It is worth to mention, however, that 
the lattice QCD calculations 
for the disconnected  diagrams contributions
are at the moment restricted to the pion masses around 370 MeV 
and the statistics is poor~\cite{ARehim15}.

\subsection{Octet baryons}

The axial form factors for the transition  
between the octet baryon members $B$ and $B'$  
can be generalized taking into account 
the quark axial current 
$\bar u \gamma^\mu \gamma_5 d$ ($d \rightarrow u$) 
and $\bar u \gamma^\mu \gamma_5 s$ ($s \rightarrow u$) 
in the 
$SU(3)$-flavor quark model [$SU_F(3)$ symmetry].
They are defined 
in terms of the weak-axial current 
as~\cite{Bernard94}
\ba
& &
%\hspace{.5cm}
J_5^\mu = \nonumber \\
& &
%\hspace{.5cm}
\frac{1}{2}
\bar u_{B'}(P_+)
\left[
G_A (Q^2) \gamma^\mu + G_P(Q^2) \frac{q^\mu}{2 M_{BB'}}
\right] \gamma_5 u_B(P_-) \,
, \nonumber \\
\label{eqOctetAxial}
\ea
where $u_{B'}$ and $u_B$ are the corresponding Dirac spinors,  
and $M_{BB'}$ is the average mass 
of the final  (mass $M_{B'}$) 
and initial (mass $M_B$) baryons:
$M_{BB'}= \sfrac{1}{2}(M_{B'} + M_B)$.
The factor $1/2$ in Eq.~(\ref{eqOctetAxial})
is included to be consistent with the nucleon case~(\ref{eqAxialFFN}).

The form factors $G_A(Q^2)$ and $G_P(Q^2)$ 
defined by Eq.~(\ref{eqAxialFFN}) 
are dependent on the octet baryon indices $B$ and $B'$,
similar to the nucleon case 
($p \to p$, $n \to n$ and $n \leftrightarrow p$).
However, for simplicity we omit the baryon
indices along the paper.

As in the case of the nucleon, weak transitions 
between the octet baryons can occur by the neutral current 
($Z$-boson) and by the charged current ($W^\pm$ boson).
The neutral transitions, 
that do not change their charges or isospin states 
are $N \to N$,
$\Sigma \to \Sigma$, $\Xi \to \Xi$, 
$\Sigma \to \Sigma$ and  $\Xi \to \Xi$.
The charged current transitions can be divided 
in two kinds, 
the cases with $\Delta I=1$,
and the cases with $\Delta S=1$.
Here, $\Delta I=1$ and $\Delta S=1$ represent variations 
of $\pm 1$ (variation of 1 in absolute value) 
of the isospin and strangeness, respectively.
The transitions $\Delta I=1$ are associated 
with the $u \leftrightarrow d$ transitions.
Examples of those transitions are:
$ n \to p$, $\Sigma^+ \to \Lambda$, $\Sigma^- \to \Lambda$,
$\Sigma^- \to \Sigma^0$ and $\Xi^- \to \Xi^0$.
The transitions  $\Delta S=1$  are associated 
with the $u \leftrightarrow s$ transitions.
In this case one has
$\Lambda \to p$, $\Sigma^- \to n$,  $\Sigma^0 \to p$,
$\Xi^- \to \Lambda$, $\Xi^- \to \Sigma^0$ and $\Xi^0 \to \Sigma^+$.

It is important to note that the 
form factors associated with the $\Delta I=1$ 
and $\Delta S=1$ transitions 
contribute differently for the transition
crossed sections~\cite{Gaillard84}.
The $\Delta I=1$ form factors are 
multiplied by the factor $\cos \theta_C \simeq 0.97$, 
while the  $\Delta S=1$ form factors are 
multiplied by the factor $\sin \theta_C \simeq 0.23$.
In the neutral current transitions the factor is 1.

The transition current between the octet baryon members 
can also be represented by an $SU_F(3)$    
extension of $SU_F(2)$ using the 
Gell-Mann matrices $\lambda_a$ ($a=1,...,8$)
instead of the Pauli matrices $\tau_a$ ($a=1,2,3$)~\cite{QCDbook}.
In this case the transition currents are expressed in terms of 
the octet baryon states, or  
by the $3 \times 3$ baryon matrix 
and flavor-transition operators 
in the corresponding 
octet vector space~\cite{Bernard08,Jenkins91,Savage96,Gasiorowicz}.

\section{Methodology}
\label{secMethodology}

We discuss now the method used to calculate 
the axial form factors of the nucleon 
and the other members of octet baryons 
within the framework of 
the covariant spectator quark model.
We start by the nucleon case.
Later,  we extend the framework for the other octet baryons.
The formalism of the 
covariant spectator quark model 
is reviewed briefly in the next section.

In the covariant spectator quark model 
the electromagnetic structure 
of the baryons is described based on 
the valence quark structure of the 
baryon wave functions, and the 
electromagnetic structure of the constituent quarks.
The electromagnetic structure of the baryons 
is parametrized by the quark electromagnetic
form factors which simulate effectively the internal structure
of the constituent quarks
resulting from the interactions with quark-antiquark pairs
and from the gluon dressing~\cite{Nucleon,Omega}. 
Of particular relevance for the present work
is  the quark isovector form factors $f_{1-}$ and $f_{2-}$, 
associated respectively with 
the Dirac and Pauli  components 
of the constituent quark current
[see the following Sec.~\ref{secQuarkEMFF} for the details].
In the present study we define two new quark
form factors, $g_A^q$ and $g_P^q$, 
respectively the quark axial-vector   
and quark induced pseudoscalar form factors.
The details are discussed in Sec.~\ref{secQuarkAFF}.

In order to calculate the 
axial form factors of the nucleon,  
we need a model for the wave function of the nucleon.
We start assuming that we can describe the nucleon 
as a quark-diquark system with an $S$-state configuration
following Ref.~\cite{Nucleon}.
Previous works had shown that an $S$-state quark-diquark system 
is a good approximation for the 
nucleon~\cite{Lichtenberg67,Santopinto05,Santopinto15}.

Since the structure based only on an $S$-state 
is quite poor as we will show in Sec.~\ref{secValence},
we consider a possibility of adding 
a $P$-state mixture to the
nucleon wave function.
The motivation to include this new component 
comes from nonrelativistic quark models,
QCD sum rules, and also from CBM~\cite{Thomas84,Chang13,Wang06}.
In some models the $P$-state mixture corresponds 
to the lower component of the 
nucleon Dirac spinor, which becomes very important for 
the axial form factors in a 
relativistic treatment~\cite{Thomas84,Chang13,NRQM}.
In the covariant spectator quark model 
the $P$-state quark-diquark wave function
is generated by the integration on  
the quark-pair internal degrees of freedom 
in the three-quark wave function.
The quark-diquark wave function contains 
all the information originally 
included in the three-quark wave function,
as discussed in Ref.~\cite{Nucleon2}.
This $P$-state quark-diquark wave function
appears as the consequence of the relativity 
and vanishes in the nonrelativistic limit~\cite{Nucleon2}.

We consider then a nucleon wave function
composed of a combination of the $S$- and $P$-state components,
parametrized by the $P$-state mixing coefficient $n_P$
($n_P^2$ gives the $P$-state probability in the nucleon wave function).
The coefficient of the  $S$-state, $n_S$,
is expressed by $n_S= \sqrt{1-n_P^2}$.
When $n_P=0$  ($n_S=1$), we recover the $S$-state wave function.

The discussion about how the $P$-state can be built 
within the covariant spectator quark model 
is presented in Ref.~\cite{Nucleon2}.
The radial wave function
associated  with  the $S$-state
was already determined by the study
of the electromagnetic structure of nucleon.
For the $P$-state component,
the corresponding radial wave function
can be defined in terms of the 
radial wave function of the $S$-state,
without introducing any extra parameter~\cite{Nucleon,NucleonDIS}.

As a consequence of the underlying internal structure
of the nucleon based on the valence quark degrees of freedom, 
we decompose the nucleon axial form factors,
admitting the possibility of meson excitations of the core, as
\ba
& &
\hspace{-1cm}
G_A(Q^2)= G_A^B(Q^2) + G_A^{MC} (Q^2), 
\label{eqGAdecomp}
\\
& &
\hspace{-1cm}
G_P (Q^2)=  G_P^{\rm pole} (Q^2) + G_P^B (Q^2) + G_P^{MC} (Q^2),
\label{eqGPdecomp}
\ea
where $G_A^B$ and $G_P^B$ are the contributions from
the bare core (valence quark contribution), while
$G_A^{MC}$ and $G_P^{MC}$ are those from the meson cloud.
The meson pole term $G_P^{\rm pole}$ is an additional 
contribution that is the result 
of a meson creation by the baryon transition 
that decays by the weak interaction, into a
lepton-neutrino pair
\cite{Bernard02,Gorringe05,Schindler07a,Bernard94}.

For the nucleon and other non-strangeness 
changing transitions ($\Delta I = 1$), the pion 
(mass $m_\pi$) is expected to give a dominant contribution 
in the meson pole contributions,
which is determined by the partial conservation 
of the axial current 
(PCAC)~\cite{Bernard02,Gorringe05,Schindler07a,Bernard94,Thomas84,Kubodera88,Sasaki08}
\ba
G_P^{\rm pole}(Q^2) = 
\frac{4 M^2}{m_\pi^2+Q^2} G_A^B(Q^2).
\label{eqGPpole}
\ea
In the study of the $Q^2$-dependence of the form factors $G_P$,
the pole term is very important,  
especially in the timelike region 
($Q^2 < 0$)~\cite{Bernard02}.
Note that, contrarily to most of the works in literature,
in r.h.s.~of Eq.~(\ref{eqGPpole}),  we use 
$G_A^B(Q^2)$ the bare contribution, instead 
of the function  $G_A(Q^2)$, the total result 
for the axial-vector form factor, which includes  
also the meson cloud contribution.
We replace $G_A$ by $G_A^B$ because 
we want to use the relation~(\ref{eqGPpole})
also in the lattice QCD regime, 
in the limit where the meson cloud effects are small.
In the cases where meson cloud effects 
are significant for $G_A$, as in the physical limit, 
the term $\frac{4M^2}{m_\pi^2 + Q^2} G_A^{MC}$ 
can be interpreted as a meson cloud contribution for $G_P$.
In the literature $G_A$ is often replaced 
by parametrizations of the experimental data 
labeled here as $G_A^{\rm exp}(Q^2)$~\cite{Bernard02,Choi93}.

The factor $G_P^{\rm pole}$ can be connected
with the strong pion-nucleon coupling 
at $Q^2=0$ in the chiral limit,
via the Goldberger-Treiman relation~\cite{Goldberger58a}.
For a more complete discussion 
see Refs.~\cite{Bernard02,Gorringe05,Schindler07a,Thomas84,Kubodera88}.

Note that, since we have contribution 
from the valence quark core for $G_P$,  
we are including at $Q^2=0$ 
corrections to the Goldberger-Treiman relation
according to (\ref{eqGPdecomp}).
This is not a problem, since the 
relation is strictly valid only in the chiral 
limit, and some corrections are expected, 
according to chiral perturbation theory and 
other frameworks~\cite{Bernard02,Bernard94,Gorringe05,Kubodera88,BCano03,Eichmann12}.
Calculations of $G_P$ can be found in 
Refs.~\cite{Schindler07b,Eichmann12,Kubodera88,Kubodera89,Glozman01,Diaz04,Boffi02,Wang06,Erkol11a}.
For a review about the theory and experimental data 
associated with $G_P$, 
see Ref.~\cite{Gorringe05}.

In the limit where the meson cloud effects are small, 
we can drop the meson cloud contributions
$G_A^{MC}$ and $G_P^{MC}$ and consider only
the contributions from the bare core, 
and the pole term in the case of $G_P$.
This situation occurs when we deal 
with lattice QCD simulations with large pion masses
(large $m_\pi$).
Along this work we will use the expression 
``the large pion masses'' 
to indicate the range $m_\pi > 350$ MeV.

If the covariant spectator quark model is 
successful in the description of 
the bare core contribution of the nucleon 
axial form factors, it should also be able to reproduce
the lattice QCD data for large $m_\pi$, 
since in the lattice QCD regime the model   
is dominated by the valence quark effects.
Therefore in this work, we use
the lattice QCD data with large $m_\pi$ to calibrate 
the valence quark contributions of the model for
the axial form factors. 
At the end the model will be extrapolated 
to the physical regime ($m_\pi = m_\pi^{\rm phys} \simeq 138$ MeV)
and will be used to estimate the contributions 
of the meson cloud to the nucleon form factors.

An important step is the extension 
of the covariant spectator quark 
model to the lattice QCD regime.
This will be done taking advantage
of our parametrization for the quark form factors, 
both electromagnetic and axial currents,
which are defined based on 
vector meson dominance (VMD) parametrizations.
The extension of the model to the lattice QCD
regime will be discussed in Sec.~\ref{secLattice}.

A model based exclusively  
on the valence  quark degrees of freedom 
is particularly convenient to compare with the  
lattice QCD data with large $m_\pi$.
In this case we have a more clean parametrization
(free of meson cloud effects)
for the valence quark effects.
The same method was used previously and 
successfully in the studies of the electromagnetic proprieties of the nucleon,
the Roper,
the $\gamma^\ast N \to \Delta$ reaction,
as well as in the studies of the octet and decuplet
baryon properties~\cite{Lattice,LatticeD,Omega,Roper,Omega2,OctetFF,Medium}.

The methodology used in the present study 
can be summarized as follows:
\begin{itemize}
\item
First, we calibrate our model by a fit 
to the lattice QCD data for $G_A(Q^2)$.
The calculation of $G_A(Q^2)$ depends 
on the quark axial-vector form factor $g_A^q(Q^2)$
and the amount of the $P$-state mixture ($n_P$).
Since $g_A^q(Q^2)$ is associated with an isovector structure
we simply assume that the $Q^2$ dependence 
of $g_A^q(Q^2)$ can be approximated by 
the quark Dirac isovector form factor $f_{1-}$.
Under this assumption we try to find if 
there are  proper values for $n_P$ that can describe 
the nucleon lattice QCD data for $G_A$.
We conclude that the answer is positive, 
and $n_P$ is determined by the best fit to the data. 
Up to this stage we neglect the induced pseudoscalar 
form factor of the quark by setting $g_P^q \equiv 0$.
\item
Next, we check whether or not 
the lattice data for $G_P(Q^2)$,
associated with several values of $m_\pi$,  
can be described by a simple 
model for the quark induced pseudoscalar form factor $g_P(Q^2)$,
parametrized  by a VMD form.
Again, the answer turns out to be positive,  
and we use the lattice QCD data to 
estimate the shape of $g_P^q$, 
fixing the two parameters of the VMD expression.
With the determination of $n_P$ and $g_P^q(Q^2)$
by the fit to the lattice data, we finish the calibration of 
the valence quark component of our model.
\item
The next step is to extrapolate the model 
to the physical regime ($m_\pi \to m_\pi^{\rm phys}$). 
The extrapolation is performed in two steps.
First, we extrapolate our parametrization of 
the valence quark contributions 
(obtained from the lattice QCD data) 
to the physical regime, to get $G_A^B(Q^2)$ and $G_P^B(Q^2)$.
Next, we correct the result for the form factors including 
the normalization factor of the wave function, 
$\sqrt{Z_N}$, corresponding to the 
fraction of the three-quark valence quark system,
in the physical nucleon wave function,
redefining the effective contribution
of the valence quarks by effectively taking 
into account the meson cloud effects.
With this procedure $G_A^B$ is 
modified according to
\ba
G_A^B (Q^2) \to Z_N G_A^B (Q^2).
\ea
We estimate $Z_N$ by comparing our valence quark model result with 
a parametrization extracted from the physical data $G_A^{\rm exp}(Q^2)$ 
in the the region $Q^2  \gtrsim 1$ GeV$^2$, where
the meson cloud effects are expected to be small. 
With this procedure we determine 
a parametrization for the valence quark contribution, 
for the nucleon axial-vector form factor in the physical regime.
After this the meson cloud effects 
can be estimated using
$G_A^{\rm exp}(Q^2) -Z_N G_A^B(Q^2)$.
Also for the induced pseudoscalar form factor $G_P$, 
the contribution of the valence quarks $G_P^B$ 
is corrected by the factor $Z_N$, in the physical limit.
Details of the procedure 
are discussed in Sec.~\ref{secMesonCloud}.
\item
With the model calibrated for the nucleon  
axial form factors ($n \to p$ transition),
we use $SU_F(3)$ symmetry at the quark level 
to extrapolate the results of the nucleon 
to make predictions for the other octet  baryon axial form factors.
In Sec.~\ref{secValenceOctet} we discuss 
our extrapolation from $SU_F(2)$ to $SU_F(3)$.
To obtain the final result for 
the octet baryon  axial form factors, 
we need also to take into account the meson cloud 
effects for the other octet baryon members,
which can be done making some assumptions 
about the amount of the meson cloud contribution 
based on $SU(3)$ and/or $SU(6)$ symmetries.
Finally, we compare the results with 
those of the $SU(6)$ baryon-meson model 
discussed in Sec.~\ref{secSU6}. 
\end{itemize}

\section{Covariant spectator quark model}
\label{secCSQM}

We discuss now the covariant spectator quark model.
The covariant spectator quark model was first developed
for the study of the electromagnetic 
properties of nucleon~\cite{Nucleon,Nucleon2,FixedAxis},
and subsequently extended for the studies 
of the electromagnetic properties of several resonances,  
and electromagnetic transitions 
between baryon states~\cite{NucleonR,Delta,NucleonR2},
including the octet and decuplet 
baryons~\cite{OctetFF,Omega,DecupletDecays,Medium,OctetMM,LambdaSigma}.

In the following, we review the formalism of the 
covariant spectator quark model
related with the electromagnetic structure of the quarks
and baryons.
Next, we introduce the quark axial form factors, 
and explain how the axial current between 
the baryon states is calculated. 
Later, we describe the structure of the nucleon wave function 
in term of the valence quark structure, 
and explain how it can be extended for the octet baryons.
Finally, we show how the model can be 
generalized to the lattice QCD regime.

\subsection{Electromagnetic form factors}
\label{secQuarkEMFF}

In the covariant spectator quark model 
the electromagnetic transition current
is calculated in a relativistic impulse
approximation using
the nucleon wave function $\Psi_N$
expressed in terms of the states of the quark 3
and the quark current $j_q^\mu$~\cite{Nucleon,Nucleon2,Omega}:
\ba
J^\mu
=
3 \sum_{\Gamma}
\int_k \overline \Psi_N(P_+,k) j_q^\mu \Psi_N(P_-,k).
\ea
In the above 
the integral symbol represents the 
covariant integration associated with  
the diquark three-momentum,
$\Gamma$ represents the diquark polarizations
(scalar and axial-vector),
and $k$ the diquark momentum.
As before $P_+$ and $P_-$ are respectively
the final and initial nucleon momenta.

The quark electromagnetic
form factors are defined 
by the quark electromagnetic current $j_q^\mu$ as~\cite{Nucleon,NDelta}:
\ba
j_q^\mu &=& \left(\frac{1}{6}f_{1+}  +
\frac{1}{2} f_{1-} \tau_3\right) \gamma^\mu + \nonumber \\
& &
\left(\frac{1}{6}f_{2+}  +
\frac{1}{2} f_{2-} \tau_3\right) \frac{i \sigma^{\mu \nu} q_\nu}{2M}.
\label{eqJQ}
\ea
The functions $f_{1\pm}$ define the Dirac 
isoscalar/isovector form factor 
and $f_{2\pm}$ define the Pauli 
isoscalar/isovector form factor.

For the present discussion it is sufficient to mention
the isovector form factors, $f_{1-}$ and $f_{2-}$.
These form factors were defined 
in previous works using 
a parametrization motivated by vector
meson dominance~\cite{Nucleon,NDelta}:
\ba
& &
\hspace{-1cm}
f_{1-}= \lambda + (1-\lambda)\frac{m_\rho^2}{m_\rho^2+ Q^2}
+ c_-  \frac{ Q^2 M_h^2}{(M_h^2+ Q^2)^2}, \\
& &
\hspace{-1cm}
f_{2-}=
\kappa_- \left\{
d_- \frac{m_\rho^2}{m_\rho^2+ Q^2} +
(1- d_- )  \frac{M_h^2}{M_h^2+ Q^2} \right\},
\ea
where $m_\rho$ is the $\rho$ meson mass and
$M_h$ represents a mass of an effective heavy meson
that simulates the structure of all the 
high mass resonances.
The parameters $\lambda$, $\kappa_-$, $c_-$ and $d_-$ are
coefficients calibrated by the nucleon form factor data
and deep inelastic scattering (for $\lambda$)~\cite{Nucleon}.
We choose in particular the model II in Ref.~\cite{Nucleon},
where $\lambda= 1.21$, $\kappa_-= 1.823$,
$c_-= 1.16$ and $d_-=-0.686$.
As for $M_h$ we use $M_h=2M$ 
(twice the nucleon mass)
in order to simulate the short range structure
of the current.

\subsection{Axial form factors}
\label{secQuarkAFF}

Similarly to the electromagnetic form factors, 
the nucleon axial current can be written as 
\ba
J_5^\mu =
3 \sum_{\Gamma}
\int_k \overline \Psi_N(P_+,k) (j_{Aq}^\mu) \Psi_N(P_-,k).
\label{eqJ5}
\ea
The constituent 
quark axial current, $j_{Aq}^\mu$ is defined 
in terms of the quark 
axial form factors $g_A(Q^2)$
and $g_P(Q^2)$ as
\ba
j_{Aq}^\mu &=&
\left(
g_A^q \gamma^\mu +
g_P^q \frac{q^\mu}{2M}
\right)
\gamma_5 \frac{\tau_a}{2}.
\label{eqJAq}
\ea
As for the nucleon, we can add
an isospin label $a$ to the quark form factors $g_A^q$ and $g_P^q$, 
but only one function is relevant due to the isospin symmetry.
The form factor $g_A^q$ is associated 
with the $u \leftrightarrow d$ quark transitions  
($W$-boson emission or absorption)
responsible by the $\Delta I = 1$ transitions.

As for $g_A^q$,  we assume that it is the same as the 
isovector component of the Dirac form factor
defined by Eq.~(\ref{eqJQ}), due to its isovector character:
\ba
g_A^q (Q^2)\equiv f_{1-} (Q^2).
\label{eqQuarkGA}
\ea
Note that, then 
$g_A^q(0)=1$ \cite{Thomas84,Weinberg90}.

As for $g_P^q$ we may be tempted to relate it
with $f_{2-}$, because we expect a falloff,   
$g_P^q \propto 1/Q^2$.
However, since the structure of the 
Pauli term and the term 
associated with the induced 
pseudoscalar current are very different, 
we choose instead only a form inspired by $f_{2-}$,
given by 
\ba
g_P^q (Q^2)= \alpha  \frac{m_\rho^2}{m_\rho^2+ Q^2} +
\beta  \frac{M_h^2}{M_h^2+ Q^2},
\label{eqQuarkGP}
\ea
where the coefficients $\alpha$ and $\beta$ 
will be determined by a fit to the lattice QCD data
obtained with large $m_\pi$  
(small meson cloud contamination).

To summarize, we choose parametrizations 
for the quark axial form factors 
$g_A^q$ and $q_P^q$, motivated by VMD,
similarly to what was done previously 
for the quark electromagnetic form factors, $f_{i\pm}$ ($i=1,2$).

\subsection{Nucleon wave function}

For the nucleon wave function, we consider 
a mixture of the $S$- and $P$-states 
as suggested by Ref.~\cite{Nucleon2},
\ba
\Psi_N(P,k) =
%\sqrt{1-n_P^2} 
n_S
\Psi_S(P,k) + n_P \Psi_P(P,k),
\label{eqPsiN}
\ea
where $n_P$ is the $P$-state mixture coefficient,
and $n_S= \sqrt{1-n_P^2} $, as discussed already.

For the $S$-state $\Psi_S(P,k)$, we use~\cite{Nucleon}
\ba
\Psi_S(P,k)=
\frac{1}{\sqrt{2}}
\left[
\phi^0 u(P) - \phi^1 (\varepsilon_P^\ast)_\alpha
U^\alpha (P)
\right] \psi_S(P,k), \nonumber \\
\label{eqPsiS}
\ea
where $\phi^{0,1}$ are the isospin wave functions,
$\varepsilon_P^\alpha$ is the diquark polarization vector, 
$U^\alpha (P)$ a spin 1/2 state, to be defined next,
and $\psi_S$ is the radial wave function.

The isospin wave functions,  $\phi^{0,1}$, 
can be represented in terms of 
the isospin-0 and isospin-1 components
that are also function of 
the nucleon isospin projection $I_z$,
that labels the proton ($I_z=  + \sfrac{1}{2} $) 
and the neutron ($I_z=  - \sfrac{1}{2} $) states.
More specifically, we can write~\cite{Nucleon}
\ba
& &
\phi^0 (I_z) = \xi^{0\ast} \chi (I_z), \\
& & 
\phi^1 (I_z) = - \frac{1}{\sqrt{3}} 
(\tau \cdot \xi^{1 \ast}) \chi (I_z),
\ea 
where $\chi (I_z)$ are the 
nucleon isospin state, 
that correspond also the  
isospin state of the quark-3 ($u$ or $d$), and 
$\tau_\pm = \tau_x \pm i \tau_y$, 
are the isospin raising and lowering operators
and $\tau_0 = \tau_z$.
The operators $\xi^{0,1}$ are represented as~\cite{Nucleon}
\ba
& &
\xi^0 = \frac{1}{\sqrt{2}} (u d- d u), \\
& &
\xi^1_0 = \frac{1}{\sqrt{2}} (u d + d u) = \xi_z, \\
& &
\xi^1_+ = uu = - \frac{1}{\sqrt{2}} (\xi_x + i \xi_y), \\
& &
\xi^1_- = dd =  \frac{1}{\sqrt{2}} (\xi_x - i \xi_y).
\ea
In the next section, we use an alternative notation 
to represent the flavor states of the remaining 
octet baryon members.

The spin-1 diquark component of the wave function (\ref{eqPsiS})
includes the spin state~\cite{NDelta} 
\ba
U^\alpha (P) 
= 
\frac{1}{\sqrt{3}} \gamma_5 
\left( \gamma^\alpha - \frac{P^\alpha}{M}  \right) u(P).
\ea
The spin states are ruled by 
the $SU(2)$-spin symmetry [$SU_S(2)$ symmetry].
The spin 1/2 states $U^\alpha(P)$ is combined 
with the diquark polarization vector, 
$\varepsilon_P^\alpha (\lambda)$ ($\lambda = 0,\pm$)
defined in a fixed-axis base, 
for a three-momentum ${\bf P}$ along the $z$-axis~\cite{FixedAxis},
\ba
& &
\varepsilon_P^\alpha (\pm) = \mp \frac{1}{\sqrt{2}} (0,1,\pm 1,0), \\
& &
\varepsilon_P^\alpha (0)= \frac{1}{M}(P,0,0,E),
\ea
where $E= \sqrt{M^2 + {\bf P}^2}$.

In order to write the 
expression for the $P$-state conveniently, we define 
\ba
\tilde k= k - \frac{P \cdot k}{M^2}P.
\label{eqKtil}
\ea
Note that at the nucleon rest frame,    
\mbox{$P=(M,0,0,0)$}, $\tilde k=(0,{\bf k})$
is reduced to the diquark three-momentum,  
and $\tilde k^2 = - {\bf k}^2$.
Following Ref.~\cite{Nucleon2}
we define the $P$-state wave function as
\ba
\Psi_P(P,k)=
\frac{1}{\sqrt{2}}  \not \! \tilde k
\left[
\phi^0 u(P) - \phi^1 (\varepsilon_P^\ast)_\alpha
U^\alpha (P)
\right] \psi_P(P,k), \nonumber \\
\label{eqPsiP}
\ea
where $\psi_P(P,k)$ is the $P$-state radial wave function.
Since the wave function (\ref{eqPsiP}) 
is reduced to two {\it upper} components 
that vanishes in the nucleon rest frame,  
the  state correspond to a positive parity state
(the negative parity of the $P$-state is 
cancelled by the negative sign from 
the Dirac parity operator $\gamma^0$)~\cite{Nucleon2}.

The normalization conditions of the $S$- and $P$-state components
require that, 
\ba
& &
\int_k [\psi_S(\bar P,k)]^2 =1, \\
& &
\int_k (-\tilde k^2)[\psi_P(\bar P,k)]^2 =1,
\ea
where $\bar P=(M,0,0,0)$ is the nucleon momentum
in its rest frame.
The above conditions, derived from
the \mbox{$Q^2=0$} limit, ensures that the charge 
of the nucleon state  is 
$e_N=\frac{1}{2}(1 + \tau_3)$.
[The operator $e_N$ acts on the isospin states 
of the nucleon.]

The radial wave function for the $S$-state, $\psi_S(P,k)$,  
can be defined
in terms of the dimensionless variable~\cite{Nucleon,Omega} 
\ba
\chi= \frac{(M-m_D)^2-(P-k)^2}{M m_D}.
\label{eqChiN}
\ea
As in previous works, we consider the form~\cite{Nucleon,OctetFF}, 
\ba
\psi_S (P,k)= \frac{N_S}{m_D (\beta_1 + \chi)(\beta_2 + \chi)}, 
\label{eqPsiRadS}
\ea
where $N_S$ is a normalization constant, and
$\beta_1,\beta_2$ are momentum 
scale parameters in units of $M m_D$.
In the present work  we use 
the values $\beta_1=0.049$ and $\beta_2=0.717$~\cite{Nucleon}.

As for the $P$-state, we define $\psi_P(P,k)$ as
\ba
\psi_P(P,k) =  \frac{\psi_S(P,k)}{\sqrt{-\tilde k^2}}.
\label{eqPsiPscalar}
\ea
In this case both the $S$- and $P$-state components 
of the nucleon wave function 
are defined by the $S$-state parametrization
established in previous works~\cite{Nucleon,NDelta}.

\subsection{Extension of the valence quark model 
for the octet baryons}
\label{secOctetWF}

%%%%%%%%%%%{tablePHI}

\begin{table*}[t]
\begin{center}
\begin{tabular}{l c c c}
\hline
\hline
$B$   & $\ket{M_A}$  & &  $\ket{M_S}$  \\
\hline
$p$     &  $\sfrac{1}{\sqrt{2}} (ud -du) u$ & &
        $\sfrac{1}{\sqrt{6}} \left[
        (ud + du) u - 2 uu d \right]$ \\
$n$     & $\sfrac{1}{\sqrt{2}} (ud -du) d$  & &
         $-\sfrac{1}{\sqrt{6}} \left[
         (ud + du) d - 2 ddu \right]$  \\
\hline
$\Lambda^0$ &
$\sfrac{1}{\sqrt{12}}
\left[
s (du-ud) - (dsu-usd) + 2(ud -du)s
\right]$
& &
$\sfrac{1}{2}
\left[ (dsu-usd) - s (ud-du)
\right]$ \\
\hline
$\Sigma^+$  &  $\sfrac{1}{\sqrt{2}} (us -su) u $ & &  
$\sfrac{1}{\sqrt{6}} \left[(us + su) u - 2 uu s \right]$
\\
$\Sigma^0$ &
$\sfrac{1}{2}
\left[ (dsu+usd) -s (ud+du)
\right]$
& &
$\sfrac{1}{\sqrt{12}}
\left[
s (ud + du ) +(dsu+usd) -2(ud+du)s
\right]$ \\
$\Sigma^-$ & $\sfrac{1}{\sqrt{2}} (ds -sd) d$ & &
$\sfrac{1}{\sqrt{6}}\left[ (sd + ds) d - 2 dd s \right]$ 
              \\
\hline
$\Xi^0$ & 
$\sfrac{1}{\sqrt{2}} (us -su) s$ 
& &
$-\sfrac{1}{\sqrt{6}} \left[(us + su) s - 2 ss u\right]$   \\
$\Xi^-$ & $\sfrac{1}{\sqrt{2}} (ds -sd) s$
& &
$-\sfrac{1}{\sqrt{6}} \left[(ds + sd) s - 2 ss d\right]$  \\
\hline
\hline
\end{tabular}
\end{center}
\caption{ Representations 
of the flavor wave functions of the octet baryons.}
\label{tabPHI}
\end{table*}

We discuss now the extension of the model 
for the other members of the octet baryons. 
For this we need to consider 
the modifications of the quark axial current (\ref{eqJAq}) 
as well as the modifications in the 
wave functions of the baryons.

We can extend the description of the nucleon 
wave function for the $S$- and $P$-states 
given by Eqs.~(\ref{eqPsiS}) and~(\ref{eqPsiP}), 
to the octet baryons replacing the isospin wave functions of the nucleon
$\phi^0$ and $\phi^1$ by the mixed anti-symmetric   
and mixed symmetric $SU(3)$ flavor wave functions, 
respectively $\left|M_A \right>_B$ 
and $\left|M_S \right>_B$ associated with the baryon $B$.
The flavor wave functions, $\left|M_A \right>_B$, 
$\left|M_S \right>_B$ are presented in Table~\ref{tabPHI}.

As for the radial wave functions 
we follow the study of the electromagnetic 
proprieties of the octet baryons $\Lambda, \Sigma$
and $\Xi$ based on the $S$-state~\cite{Medium}
\ba
& &
\psi_{\Lambda,S}(P,k) = 
\frac{N_\Lambda}{m_D(\beta_1 + \chi_{_\Lambda})(\beta_3 + \chi_{_\Lambda})},
\label{eqPsiR-Lambda}
 \\
& &
\psi_{\Sigma,S}(P,k) = 
\frac{N_\Sigma}{m_D(\beta_1 + \chi_{_\Sigma})(\beta_3 + \chi_{_\Sigma})}, \\
& &
\psi_{\Xi,S}(P,k) = 
\frac{N_\Xi}{m_D(\beta_1 + \chi_{_\Xi})(\beta_4 + \chi_{_\Xi})},
\label{eqPsiR-Xi}
\ea
where $\chi_{_B}$ is defined by Eq.~(\ref{eqChiN}) 
with $M$ replaced by $M_B$,
$N_B$ are normalization constants, and 
$\beta_3,\beta_4$ are new momentum range parameters 
(in units of $M_B m_D$).
The motivation for the above expressions 
is to modulate the short range behavior $\beta_2$, 
defined in the nucleon radial wave function  
by a different parameter 
(smaller value for $\beta_3$ and $\beta_{4}$)
according to the number of strange quarks.
We take the values from Ref.~\cite{Medium}:
$\beta_3=0.603$ and $\beta_4=0.381$. 
Similarly to the nucleon,  
we can also define the $P$-state radial wave functions as
$\psi_{B,P}= \psi_{B,S}/\sqrt{-\tilde k^2}$, 
where $\tilde k$ is defined by Eq.~(\ref{eqKtil}) 
in terms of the baryon momentum $P$.

The quark axial current~(\ref{eqJAq}) 
defined so far in the $SU_F(2)$ sector,
is extended to the $SU_F(3)$ sector
for transitions between the other octet baryons  
replacing the Pauli matrices $\tau_a$ ($a=1,2,3$)
by the Gell-Mann matrices $\lambda_a$ ($a=1,...,8$).

Using the  Gell-Mann matrices we can 
describe the neutral current transitions 
($B \to B$) when the operator is $I_0= \lambda_3$,
and also the transitions with  $\Delta I=1$ or $\Delta S=1$.
The transitions with $\Delta I= 1$ 
are associated with the operator 
$I_\pm= \frac{1}{2}(\lambda_1 \pm i \lambda_2)$
which correspond to the $d \rightarrow u$ ($I_+$)
and $u \rightarrow d$ ($I_-$) transitions.
The transitions with  $\Delta S=1$
are associated with the operator 
$V_\pm= \frac{1}{2}(\lambda_4 \pm i \lambda_5)$
which correspond to the $s \rightarrow u$ ($V_+$)
and $u \rightarrow s$ ($V_-$) transitions.

According to the $SU_F(3)$ symmetry 
the quark form factors $g_A^q$ and $g_P^q$ are the same
as in the $SU_F(2)$ sector.
Therefore, once the model is fixed in the $SU_F(2)$ sector,
the results for the $SU_F(3)$ sector represent 
predictions of the model.

\subsection{Lattice QCD regime}
\label{secLattice}

We discuss here how we can perform 
the extension of the covariant spectator 
quark model to the lattice QCD regime.
This extension was already done in the past 
for electromagnetic transitions~\cite{Lattice,LatticeD,OctetFF,Medium}.

In the previous sections we have shown 
that the wave functions 
of the baryons, including the radial part    
$\psi_S$ and $\psi_P$, can be written 
in terms of the baryon mass $M_B$.
In Eq.~(\ref{eqChiN}) we have presented 
the  parametrization for the  
nucleon, but the generalization for 
other baryons can be done by replacing 
$M$ by the corresponding baryon mass $M_B$.
We have also discussed  how the quark axial current 
$j_{Aq}^\mu$ given by Eq.~(\ref{eqJAq})
can be defined in terms of the axial form factors 
$g_A^q$ and $g_{P}^q$, and the nucleon mass $M$.
We have also concluded that the quark axial form factors can be 
represented by a VMD parametrization 
in terms of the mass of the vector meson mass ($\rho$ meson), 
and an effective heavy meson 
with mass $M_h = 2 M$.

Since the bare contribution for 
the electromagnetic and the axial form factors can be 
completely determined by the masses of the baryon ($M_B$),  
the $\rho$ mass ($m_\rho$) and the nucleon mass ($M$),
we extend the model for the lattice QCD regime 
replacing these masses by the 
corresponding masses in the lattice QCD simulations.
The remaining parameters in the quark current 
and in radial wave functions
are the same as those used in the physical limit.
As for $m_\rho$,  since the value is not always provided 
in the lattice QCD simulations,
we use the following expression based on the lattice studies
made in Ref.~\cite{Leinweber01},
\be
m_\rho= a_0 + a_2 m_\pi^2,
\label{eqMrho}
\ee
where $m_\pi$ is the value of the 
pion mass   used in the lattice QCD simulation, 
$a_0=0.766$ GeV and $a_2= 0.427$ GeV$^{-1}$.

With the procedure explained above, we can associate our model 
with a lattice QCD simulation with the same $m_\pi$ 
(lattice QCD regime).

\section{Valence quark contributions for the nucleon}
\label{secValence}

We present in this section the expressions for the nucleon
axial-vector and induced pseudoscalar form factors
associated with the different valence quark contributions of    
the nucleon wave function.
Since the nucleon wave function~(\ref{eqPsiN})
is a combination of the $S$- and $P$-states,  
the contributions for the axial current~(\ref{eqJ5})
can be decomposed into an $S$-state term ($\propto n_S^2$),
an $S \to P$ term ($\propto n_S n_P$)
and a $P$-state term ($\propto n_P^2$),
as presented in the next sections.
The individual contributions for the form factors 
associated with the transitions between the $S$- and $P$-states, 
$S \to S$, $S \leftrightarrow P$ and $P \to P$,  
will be represented by the upper indices
$SS$, $SP$ and $PP$, respectively.
Note that, the transition between
the $S$- and $P$-states is possible
due to the structure of the axial current, 
$\gamma^\mu \gamma_5$.   
However, in the limit  $Q^2=0$ 
the $SP$ contribution vanishes
(the same happens for the current given
by a Dirac term $\gamma^\mu$).

In order to present the results in a covariant form, 
we introduce some useful notation below.
For the average momentum between the initial
and final momenta we use
\ba
P^\prime = \frac{1}{2}(P_+ + P_-). 
\ea
Then $(P')^2$ can be written as
$(P')^2 =M^2 (1 + \tau)$, with 
$\tau= \frac{Q^2}{4M^2}$.
It is also convenient to define 
\ba
\tilde k^\prime = k - \frac{P' \cdot k}{(P')^2}P^\prime.
\ea
In the Breit frame $\tilde k'$ is reduced
to the spacial component, $\tilde k'= (0,{\bf k})$,
and $(\tilde k')^2=- {\bf k}^2$.

The analytic expressions for the 
transition form factors, to be given next,
can be expressed in terms of a few invariant 
integrals defined by the factors:
\ba
& &a= \frac{P' \cdot k}{M},
\label{eqA} \\
& &c_0=    \frac{(P^\prime  \cdot k)^2}{(P')^2}, \\
& &c_1=   - (\tilde k^\prime )^2, \\
& &c_2 = %- \frac{( q \cdot k^\prime)^2}{q^2} =
- \frac{( q \cdot k)^2}{q^2}.
\label{eqC2}
\ea
In the Breit frame one has $a= \sqrt{1 + \tau} E_D$,
$c_0= E_D^2$, 
$c_1 = {\bf k}^2$ and $c_2= k_z^2$,
where $E_D=\sqrt{m_D^2 + {\bf k}^2}$ and $k_z$
is the $z$-component of
the three-momentum ${\bf k}$.

\subsection{$S$-state}

The $S$-state contribution for the form factors 
can be expressed as
\ba
%\hspace{-1cm}
& &
G_A^{SS}(Q^2) = \frac{5}{3} n_S^2 g_A^q(Q^2) B_0(Q^2),\\
%\hspace{-1cm}
& &
G_P^{SS}(Q^2) = \frac{5}{3} n_S^2  g_P^q(Q^2) B_0(Q^2),
\ea
where $n_S^2= 1-n_P^2$, and
\ba
B_0(Q^2)=  \int_k \psi_S(P_+,k) \psi_S(P_-,k),
\ea
is the nucleon {\it Body form factor}.
This calculation can be done using 
the $S$-state model from Ref.~\cite{Nucleon}.

In the limit $n_S^2=1$ and $g_A^q(0)=1$ we obtain the
result of the static quark model
(or naive quark model) $G_A(0)=\frac{5}{3}$
\cite{Jenkins91,Thomas84,Hayne82,Isgur88,NRQM,Chang13,BCano06,Lorce08}.
The inclusion of relativistic corrections 
on  nonrelativistic models 
reduces the value of $G_A(0)= \frac{5}{3}$,
to a value closer to the 
experimental value 
$G_A(0) \simeq 1.27$~\cite{Chang13,Hayne82,Isgur88,Boffi02}.

The result $G_A(0)= \frac{5}{3} g_A^q(0)$, implies  
that if we want to explain the 
experimental value within a  
simplified model, 
we need to admit  
that the axial charge of the quark 
is smaller than  the unit, $g_A^q(0) < 1$,
breaking the connection with the electromagnetic 
isovector current [$g_A^q(0) \ne f_{1-}(0)$].
Similar effects were already observed 
in other frameworks~\cite{Chang13,Liu14,Hayne82,BCano06,BCano03,Yamanaka14,Eichmann12}.
Calculations based on the  Dyson-Schwinger formalism
suggests that the quark axial-vector coupling $g_A^q(0)$ 
is reduced relatively to $g_A^q(0)=1$, 
due to the gluon dressing of the quarks~\cite{Chang13,Yamanaka14,Eichmann12}.

\subsection{Transition between $S$- and $P$-states}

For the $S$- to $P$-state and the $P$- to $S$-state transitions 
we obtain, using %$n_{SP}=n_P \sqrt{1- n_P^2}$, 
$n_{SP}= n_S n_P$,
\ba
G_A^{SP} (Q^2) &=& - \frac{10}{3}n_{SP} \frac{\tau}{1+\tau}
g_A^q(Q^2) B_1(Q^2), \\
G_P^{SP} (Q^2) &=& - \frac{10}{3} n_{SP}
%\nonumber \\ %& & \times
\left[ \frac{1}{1+\tau} g_A^q(Q^2) + g_P(Q^2)  
\right] B_1(Q^2), \nonumber \\
\ea
where 
\ba
B_1(Q^2)=  \int_k \frac{P' \cdot k}{M}
\psi_P(P_+,k) \psi_S(P_-,k).
\ea
Note that in the Breit frame
$\frac{P'  \cdot k}{M} = E_D \frac{E_N}{M}$,
with \mbox{$E_N = M \sqrt{1 + \tau}$}
being the nucleon energy (initial or final state).

\subsection{$P$-state}

The results for the $P$- to $P$-state transition is given by 
\ba
G_A^{PP} (Q^2) &=&  \frac{4}{3}
n_{P}^2 \; g_A^q(Q^2) \nonumber \\
& & \times
\left[ \tau B_2(Q^2) - (1+ \tau) B_4(Q^2)
\right], \nonumber \\
\\
G_P^{PP} (Q^2) &=&  \frac{5}{3} n_{P}^2  \; g_A^q(Q^2)
\nonumber \\
& & \times
\left[ \frac{B_5(Q^2)}{\tau} +  2 B_2(Q^2) - 2 B_4 (Q^2) 
\right] \nonumber \\
& & +
\frac{5}{3} n_{P}^2  \; g_P(Q^2)
\nonumber \\
& & \times
\left[  \tau B_2(Q^2) +B_3(Q^2) - (2+ \tau)B_4(Q^2)
\right], \nonumber \\
\ea
where
\ba
& &
B_2(Q^2)= \int_k \frac{(P' \cdot k)^2}{(P')^2} \psi_P(P_+,k) \psi_P(P_-,k),
\label{eqInt2}
\\
& &
B_3(Q^2)=  \int_k
(- \tilde k^{\prime 2}) \psi_P(P_+,k) \psi_P(P_-,k), 
\label{eqInt3}
\\
& &
B_4 (Q^2)=  \int_k \frac{ (q \cdot k)^2}{Q^2} \psi_P(P_+,k) \psi_P(P_-,k),
\label{eqInt4} \\
& &
B_5 (Q^2)=  \int_k %\frac{ (q \cdot k)^2}{Q^2} 
(2 S_3)
\psi_P(P_+,k) \psi_P(P_-,k).
\label{eqInt5}
\ea
In the last equation,   
$2  S_3 = \frac{q^2 \tilde k^{\prime 2} - 3(q \cdot k)^2 }{q^2}$.
Note that the integrals~(\ref{eqInt2})-(\ref{eqInt5}) can be reduced to  
simpler forms in the Breit frame,
according to Eqs.~(\ref{eqA})-(\ref{eqC2}).

The function $B_5$ can also be represented as 
$B_5= B_3 - 3 B_4$. 
However, it is convenient 
to define $B_5$ as an independent function, 
since $B_5 \propto \tau$ for small $\tau$,
which implies that $\frac{B_5}{\tau} \to$ constant  
when $Q^2 \to 0$. 
This property is the consequence 
of the result, $S_3= |{\bf k}| \sqrt{\frac{16 \pi}{5}}Y_{20}(\hat k)$
when $Q^2 \to 0$.
Note that the factor $Y_{20}({\hat k})$ 
is  associated with $L=2$ transitions 
between the two $P$-state components in the nucleon wave function.

\subsection{Summary of the valence quark contributions}
\label{secSummary}

We discuss now the total contribution
from the baryon core (bare) 
given by the sum of the components 
presented below:
\ba
\hspace{-1cm}
& &
G_A^B(Q^2)=
G_A^{SS}(Q^2) + G_A^{SP}(Q^2) +G_A^{PP}(Q^2), 
\label{eqGAspec}\\
\hspace{-1cm}
& &
G_P^B(Q^2)=
G_P^{SS}(Q^2) + G_P^{SP}(Q^2) +G_P^{PP}(Q^2).
\label{eqGPspec}
\ea

Using the result for $G_A^B$,
we can estimate the amount of $P$-state mixture $n_P$
in terms of the value of $G_A^B(0)$,  
in the case where there are no meson cloud contributions.
From the normalization of the radial wave functions,  
we can conclude that $B_1(0)=B_3(0)=3 B_4(0)$.
Since at $Q^2=0$ the $SP$ term vanishes,
we can conclude that,
\ba
G_A^B(0)= \frac{15-19 n_P^2}{9} g_A^q(0).
\ea
Then, if $g_A^q(0)=1$ as 
already discussed,
we may estimate the $P$-state mixture
in terms of the valence quark contribution $G_A^B(0)$ as 
\mbox{$n_P= \pm \sqrt{[15 -9 G_A^B(0)]/19}$.}
When \mbox{$G_A^B(0) \ =1.1,...,1.2$} we obtain 
\mbox{$n_P= \pm 0.47,...,\pm 0.52$.}

The improvement of the agreement 
with the data due to the inclusion 
of angular momentum components 
beyond the $S$-state approximation, 
was observed long time ago 
in the context of nonrelativistic quark models~\cite{Hayne82}
and in CBM~\cite{Thomas84}.
In CBM the reduction of $G_A(0)$ from the value  5/3
is also a consequence of $P$-state,
associated with the lower components of 
the quark Dirac spinors~\cite{Thomas84}.

It is important to recall at this point 
that in lattice QCD simulations 
with large $m_\pi$, the contribution 
of the meson cloud effects to the form factors 
are expected to be small, therefore $G_A \simeq G_A^B$.
It can be then very interesting to compare
the results of the extended model for the lattice QCD regime
(without meson cloud)
as discussed in Sec.~\ref{secLattice},
directly with the lattice QCD data.

\begin{figure}[t]
\vspace{.5cm}
\centerline{
\mbox{
\includegraphics[width=3.0in]{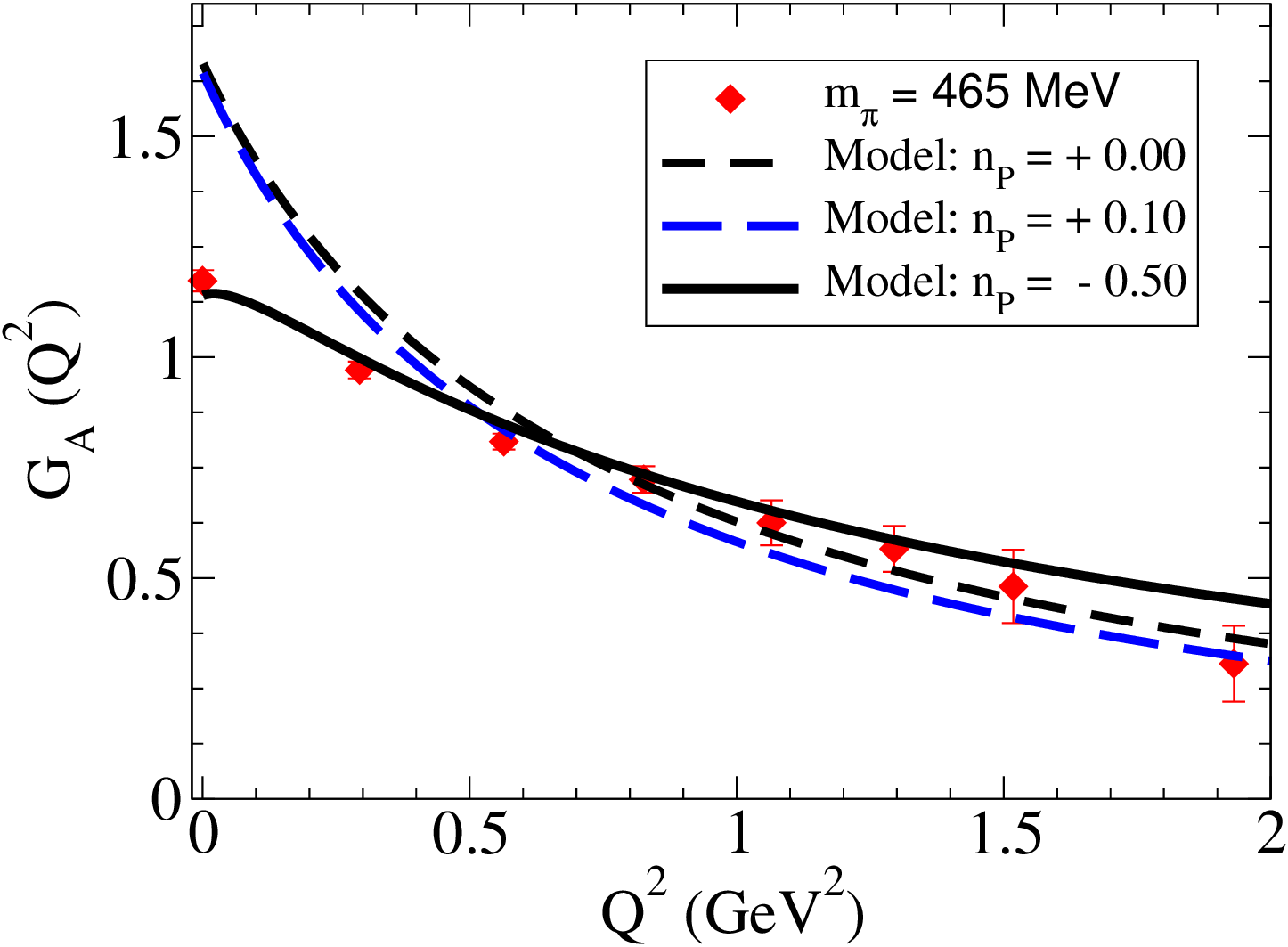}
}}
\caption{\footnotesize{
Model result of $G_A^B$ in the lattice QCD regime
($m_\pi = 465$ MeV) for values of $n_P=0.0, \, 0.1$, and $-0.5$.
Lattice data are from Ref.~\cite{Alexandrou11a}.
}}
\label{figGAmod1_lat}
\end{figure}

In Fig.~\ref{figGAmod1_lat} we compare
the results of our model extended for the lattice QCD regime  
with the lattice QCD simulation for  
$m_\pi = 465$ MeV from Ref.~\cite{Alexandrou11a}.
We can see that the pure $S$-state ($n_P=0$; short-dashed line),
fails to describe the data for the small $Q^2$ region,
although it approaches the lattice data for large $Q^2$.
The result for $n_P=0.1$ (long dash line)
overestimates the data in the small $Q^2$ region, 
while in the large $Q^2$ region it underestimates.
Finally, the result for $n_P=-0.5$ (solid line)
gives an excellent description of the lattice QCD data.
This suggests that a mixture between the $S$- and $P$-states
of about  25\%  with a negative coefficient  ($n_P \simeq -0.5$)
is adequate to describe the lattice QCD data for 
$G_A$ obtained with large $m_\pi$.

The result of the systematic study of the
lattice QCD data for the range $m_\pi=350$--$500$ MeV,
for both form factors, $G_A$ and $G_P$,  
will be presented in Sec.~\ref{secResults-Nucleon}.
In the next section we extend the formalism 
developed here for the nucleon to the 
octet baryons using the $SU_F(3)$ flavor symmetry  
at the quark level.

Since in the physical regime, contrarily 
to the lattice QCD regime with large $m_\pi$,
the effect of the meson cloud 
(in particular the pion cloud) 
cannot be ignored for the nucleon as well as  for the other 
octet baryon members, in Sec.~\ref{secMesonCloud} we discuss how the
results for the physical case can be corrected 
by the effect of the meson cloud on the octet baryon wave functions.

\section{Valence quark contributions for the octet baryons}
\label{secValenceOctet}

As in the case of the nucleon, 
we can calculate the axial form factors  
involving the other octet baryon members 
using the wave functions given in Sec.~\ref{secOctetWF}.
There are four main differences 
relatively to the nucleon case:
\begin{itemize}
\item
we have now the $u \leftrightarrow s$ transitions,
\item
the nucleon isospin wave function, $\phi_I^0$ and $\phi_I^1$, 
 are replaced by the anti-symmetric $\left| M_A\right>_B$ 
and symmetric $\left| M_S\right>_B$ flavor 
wave functions in $SU(3)_F$, 
as displayed in Table~\ref{tabPHI},
\item
the radial wave functions $\psi_S$ 
and $\psi_P$ are replaced by   
functions with different momentum range parameters, 
\item
there are in general a mass difference 
between the initial ($M_B$) and the final ($M_{B'}$) 
baryon states 
(for the nucleon the difference between 
the proton and neutron mass is negligible).
\end{itemize}

Contrarily to the nucleon case ($n \to p$ transition)
the difference in masses between the initial and final states
can originate in principle 
additional terms to the structure of 
transition axial current (\ref{eqOctetAxial}),
besides corrections dependent on the mass difference
to the form factors $G_A$ and $G_P$. 
In this work as a first approximation 
we consider the limit $M_{B'}= M_B$ 
and replace those masses by the average $M_{BB'}$.

Except for the value of the mass ($M$ or $M_{BB'}$)
we can calculate the results of the axial form factors 
for the octet, using the results for 
the nucleon modified by the flavor wave functions.
Since the results for the nucleon can 
be divided into the diquark spin-0 contribution 
that goes with the mixed anti-symmetric flavor wave function 
$\left| M_A\right>_B$, and the  diquark spin-1 contribution 
that goes with the mixed symmetric flavor wave function 
$\left| M_S\right>_B$.  
We define
the symmetric ($S$) and anti-symmetric ($A$) transition coefficients,
\ba
& &
f_{X}^A = 
{_{B'}}\!\left< M_A | X | M_A \right> _B, \\
& &
f_{X}^S = 
{_{B'}}\!\left< M_S | X | M_S \right> _B, 
\ea 
where the operator $X$ can be either $I_0$, $I_\pm$ or  $V_\pm$. 
The results of the octet transition coefficients are 
presented in Table~\ref{tabFAS}. 
The results for the neutral transitions  $B \to B$
defined by $X = I_0 \equiv \lambda_3$ 
are presented in Table~\ref{tabElastic} 
for $B=N,\Sigma$, and $\Xi$.
For $\Lambda$ and $\Sigma^0$ 
the coefficients are both zero,
therefore $G_A(Q^2)=0$.
As explained in caption of Table~\ref{tabElastic},  
the values are 
redefined to be independent of the charges of the 
baryons as in the nucleon case.

\begin{table*}[t]
\begin{center}
\begin{tabular}{c c | c c c c c}
\hline
\hline
  &   & $B \to B'$&   $f_{X}^A$ & $f_{X}^S$ & $[G_A^B(0)]_{n_P=0}$ & $G_A^B (Q^2)$\\
\hline 
$\Delta I = 1 \hspace{1em}$ 
&  ($I_+$) & \hspace{2em}$n \to p$\hspace{2em}                  
      &$1$ &$-\frac{1}{3}$ &  $\frac{5}{3}$ &   $G_{A,N}^B$\\
   &  $(I_\mp)$          &\hspace{2em}$\Sigma^{\pm} \to \Lambda$\hspace{2em}  
                         &$\pm\frac{1}{\sqrt{6}}$ &$\mp\frac{1}{\sqrt{6}}$ 
&  $\pm \frac{\sqrt{6}}{3}$ &  $\pm \frac{\sqrt{6}}{5} G_{A,N}^B$\\
   &   $(I_+)$             &\hspace{2em}$\Sigma^{-} \to \Sigma^0$\hspace{2em} 
                                &$\frac{1}{\sqrt{2}}$ &$\frac{1}{3\sqrt{2}}$
& $\frac{2 \sqrt{2}}{3}$   & $\frac{2 \sqrt{2}}{5} G_{A,N}^B$\\
   &  $(I_+)$            &\hspace{2em}$\Xi^- \to \Xi^0$\hspace{2em} 
  &$0$   &$\frac{2}{3}$ &   $- \frac{1}{3}$
& $- \frac{1}{5} G_{A,N}^B$\\
\hline
$\Delta S = 1\hspace{1em} $ & 
($V_+$)  &\hspace{2em}$\Lambda \to p$\hspace{2em} 
                                & 
  $-\frac{2}{\sqrt{6}}$ & $0$ &  $-\sqrt{\frac{3}{2}}$  & 
$-\frac{3\sqrt{3}}{5\sqrt{2}} G_{A,N}^B$\\
   & $(V_+)$                      &\hspace{2em}$\Sigma^{-} \to n$\hspace{2em} 
          &$0$ &  $-\frac{2}{3}$ & $\frac{1}{3}$ &
$ \frac{1}{5} G_{A,N}^B$ \\
   & $(V_+)$                      &\hspace{2em}$\Sigma^{0} \to p$\hspace{2em} 
          &$0$ &  $-\frac{\sqrt{2}}{3}$ & $\frac{1}{3\sqrt{2}}$ &
$ \frac{1}{5 \sqrt{2}} G_{A,N}^B$ \\

   & $(V_+)$                 &\hspace{2em}$\Xi^- \to \Lambda$\hspace{2em} 
                                &$-\frac{1}{\sqrt{6}}$ &$-\frac{1}{\sqrt{6}}$ 
&  $- \frac{1}{\sqrt{6}}$ & $- \frac{\sqrt{3}}{5 \sqrt{2}}  G_{A,N}^B$ \\
   &  ($V_+$)                 &\hspace{2em}$\Xi^- \to \Sigma^0$\hspace{2em} 
           &$\frac{1}{\sqrt{2}}$ &$-\frac{1}{3\sqrt{2}}$ 
&  $\frac{5}{3 \sqrt{2}}$   &  $\frac{1}{\sqrt{2}}  G_{A,N}^B$ \\
    & ($V_+$)                 &\hspace{2em}$\Xi^0 \to \Sigma^+$\hspace{2em} 
                 &$1$ &$-\frac{1}{3}$  &
$\frac{5}{3}$ & $ G_{A,N}^B$ \\
\hline
\hline
\end{tabular}
\end{center}
\caption{Coefficients $f_{I}^{S,A}$ and $f_{V}^{S,A}$ 
for the octet baryon transitions.}
\label{tabFAS}
\end{table*}

\begin{table*}[t]
\begin{center}
\begin{tabular}{c |c c c c}
\hline
\hline
$B$ \hspace{2em} & \hspace{2em}$f_X^A$ &\hspace{2em} $f_X^S$ & \hspace{2em} 
$[G_A^B(0)]_{n_P=0}$ &  \hspace{2em}$G_A^B (Q^2)$\\
\hline
$N$ \hspace{2em}   & \hspace{2em} $1$ & \hspace{2em} $- \frac{1}{3}$ & 
\hspace{2em}$\frac{5}{3}$ &  \hspace{2em}$G_{A,N}^B$ \\                    
%%$\Lambda$ & $0$ & $0$ & $0$ \\ 
$\Sigma$ \hspace{2em}  & \hspace{2em} $1$ &  \hspace{2em}$\frac{1}{3}$  &  
\hspace{2em} $\frac{4}{3}$ & \hspace{2em} $\frac{4}{5} G_{A,N}^B$\\
$\Xi$ \hspace{2em}  &  \hspace{2em} $0$ &  \hspace{2em} $\frac{2}{3}$  &   
\hspace{2em}$-\frac{1}{3}$ &   \hspace{2em} $- \frac{1}{3} G_{A,N}^B$\\
\hline
\hline
\end{tabular}
\end{center}
\caption{
Coefficients $f_{I_0}^{S,A}$ for the neutral transitions. 
In order to compare with the literature 
we correct the results for $f_{I_0}^{S,A}$
by an isospin factor.
In the case of the nucleon and $\Xi$ the factor 
is the isospin projection 
of the baryon ($+$ for $p$, $\Xi^0$ and $-$ for $n$, $\Xi^-$).
For the  $\Sigma$ case the factor is taken as the $\Sigma$ charge ($+,0,-$).}
\label{tabElastic}
\end{table*}

Using the new notation we can re-write 
the results for the nucleon 
in terms of the factor
\ba
\frac{3}{2} \left( f_{X}^A - \frac{1}{3} f_{X}^S \right)
= \frac{5}{3},
\ea
since the nucleon case we have 
$f_{I+}^A= 1$ 
and $f_{I+}^S= -\frac{1}{3}$.
The difference in the calculation 
relatively to the nucleon case, 
apart the mass, is that the factor due 
to the spin structure
in the nucleon case given by 
$f_{I+}^A - \frac{1}{3} f_{I+}^S= \frac{10}{9}$,  
should be replaced by the factor 
$f_{X}^A - \frac{1}{3} f_{X}^S$ in the general case.
Therefore, we can obtain the results 
of the axial form factors for the octet baryons  
multiplying the nucleon results by 
$\frac{9}{10} (f_{X}^A - \frac{1}{3} f_{X}^S)$.

Then, assuming that the baryon wave functions of 
$B$ and $B'$ are also defined with the mixture coefficients 
$n_S$ and $n_P$, we can write
the transition form factors with 
$n_S$, $n_{SP}$ and $n_P$ in general: 
\ba
G_A^B(Q^2) &=&  
g_A^q {\cal F}
\left\{
\frac{3}{2} n_{S}^{2} B_0 
- 3 n_{SP}  \frac{\tau}{1 + \tau}B_1 \right. \nonumber \\
& &
\left. + \frac{6}{5} n_P^{2} 
\left[ \tau B_2 -(1+ \tau) B_4 \right]
%\left[ \frac{B_5}{\tau} + 2(B_2-B_4) \right]
\right\},
\label{eqGAoct}
\ea
\ba
G_P^B(Q^2)&=& 
g_A^q {\cal F}
\left\{
 -  3 n_{SP} \frac{1}{1 + \tau} B_1 \right. \nonumber \\
& &
\left. + \frac{3}{2} n_P^{2} 
\left[ \frac{B_5}{\tau} + 2(B_2-B_4) \right]
\right\} \nonumber \\
& &
+ \frac{M_{BB'}}{M}g_P^q {\cal F}
\left\{
\frac{3}{2} n_{S}^{2} B_0 - 
3 n_{SP} B_1 \right. \nonumber \\
& &
\left. + \frac{3}{2} n_P^{2} 
\left[ \tau B_2 +  B_3   - (2 + \tau)B_4 \right]
\right\},
\label{eqGPoct}
\ea
where one has now 
$\tau= \frac{Q^2}{4 M_{BB'}^2}$, and  
\ba
{\cal F}= \left(f_{X}^A - \frac{1}{3} f_{X}^S \right).
\ea
The effect of the mass ($M$ or $M_{BB'}$) 
appears in the functions $B_i$ ($i=0,...,5$).

Note in particular 
in Eq.~(\ref{eqGPoct}) the factor $\frac{M_{BB'}}{M}$ 
which corrects the quark form factor $g_P^q$
relatively to the nucleon case.
This factor is the consequence 
of the definition of the quark axial current 
given by Eq.~(\ref{eqJAq}), in terms of 
the nucleon mass $M$ for {\it all the baryons}, in contrast   
to the baryonic transition 
currents~(\ref{eqOctetAxial}) that depend on $M_{BB'}$.

For a future discussion it is important to note 
that the second term  in the r.h.s.~in~Eq.~(\ref{eqGPoct}) 
for $G_P^B$, the term proportional to $g_A^q n_P^2$, 
has a dependence on $Q^2$ 
that is similar to the first term  
of the r.h.s.~in Eq.~(\ref{eqGAoct}) for $G_A^B$, for large $Q^2$.
Both terms scale as $1/Q^4$ for very large $Q^2$.
This behavior is a consequence 
of the choice made for the radial wave functions,
that enter in the definitions of the functions $B_i(Q^2) (i=0,...,5)$,
and also from the result $g_A^q(Q^2) \simeq \mbox{constant}$ 
in the large $Q^2$ limit.
Recall that the radial wave functions 
are chosen with the form of the nucleon radial wave function,
and that the nucleon radial wave function 
is parametrized in order to 
describe the nucleon electromagnetic form factors for large $Q^2$.
The consequence of this choice is that 
one has $B_0(Q^2) \propto 1/Q^4$ for large $Q^2$,
apart from some logarithmic corrections~\cite{NDelta}.

In the third columns of Tables~\ref{tabFAS} and \ref{tabElastic}  
we present the result for $G_A^B(0)$ for $n_P=0$.
It is interesting to look to those values,  
because they agree with the results 
of the static quark model (naive quark model),
obtained also in the $S$-state limit.
As already discussed, the covariant spectator 
quark model improves the result for the nucleon at $Q^2=0$ 
when we include a $P$-state mixture with the value $n_P  \simeq -0.5$.

In the last columns of Tables~\ref{tabFAS} 
and \ref{tabElastic} 
we write the expressions 
obtained for the form factor $G_A^B(Q^2)$ 
in terms of the result for the nucleon 
($n \to p$ transition) in the limit $M=M_{BB'}$,
represented by the function $G_{A,N}^B(Q^2)$.
Since the results for the octet baryons   
for finite $Q^2$ are calculated 
under the $SU_F(3)$ symmetry,
but the symmetry is broken by the mass $M_B$
(dependent on  the isospin of the baryon), 
it will be interesting to see if the relations 
from Tables~\ref{tabFAS} and~\ref{tabElastic} 
expressed in terms of the result for the nucleon, $G_{A,N}^B$,
are a good approximation for the octet baryons or not.
(The parametrizations of the radial wave 
functions are also different for the octet baryons.)
These issues will be discussed in Sec.~\ref{secResults-Octet}.

Note about the function $G_A^B(Q^2)$  that 
the results are no longer related with 
the static quark model limit,
since we now include a $P$-state mixture.
In  order to better understand the difference 
between the two models,  
the static quark model and a model with the $P$-state, 
we compare the results for the case where $n_P  \simeq -0.5$.
In the static quark model we obtain 
for the nucleon 
$G_A^B(0) = \frac{5}{3} \simeq 1.67$.
In this case the  expected result for the $\Xi^- \to \Xi^0$ 
form factor is $G_A^B(0)=-0.33$.
If we consider instead the model with the $P$-state  
discussed in Sec.~\ref{secValence}, assuming  
$G_A^B(0) \simeq 1.1$ (lattice case),
we obtain $G_A^B(0)= -0.22$ for 
%$\Xi^- \to \Xi^0$.
the same transition.
There is therefore a significant 
deviation from the $SU(6)$.

In Sec.~\ref{secSU6} we compare 
the results of our valence quark model with those of the 
$SU(3)$ baryon-meson model.

\section{Meson cloud effects for the  
octet baryon axial form factors}
\label{secMesonCloud}

We discuss in this section how the meson cloud effects 
can be taken into account 
in the octet baryon axial form factors.
We start the nucleon case. 
Next, we explain how the method can be extended 
for the other octet baryon members.

\subsection{Meson cloud dressing of the nucleon}

Since the pion cloud is expected to be 
the dominant contribution in the meson cloud,
we could in principle replace meson cloud  by pion cloud.
We will keep however the discussions general,
aiming for the generalization for the octet baryons.

The meson cloud contribution can be included 
in the nucleon wave function using the 
nucleon state expanded as   
\ba
\left| N \right> =
\sqrt{Z_N} \large[
\left| 3q \right>  +  c_N
\left| {\rm MC} \right>  
\large],
\label{eqNucleonWF}
\ea
where  $\sqrt{Z_N}$ is the normalization constant, 
$\left| 3q \right> $ gives the nucleon bare 
(three valence quark)  wave function part, 
and $c_N \left| {\rm MC} \right>$ 
represents the meson cloud component 
associated with baryon-meson states.
The coefficient $c_N$ is determined by 
the normalization of the state
[$Z_N( 1 + c_N^2)=1$, if the meson cloud component is 
normalized to the unit].
The component $\left| 3q \right> $ 
has already been discussed in the previous section.

Since the nucleon wave function associated with the 
state (\ref{eqNucleonWF}) includes states beyond the valence quark core, 
we simply refer the state $\left| N \right>$ as the 
physical nucleon state~\cite{Thomas84}.
Note however, that although the higher 
order states such as baryon-meson-meson states  
may also be included in the nucleon wave function 
by including the corresponding states in Eq.~(\ref{eqNucleonWF}),
we assume that the baryon-meson states give the more 
relevant corrections to the valence quark core, and 
ignore the higher states in this work.

We discuss now how the results obtained 
for the form factors due to the valence quark component 
are modified by the existence of 
the component $\left| {\rm MC} \right>$.
In a framework where the baryon-meson interactions 
are defined by an underlying theory,  
we can calculate the normalization constant 
${Z_N}$ using 
the derivative of the nucleon self-energy
(nucleon dressed by the meson cloud)~\cite{OctetFF,Medium,OctetMM}.

Once determined ${Z_N}$, we can calculate 
the effective contribution of the valence quark component
for a given process, 
including the factor $\sqrt{Z_N}$ 
associated with the component $\left| 3q \right> $.
Since the valence quark component itself 
is normalized to unity
and there is a meson cloud component, 
we need to correct the 
bare contribution when we compute the effect of the valence quarks 
by the probability of finding a bare nucleon state (three valence quark) 
in the physical nucleon  state $\left| N \right>$,
which will reduce the bare contribution by the factor 
$(\sqrt{Z_N})^2$ due to the presence of the meson cloud.

For an easier understanding of this 
normalization procedure, 
we discuss the calculation of the 
proton charge $e$, defined by the  
proton Dirac form factor 
in the limit $Q^2=0$, $F_1(0)$.
Since the determination of the nucleon 
elastic form factors depends on the two 
nucleon wave functions associated 
with the initial and the final states,
the valence quark contribution for $F_1(0)$ which is unity, 
should be modified by $Z_N=\sqrt{Z_N} \sqrt{Z_N}$
due to the normalization of the valence 
quark component of the wave function.
Therefore only the fraction $Z_N$
contributes to the proton charge.
The remaining contribution $(1-Z_N)e$ is 
due to the meson cloud.
Considering as an example the model that we will 
present in Sec.~\ref{secResults-Nucleon}
with $Z_N=0.73$, we can conclude that only about 73\%  
of the proton's charge is due to the valence quarks 
(27\% of meson cloud).

To summarize, the contribution 
of the valence quarks for the 
axial form factor $G_A$ can be determined 
from the result obtained by the bare 
contribution $G_A^B$, multiplied by $Z_N$:
\ba
G_A^B (Q^2) \to Z_N G_A^B (Q^2).
\label{eqZNestimate}
\ea
See Refs~\cite{OctetFF,Medium,OctetMM} 
for more details.

To avoid confusion between the result for $G_A$ obtained 
in the case where only the valence quarks are relevant 
(so far represented as $G_A^B$),
and the case with the meson cloud 
as in the physical case, we define 
\ba
\tilde G_A^B (Q^2) = Z_N G_A^B(Q^2), 
\label{eqBarePhys}
\ea  
as the effective
contribution of the bare core 
for the axial-vector form factor.
Using this notation we can rewrite Eq.~(\ref{eqGAdecomp}) as  
\ba
G_A(Q^2)= \tilde G_A^B (Q^2) + G_A^{MC} (Q^2),
\ea
where $G_A^{MC}$ represents, as before, the contribution 
of the meson cloud.
Note that, as mentioned earlier, if we have a model for 
$\tilde G_A^B$, we can estimate $G_A^{MC}$
phenomenologically replacing $G_A$ by some 
parametrization of the experimental data.

In this work, instead of calculating $Z_N$
from an underlying theory we chose to 
use the experimental data for $G_A$ to estimate the amount 
of the meson cloud in the nucleon system.

Our method to estimate $Z_N$ is the following:
i) first, we calibrate our valence quark model 
by the lattice QCD data with large $m_\pi$,  
ii) next, we extrapolate the result
for the physical limit to obtain $G_A^B$,
iii) finally, we use Eq.~(\ref{eqBarePhys}) 
with $\tilde G_B$ replaced by a phenomenological 
parametrization of the data for $Q^2 > 1$ GeV$^2$,
a region where the meson cloud effects are small  
to calculate $Z_N$.

Note that the estimate of the factor $Z_N$ 
by the nucleon $G_A$ form factor data,
instead  of just by the 
nucleon electromagnetic form factor data,
provides in principle a more consistent 
estimate of the meson cloud component
in the physical nucleon state $\left| N\right>$.

\subsection{Meson cloud dressing of octet baryons}

We assume that for the other octet baryon members 
we can write also 
an equation similar to Eq.~(\ref{eqNucleonWF}),
that includes a coefficient $c_B$, 
that is related with the normalization constant $\sqrt{Z_B}$.
For the other octet baryon members, however, 
it is not possible to estimate $Z_B$ directly from 
the data, since there are no data for finite $Q^2$.
We cannot use the method based on Eq.~(\ref{eqZNestimate})
to estimate $Z_B$.

Therefore, we rely on an alternative method to 
estimate the normalization factor $Z_B$
for the other octet baryon members.
The method is based on the similarity 
between the contribution of 
the meson cloud for the nucleon 
in our model and CBM~\cite{Shanahan13}.

In the formalism of the 
covariant spectator quark model we can 
represent~\cite{OctetFF,OctetMM,Medium,LambdaSigma}
\ba
Z_B= \frac{1}{1 + a_B b_1},
\ea
where $b_1$ is a parameter that 
establishes the magnitude of the 
meson cloud in the nucleon system, 
and $a_B$ is a factor dependent on the baryon flavor,
constrained by $a_N=1$.
The coefficient $a_B$ is defined by $c_B^2= a_B b_1$,
according to Eq.~(\ref{eqNucleonWF}).
Since $a_N=1$, in the case of the nucleon, 
$Z_N$ is determined directly by $b_1$ and vice versa.

We compare our result for $Z_N$
with the result of CBM~\cite{Shanahan13}.
Our result for $Z_N$, presented in Sec.~\ref{secResults-Nucleon}
is $Z_N=0.7343$.
The results from CBM is  $Z_N= 0.7114$.
We conclude therefore that
the effect of the meson cloud 
in the nucleon wave function is very similar 
in both models.

Assuming that the meson cloud contribution  
for the octet baryon members keep the same proportion 
for the nucleon as in CBM, we can determine 
$a_B$, and consequently calculate $Z_B$.
Since in the weak transitions 
between octet baryon members 
may involve  different isospin multiplets  
in the initial and in final states,  
it is more convenient to  present the results for $\sqrt{Z_B}$.
The values of $\sqrt{Z_B}$ determined by the method 
described above are presented in Table~\ref{tabZB}.

%%%%%%

\begin{table}[t]
\begin{center}
\begin{tabular}{c  c }
\hline
\hline
$\sqrt{Z_N}=$ & 0.8569 \\
$\sqrt{Z_\Lambda}=$ & 0.8822 \\
$\sqrt{Z_\Sigma}=$ & 0.8751 \\
$\sqrt{Z_\Xi}=$ & 0.9019 \\
\hline
\hline
\end{tabular}
\end{center}
\caption{Normalization factors $\sqrt{Z_B}$
of the octet baryon wave functions as 
the result of the meson cloud dressing.}
\label{tabZB}
\end{table}

\section{$SU(3)$ baryon-meson model}
\label{secSU6}

In Secs.~\ref{secCSQM}, 
\ref{secValence} and  \ref{secValenceOctet} 
we have discussed the covariant spectator quark model
which is based on the wave functions in flavor-spin space  
determined by the 
$SU_F(3) \otimes SU_S(2)$ symmetries.
In the following for simplicity we 
use $SU(6)$ to represent $SU_F(3) \otimes SU_S(2)$.

We discuss here the results obtained 
by the $SU_F(3)$ symmetry model 
for the hadronic weak-axial current 
modified by the strong interaction according to PCAC.
Using the $SU_F(3)$ symmetry we can represent 
the  baryon-meson interactions 
in terms of an $SU(3)$ chiral perturbation theory Lagrangian 
parametrized by three quantities, namely, the Cabibbo angle ($\theta_C$), 
and the anti-symmetric $F$ and the symmetric   
$D$ couplings~\cite{Bernard08,Gaillard84}.
Some of those models may also be refereed to as 
the Cabibbo theory~\cite{Swart63,Gaillard84}
or the heavy baryon chiral perturbation 
theory~\cite{Bernard08,Jenkins91,Dai96,Savage96,FMendieta98}.

In the $SU(3)$ baryon-meson approach, 
the properties of the 
beta decays of the octet baryons  
can be characterized by the couplings  $F$ and $D$.
In particular the results for $G_A(0)$ associated 
with the $\Delta I=1$ and $\Delta S=1$ octet baryon decays
can be expressed in terms of $F$ and $D$.
The expressions for $G_A(0)$ are presented 
in the column labeled as $SU(3)$ in Table~\ref{tabGA}.
In Table~\ref{tabGA} the signs are adjusted according 
to our results for $G_A^B(0)$ from Table~\ref{tabFAS}. 
For the  discussions about the sign conventions 
see Refs.~\cite{Gorringe05,PDG2014,Gaillard84,Bender68,Linke69}.
Also in Table~\ref{tabGA}  one can see, 
for instance, that  $G_A(0)= F + D$ for the  $n \to p$ transition 
and  $G_A(0)=\sqrt{2} F$ for the $\Sigma^- \to \Sigma^0$ transition.

If we resort furthermore  
on the $SU_S(2)$ symmetry we obtain the  
$SU(6)$ flavor-spin symmetry.
In this case:
$F=0.4(F+D)$ and $D=0.6(F+D)$.
Then, all the values of $G_A(0)$
can be determined by the value of $F+D$
that can be fixed by the value 
of $G_A(0)$ for the $n \to p$ transition.
The results for the $SU(6)$ case 
are also presented in Table~\ref{tabGA}
[see column labeled as $SU(6)$].

Note that, in the $SU(3)$ and $SU(6)$ 
approaches the dependence on the baryon 
masses is not reflected directly on the coupling 
constants $D$ and $F$,  
which should be valid 
for all the weak transitions 
between the octet baryons ($N,\Lambda,\Sigma,\Xi$).
It is important to mention that although 
the many successful $SU(3)$ baryon-meson models 
are close to the $SU(6)$ limit of $\alpha = F/(F+D) =0.6$, 
the two parametrizations can differ 
up to about 17-25\%. 
The  $SU(3)$ baryon-meson models generally have 
a better agreement with the data.

It is interesting to compare the results of  
the $SU(3)$ baryon-meson model in the $SU(6)$ limit 
(last column in Table~\ref{tabGA}) with 
the results of $G_A^B$ in Table~\ref{tabFAS} 
(last column).
We can conclude that  
the function $G$ in Table~\ref{tabGA}, 
and the function $G_A^B$ in Table~\ref{tabFAS}, 
are multiplied by the same (constant) factor for 
a given transition.
This result means that the covariant spectator 
quark model is equivalent to an $SU(6)$ 
baryon-meson model in the limit $Q^2=0$. 
For the convenience of future discussions  
we define the relative proportion factor, 
$\eta_{BB'}$, for the axial-vector form factor 
for the $B \to B'$,  
relative to that of the nucleon ($n \to p$).

If we estimate the contribution 
of the quarks $u$ and $d$ for the proton spin
using the static quark model, we obtain 
$\Delta \Sigma^u =4/3$ and $\Delta \Sigma^d=-1/3$,
which correspond to the total spin 
of the proton ($\Delta \Sigma=1$).
However, the experimental value is 
$\Delta \Sigma \simeq 0.33$~\cite{HERMES,NucleonDIS},  
which raised the well known 
{\it proton spin crises}~\cite{HERMES,NucleonDIS,Myhrer08,Shanahan13}. 
The above values of $\Delta \Sigma^q$ 
correspond  to    
$F=\frac{2}{3}$ and $D=1$~\cite{Hayne82,Isgur88,Jenkins91,BCano06,Lorce08}.
In this case one has an $SU(6)$ model 
($F=\frac{2}{3}D$) where $D=1$.
As $F+D = \frac{5}{3}$, 
we recover the results $[G_A^B(0)]_{n_P=0}$
discussed in Sec.~\ref{secValenceOctet}.

The $SU(3)$ baryon-meson model 
gives a good description of the data when 
the parameters $D$ and $F$
are fitted to the available $G_A(0)$ 
data for the octet baryons. 
It is important to note however 
that the results of the $SU(3)$ baryon-meson model are expected to 
be only an approximation since 
the $SU_F(3)$ symmetry is broken 
due to the  large $s$-quark mass 
compared to the $u$ and $d$ quarks.
The symmetry breaking due to the $s$-quark 
is indeed reflected on the variation 
of the octet baryon physical masses.
Thus, due to the symmetry breaking,
a deviation of about 20--30\% from the data  
can be expected~\cite{Gaillard84,Savage96,FMendieta98}.

We can extend the results of the $SU(3)$ baryon-meson model
for finite $Q^2$ replacing the constants $F$ and $D$
by two form factors dependent on $Q^2$, $F(Q^2)$ and $D(Q^2)$, 
as suggested in Ref.~\cite{Gaillard84}.
If we demand also $SU(6)$ symmetry, one gets 
$\alpha \equiv D(0)/(F(0)+D(0)) = 0.6$.
The results based on this, are presented  
in the last column of Table~\ref{tabGA}
in terms of $G(Q^2) \equiv F(Q^2) + D(Q^2)$,
that is now a function of $Q^2$.

Note that, the difference between the
calculation in Sec.~\ref{secValenceOctet}
and the $SU(3)$ baryon-meson model 
extended for finite $Q^2$, 
is that the functions $G_A^B (Q^2)$
only take into account the effect 
of the valence quark contributions,
and the calculations of $G(Q^2) = F(Q^2)+ D(Q^2)$
are parametrized by the 
octet baryon beta decay data for $Q^2=0$,
and the nucleon data, including finite $Q^2$ data.
Therefore the $SU(3)$ baryon-meson model 
takes into account effectively 
all possible physical effects 
including the meson cloud effects.

Thus, if we consider the $SU(3)$ baryon-meson model, 
with a $Q^2$ dependence 
extracted from nucleon experimental data for $G_A$, 
one can estimate all the axial-vector octet baryon form factors, 
based on the $SU(3)$ symmetry.
However, since the baryon-meson models based on the 
$SU(3)$ symmetry generally  ignore the effects 
of the octet baryons masses in the transitions, 
the estimates of the $Q^2$ dependence, 
except for the nucleon case, have to be taken with caution.
In Sec.~\ref{secResults-Octet} we will discuss 
the expected falloff with $Q^2$ according to  
the $SU(6)$ baryon-meson model, and compare it   
with the results of the covariant spectator quark model.

Recall that,
since  the $SU(3)$ and the $SU(6)$ baryon-meson models 
are based on the $SU_F(3)$ symmetry,
we obtain results only in a first order 
(equal mass limit of the octet baryons),
and a deviation from the data 
can be expected,  
even with an effective inclusion of the meson cloud effects.
Corrections due to the $SU(3)$  symmetry breaking,
as the  result of the large $s$-quark mass
as well as the corrections from meson loops 
can be calculated using 
the formalism of chiral perturbation theory
\cite{Jenkins91,Dai96,Savage96,Lin09}.
It is known that 
meson loop corrections 
do not satisfy the $SU(3)$ symmetry~\cite{Jenkins91,Dai96}.
The explicit calculation 
of the next leading order corrections
of the $SU(6)$ baryon-meson model are, however,  
beyond the scope of the present work.

\begin{table}[t]
\begin{center}
\begin{tabular}{c | c c c }
\hline
\hline
  &Process & $SU(3)$ &  $SU(6)$\\
\hline
$\Delta I = 1\hspace{2em}$ &\hspace{2em}$n \to p$\hspace{2em} &$F+D$  & $G$\\
                           &\hspace{2em}$\Sigma^{\pm} \to \Lambda$\hspace{2em} &$ \pm \sqrt{\frac{2}{3}}  D$ 
& $ \pm \frac{\sqrt{6}}{5}G$ \\
                           &\hspace{2em}$\Sigma^{-} \to \Sigma^0$\hspace{2em} &$\sqrt{2} F$ & 
 $\frac{2 \sqrt{2}}{5}G$  \\
                           &\hspace{2em}$\Xi^- \to \Xi^0$\hspace{2em} &$F-D$  &  $-\frac{1}{5}G$\\
\hline
$\Delta S = 1\hspace{2em}$ &\hspace{2em}$\Lambda \to p$\hspace{2em} &
$-\sqrt{\frac{3}{2}}  \left(F + \frac{1}{3}D\right)$ 
& $-\frac{3\sqrt{3}}{5 \sqrt{2}}G$  \\
                           &\hspace{2em}$\Sigma^{-} \to n$\hspace{2em} &$-F+D$  & 
$\frac{1}{5}G$\\
                     &\hspace{2em}$\Sigma^{0} \to p$\hspace{2em} &
                       $-\frac{1}{ \sqrt{2}} (F-D)$  & 
$\frac{1}{5 \sqrt{2}}G$\\
                         &\hspace{2em}$\Xi^- \to \Lambda$\hspace{2em} &
$- \sqrt{\frac{3}{2}} \left(F-\frac{1}{3}D\right)$  &
$- \frac{\sqrt{3}}{5\sqrt{2}}G$\\
                           &\hspace{2em}$\Xi^- \to \Sigma^0$\hspace{2em} &
$\frac{1}{\sqrt{2}} (F+D)$ & 
$\frac{1}{\sqrt{2}}G$\\
                           &\hspace{2em}$\Xi^0 \to \Sigma^+$\hspace{2em} &$F+D$ &  $G$\\
\hline
\hline
\end{tabular}
\end{center}
\caption{Octet baryon axial-vector form factors $G_A(Q^2)$, 
expressed in terms of $F$ and $D$
of $SU(3)$ scheme, 
and in the case of $SU(6)$ symmetry~\cite{Gaillard84,Yang15}. 
$F$ and $D$ here are functions depending on $Q^2$.
In the last column $G = F + D$ is also a function dependent on $Q^2$.}
\label{tabGA}
\end{table}

\begin{table}[t]
\begin{center}
\begin{tabular}{c  c c }
\hline
\hline
  & $SU(3)$ &  $SU(6)$\\
\hline
 $N$ &      $F+D$& $G$ \\
 $\Sigma$ & $-\sqrt{2}F$& $-\frac{2\sqrt{2}}{5} G$ \\
$\Xi$ & $-(F-D)$ & $- \frac{1}{5}G$  \\
\hline
\hline
\end{tabular}
\end{center}
\caption{Neutral current axial-vector form factors. 
As in Table~\ref{tabGA}, $G=F+D$.}
\label{tabGAel}
\end{table}

Here, we emphasize that, 
in general, the inclusion of the meson cloud 
contributions breaks 
the $SU(6)$ symmetry observed for the 
valence quark component of $G_A$ 
(in the equal mass limit).
However, as far as the valence quark  
contribution is dominant, 
and the coefficients $F$ and $D$ are fitted 
to the $G_A(0)$ data,  
it is expected that 
an $SU(3)$ or an  $SU(6)$ baryon-meson model 
gives a good description of the data
for small $Q^2$.
As for finite $Q^2$, in particular 
for large $Q^2$, it is still necessary to check  
if  the $SU(3)$ or $SU(6)$ description is good,  
or if the effect of the octet baryon masses 
(symmetry breaking) is important.

For simplicity and consistency 
with the approximate structure of the $SU(6)$ symmetry in  
the covariant spectator quark model, 
we will start by analyzing  the meson cloud contribution  
with an $SU(6)$ parametrization. 
This parametrization has no adjustable parameters
once the contribution of the meson cloud 
in the nucleon axial-vector form factors are fixed.
Later on we discuss an $SU(3)$ parametrization 
where we adjust one parameter by the 
$G_A(0)$ data of the octet baryons.
Our results for $G_A(Q^2)$ will be presented 
in Sec.~\ref{secResults-Octet}.

All the discussions in this section  
have been
centered on the functions $G_A$.
As for the induced pseudoscalar form factors $G_P$,  
there are no predictions from 
the $SU(3)$ baryon-meson model~\cite{Gaillard84,Kubodera88}.

\section{Results for the nucleon axial form factors}
\label{secResults-Nucleon}

In this section we present our results for the nucleon
axial form factors.
We divide the presentation in three steps:
\begin{itemize}
\item
Determination of the $P$-state mixture parameter ($n_P$)
in the nucleon wave function   
by a direct fit to the lattice QCD data for $G_A$.
\item
Determination of the quark form factor $g_P$  
by a fit to the lattice QCD data for $G_P$.
\item
Extrapolation of the model from the lattice QCD regime 
to the physical regime in order 
to obtain the valence quark contribution 
for the nucleon axial-vector form factor ($G_A^B$).
The normalization constant $Z_N$ is determined 
by the fit of the $\tilde G_A^B$ to the 
nucleon physical data $G_A^{\rm exp}$ 
according to Eq.~(\ref{eqBarePhys}).
\end{itemize}

In this section the masses $m_\pi, m_\rho$ 
and $m_N$ refer to the pion, $\rho$ and nucleon masses 
obtained in lattice QCD simulations.
In particular, for the nucleon we use 
$m_N$ to avoid the confusion with the physical 
mass of the nucleon $M$.

\begin{table}[t]
\begin{center}
\begin{tabular}{c c c c}
\hline
\hline
$m_\pi$(MeV) & $a$(fm)  & $L$(fm) & Ref.\\
\hline
373.0  & 0.082 & 2.60 & \cite{Alexandrou13} \\
377.0  & 0.089 & 2.10 & \cite{Alexandrou11a} \\
403.5  & 0.070 & 2.13 & \cite{Alexandrou11a} \\
431.9  & 0.089 & 2.10 & \cite{Alexandrou11a} \\
465.3  & 0.070 & 2.13 & \cite{Alexandrou11a} \\
467.5  & 0.089 & 2.10 & \cite{Alexandrou11a} \\
469.8  & 0.056 & 2.39 & \cite{Alexandrou11a} \\
\hline
\hline
\end{tabular}
\end{center}
\caption{Parameters associated with 
the lattice QCD data used in the fits.
$a$ is the lattice space and $L$ is the lattice length
that defines the lattice volume $L^3$. }
\label{tabLattice1}
\end{table}

\begin{table}[t]
\begin{center}
\begin{tabular}{c c c c c}
\hline
\hline
$m_\pi$(GeV) & $m_N$(GeV)  & $m_\rho$(GeV) & $\chi^2 (G_A)$  & $\chi^2(G_P)$\\
\hline
0.3730  & 1.2100 & 0.8355 & 1.699 & 2.138 \\
0.3770  & 1.2225 & 0.8367 & 0.475 & 0.489 \\
0.4035  & 1.2527 & 0.8456 & 1.038 & 2.164 \\
0.4319  & 1.2828 & 0.8557 & 1.943 & 2.277\\
0.4653  & 1.3289 & 0.8685 & 1.447 & 0.888\\
0.4675  & 1.3343 & 0.8694 & 1.650 & 1.158 \\
0.4698  & 1.3390 & 0.8703 & 0.402 & 0.318\\
\hline
        &        &        & 1.285 & 1.444\\
\hline
\hline
\end{tabular}
\end{center}
\caption{
Masses associated with the lattice QCD data 
used in the fits.
$\chi^2 (G_A)$ is the partial $\chi$ squared 
in the fit of $n_P$ ($G_A$ data).
$\chi^2 (G_P)$ is the partial $\chi$ squared 
in the fit of $\alpha$, $\beta$ to the $G_P$ data.}
\label{tabLatticeFit}
\end{table}

\subsection{Axial-vector form factor ($G_A$)}
\label{secNucleonGA}

% figure 2 and 3 here

The calculation of $G_A$ is done
using the expressions discussed in Sec.~\ref{secValence},
summarized in Eq.~(\ref{eqGAspec}),
where all the terms are function of the 
$P$- state mixing parameter $n_P$ (since $n_S=\sqrt{1- n_P^2}$).
In Sec.~\ref{secSummary} we have already shown 
that some lattice QCD data
are well described by  
a $P$-state mixture with $n_P \simeq -0.5$.

Lattice QCD simulations of the axial-vector 
form factor of the nucleon for $Q^2=0$ 
can be found in 
Refs.~\cite{Sasaki03,Edwards06,Yamazaki08,Bhattacharya14,Horsley14,Liu94,Green15,ARehim15,Bali15}.
The results from Refs.~\cite{Green15,ARehim15,Bali15}
are obtained near the physical point or at the physical point. 
Recent calculations of $G_A$
as a function of $Q^2$ can be found 
in Refs.~\cite{Bratt10,Sasaki08,Yamazaki09,Alexandrou11a,Alexandrou13,Sasaki09}.
Concerning the calculations of $G_A$
for small $Q^2$, it is important to mention that 
lattice  simulations performed with small 
volumes underestimates the value of 
$G_A$~\cite{Sasaki08,Yamazaki08,Yamazaki09}.
Therefore in our study we select datasets 
with large volumes.

The lattice QCD data included 
in our fit correspond to those from
Refs.~\cite{Alexandrou11a,Alexandrou13}, 
in the pion mass range $m_\pi= 350$--$500$ MeV
(large $m_\pi$),
where only the valence quark degrees of freedom are relevant.
The parameters used in the lattice simulations:
the lattice space ($a$) and the 
length $L$ that defines the volume,
are listed in Table~\ref{tabLattice1}.
In addition to the large volumes 
the chosen datasets have values of $G_A(Q^2)$
up to $Q^2= 2$  GeV$^2$,
which is convenient if we want to study the large $Q^2$ region.

In Table~\ref{tabLatticeFit}
we also present the results for $m_N$ 
given by the respective lattice QCD simulations,
and the values of $m_\rho$ determined by
Eq.~(\ref{eqMrho}).
These  mass  values are necessary to calculate
the quark axial current 
(\ref{eqJAq}) based on the VMD
parametrizations discussed in Sec.~\ref{secQuarkAFF} 
for the model in the lattice regime. 
Furthermore, $m_N$ is also necessary to obtain 
the nucleon radial wave functions in the lattice regime. 
In the fits, to avoid the contamination 
of lattice QCD artifacts that may appear for large $Q^2$, 
we use only the data  with $Q^2 < 1.6$ GeV$^2$.
Above this $Q^2$ range, the lattice QCD data are often 
affected by large errorbars
and may show unexpected oscillations.

\begin{figure}[t]
\vspace{.45cm}
%\vspace{2cm}
\centerline{
\mbox{
\includegraphics[width=2.8in]{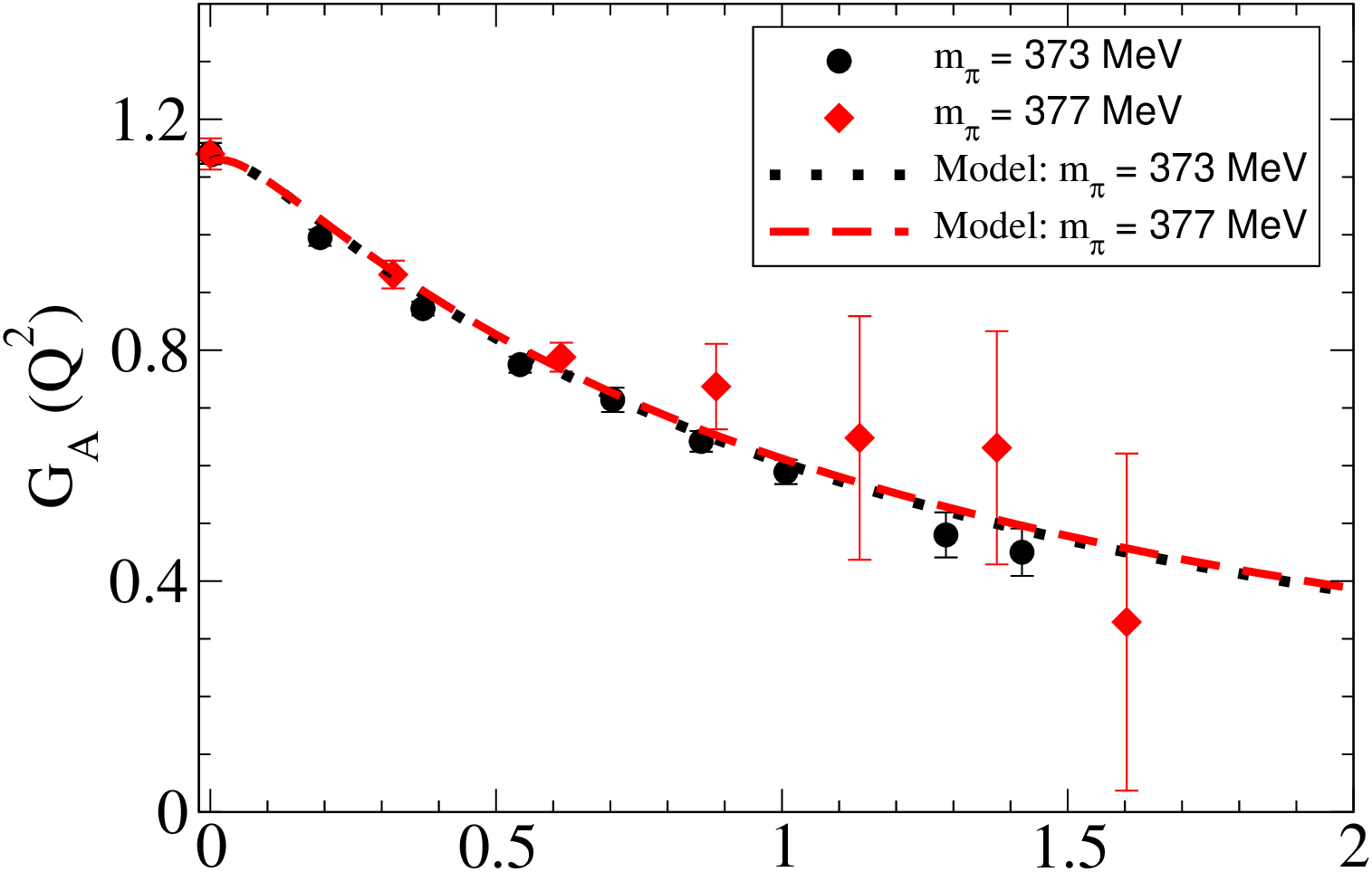}
}}
\centerline{\vspace{.5cm} }
\centerline{
\mbox{
\includegraphics[width=2.8in]{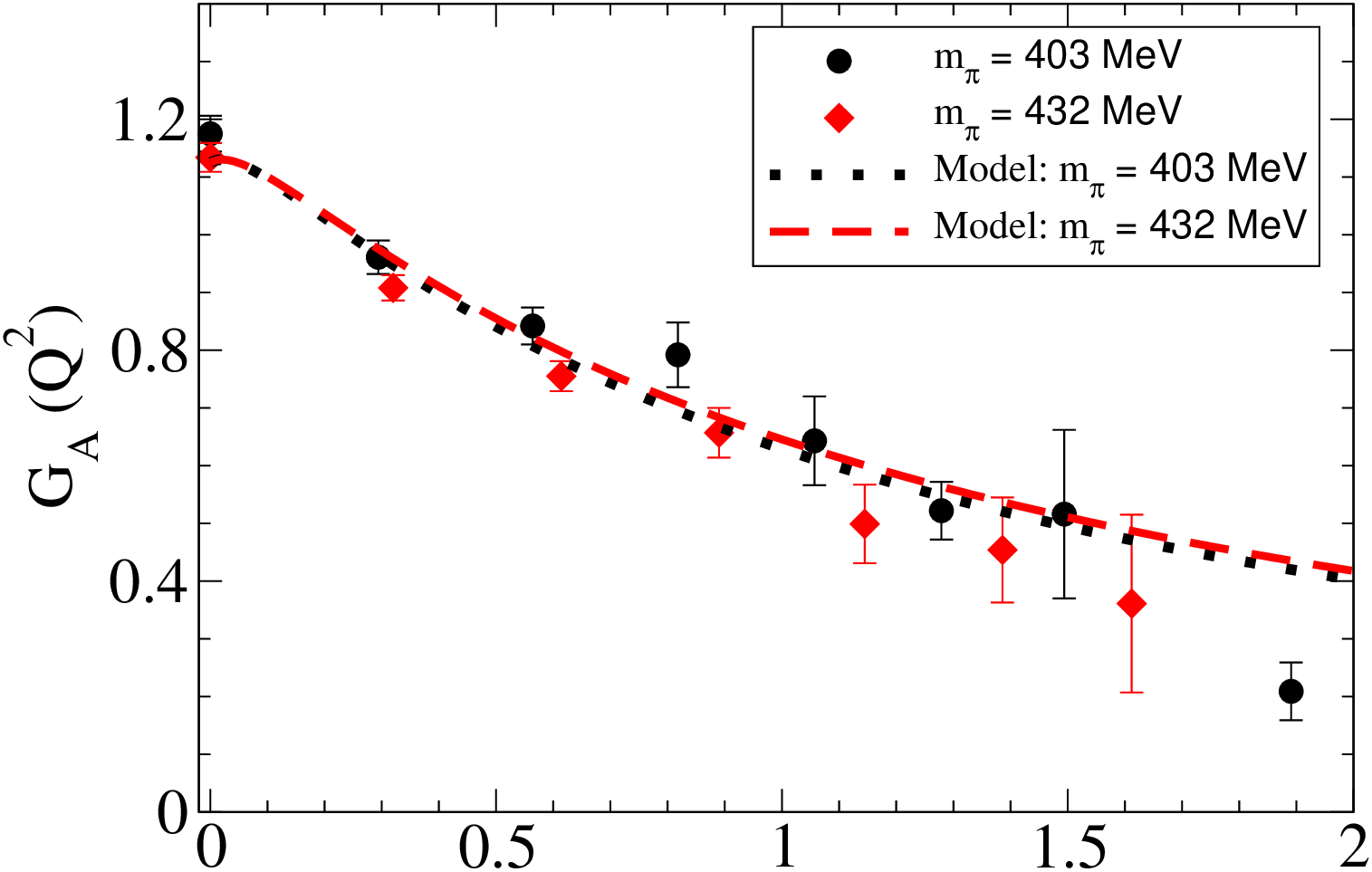}
}}
\centerline{\vspace{.5cm} }
\centerline{
\mbox{
\includegraphics[width=2.8in]{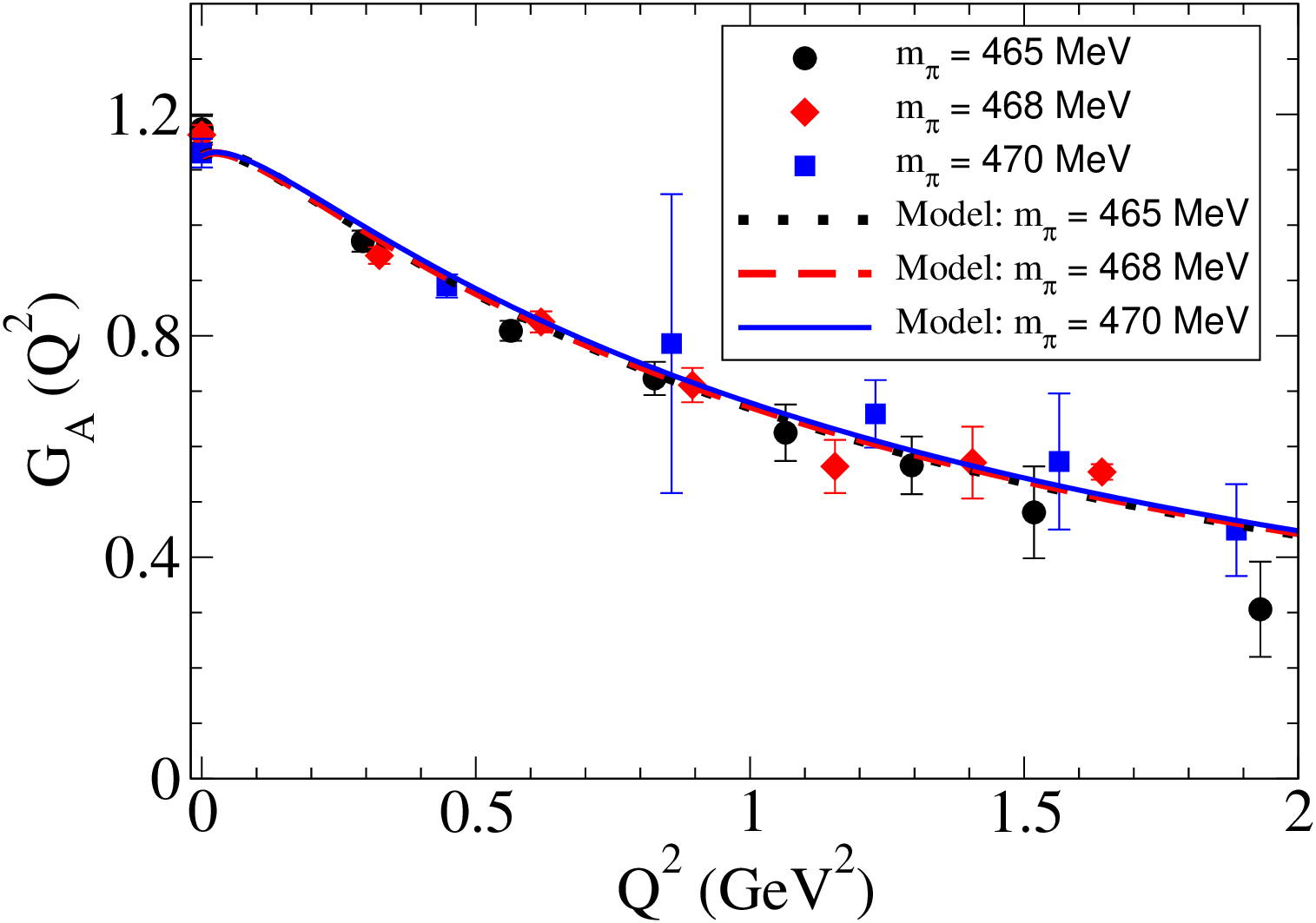}
}}
\caption{\footnotesize{
Results of the fit to the lattice QCD data
for $G_A$ with $n_P \simeq -0.5067$.
}}
\label{figGA1}
\end{figure}

The results of our $\chi^2$ per datapoint
are presented in Table~\ref{tabLatticeFit},
in the column indicated $\chi^2 (G_A)$.
By the fit we obtain the value 
\ba
n_P= -0.5067.
\ea
The results of the fit are presented in Fig.~\ref{figGA1}.
In the bottom panel we show the results for
the heavier pion cases $m_\pi= 465, \,468, \, 470$ MeV, 
while the results for $m_\pi= 403, \, 432$ MeV are in the middle panel, 
and the results for the lightest cases $m_\pi=373, \, 377$ MeV 
are in the top panel.

In Fig.~\ref{figGA1} one can confirm that the model in the lattice QCD
regime gives a very good description of the lattice data
[see also the $\chi^2(G_A)$ results in Table~\ref{tabLatticeFit}].
In the case $m_\pi = 432$ MeV, however, 
we can notice that the lattice data falloff is faster than 
the model (fit), and this is reflected in the large partial $\chi^2$ value.
Nevertheless, the data are sill consistent
with the model.

\begin{figure}[t]
\vspace{.25cm}
%\vspace{2cm}
\centerline{
\mbox{
\includegraphics[width=2.8in]{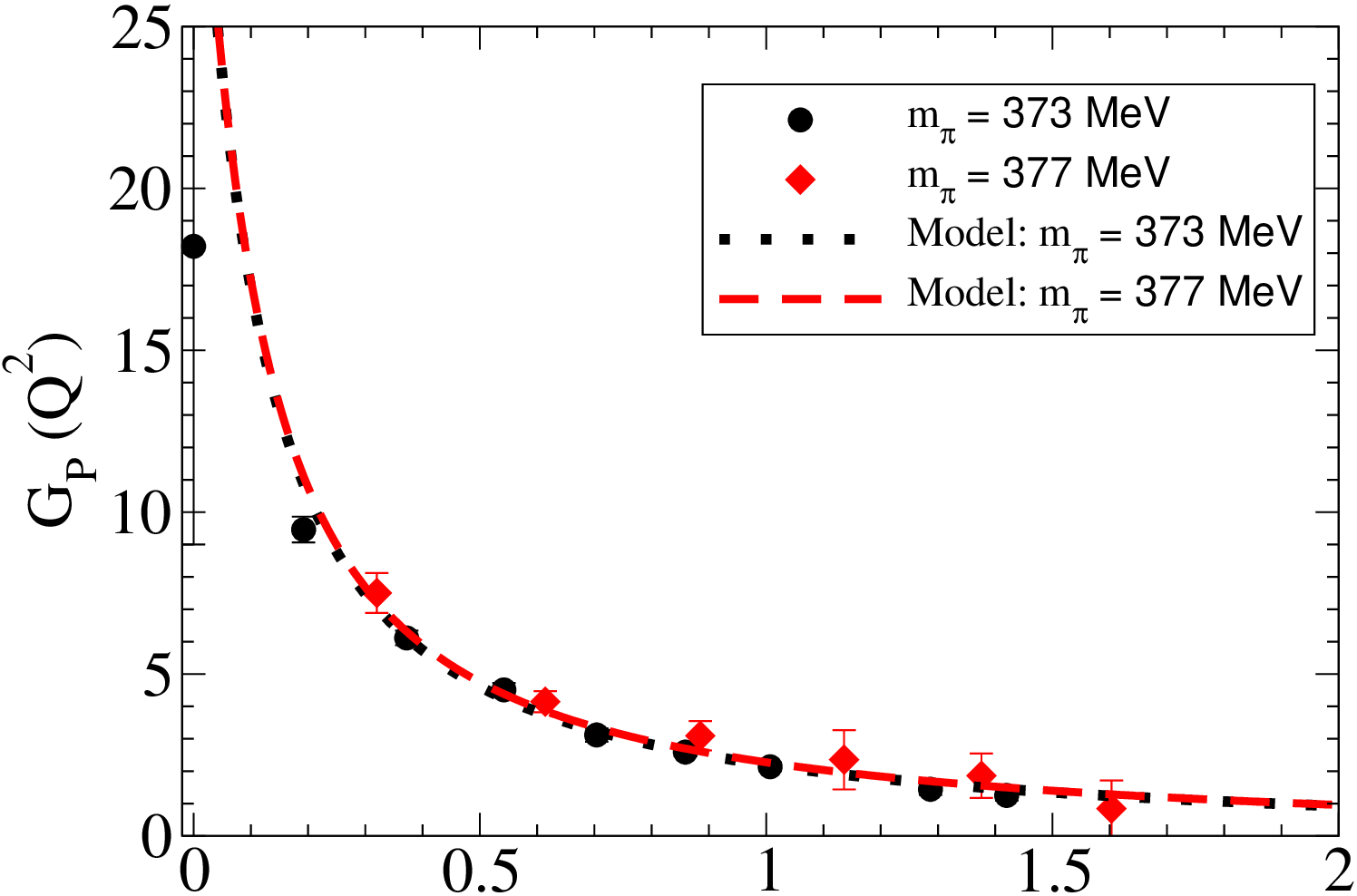}
}}
\centerline{\vspace{.35cm} }
\centerline{
\mbox{
\includegraphics[width=2.8in]{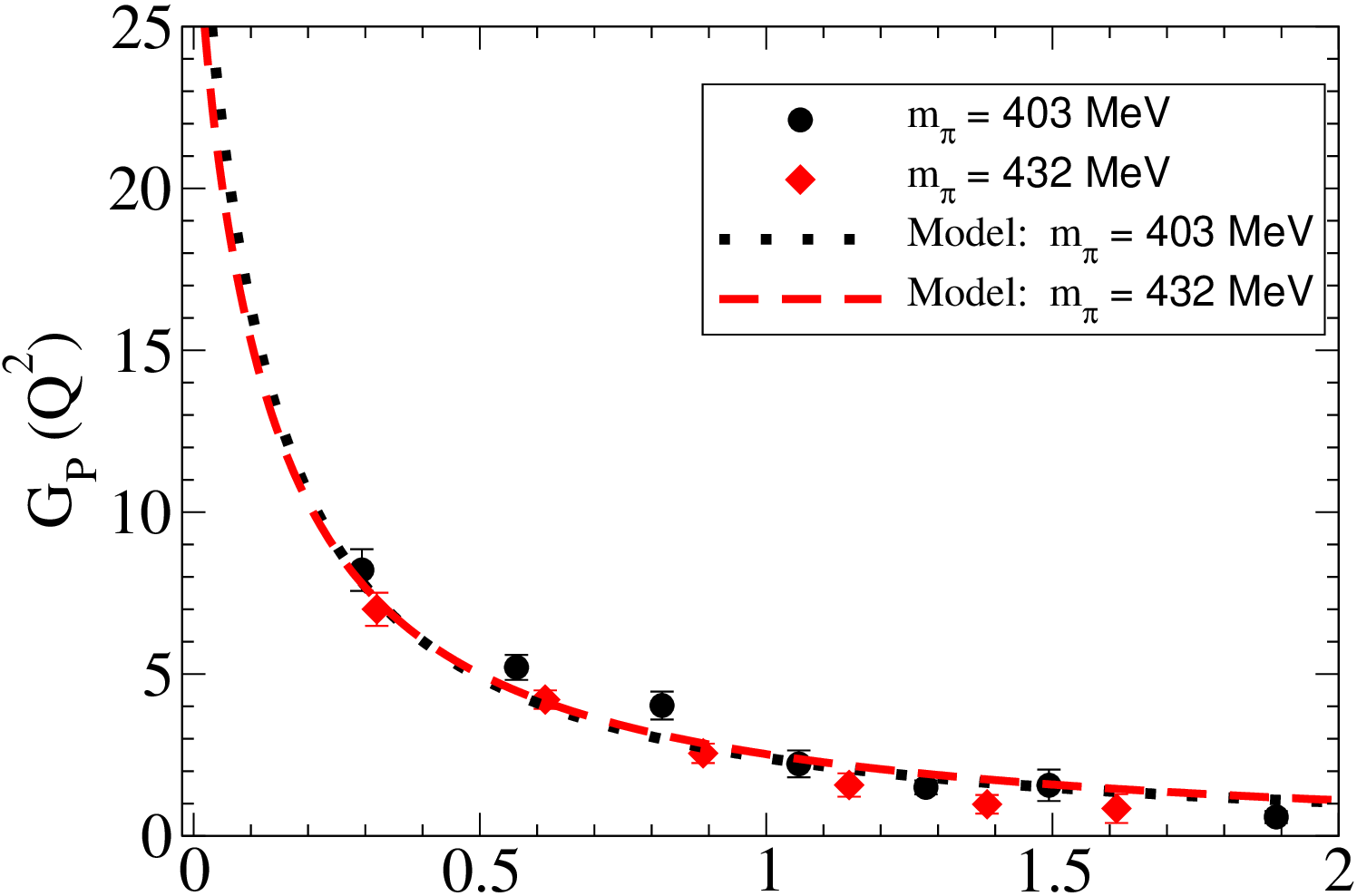}
}}
\centerline{\vspace{.35cm} }
\centerline{
\mbox{
\includegraphics[width=2.8in]{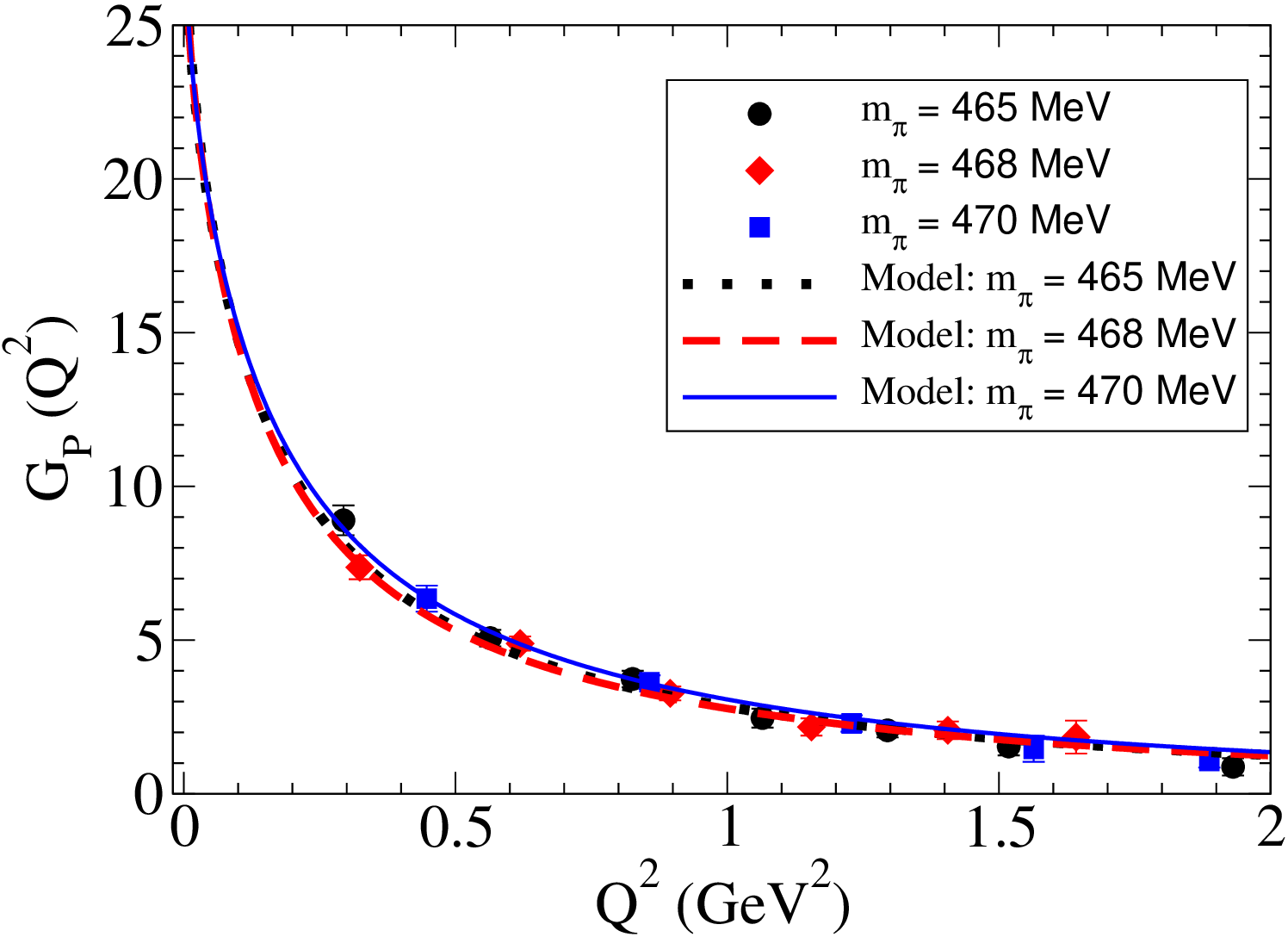}
}}
\caption{\footnotesize{
Results of the fit to the lattice QCD data
for $G_P$.
The result of $G_P(0)$ for $m_\pi=373$ MeV~\cite{Alexandrou13}
is an extrapolation from the data. 
}}
\label{figGP1}
\end{figure}

\subsection{Induced pseudoscalar form factor ($G_P$)}
\label{secNucleonGP}

With the value of $n_P$ fixed by the lattice $G_A$ data,
we can test now if the lattice $G_P$ data
can be described by a parametrization of $g_P$
based on the VMD mechanism.
We decompose $G_P$ as  
$G_P =  G_A^B +G_P^{\rm pole}$,
where the pole term is defined by 
Eq.~(\ref{eqGPpole}), and 
fit the coefficients $\alpha$ and $\beta$
in the function $g_P$ given by  Eq.~(\ref{eqQuarkGP})  
to the lattice QCD data for $G_P$.
The values obtained from the best fit are, 
$\alpha = -3.901$ and $\beta = 0.3297$.
The quality of the fit for each lattice dataset is presented
in the column $\chi^2(G_P)$ in Table~\ref{tabLatticeFit}.

In Fig.~\ref{figGP1} we show 
the results of the fit for three groups of 
pion masses discussed previously.
In any case we have a good description of the data.
(The datapoint for $Q^2=0$ for $m_\pi=373$ MeV 
is the result of an extrapolation and is not 
included in the fit).

\subsection{Extrapolation to the physical regime}

The experiments related with the form factor $G_A$
show that the value for $Q^2=0$ is 
very well determined~\cite{PDG2014}:
\ba
G_A^{\rm exp}(0) = 1.2723 \pm 0.0023. 
\ea  
The $Q^2$ dependence of $G_A$
is well approximated by a dipole form, 
$G_A(Q^2) = G_A(0)/( 1 + Q^2/M_A^2)^2$,
where the values of $M_A$ varies 
from $M_A \simeq 1.03$ GeV 
(neutrino scattering)  
to $M_A \simeq 1.07$ GeV
(electroproduction)~\cite{Bernard02}.

To represent the experimental data in a general form 
we consider the interval between the two functions, 
$G_A^{{\rm exp}-}$ and $G_A^{{\rm exp}+}$, given by 
\ba
G_A^{{\rm exp} \pm}(Q^2) =
\frac{G_A^0 (1 \pm \delta)}{\left( 1 + \frac{Q^2}{M_{A \pm}^2} \right)^2},
\label{eqGAexp}
\ea
where $\delta$ is a 
parameter that expresses the precision of the data, 
and  $M_{A-}= 1.0$ GeV and $M_{A+}= 1.1$ GeV 
are respectively the lower and upper limits for $M_A$ 
extracted experimentally.
To avoid a strong impact from the result 
for $Q^2=0$ and flexibilize the fit, we choose 
$\delta \simeq 0.03$,
a typical relative error 
(error of about 3\% and 10 times the relative error for $Q^2=0$).

As mentioned already, the prediction 
of the model for the valence quark contribution  is 
given by $Z_N G_A^B (Q^2)$, where $G_A(Q^2)$
is the extrapolation for the case $m_\pi= 138$ MeV 
of the model determined by the fit to the 
lattice QCD data (see previous section).
Since the valence quark model, 
extrapolated to the physical regime
is expected to be a good approximation
to the data only for large $Q^2$ 
(small  meson cloud effects),
we varied the value of $Z_N$ 
to see if it is possible to obtain 
a good description of the data 
%(
in the interval $Q^2=1.0,...,2.0$ GeV$^2$. 
From the best fit to the data 
we  obtain  the value of $Z_N= 0.7343$.
This result means that the 
meson cloud contribution for the 
proton charge is about 27\%.

\begin{figure}[t]
\vspace{.5cm}
%\vspace{2cm}
\centerline{
\mbox{
\includegraphics[width=3.0in]{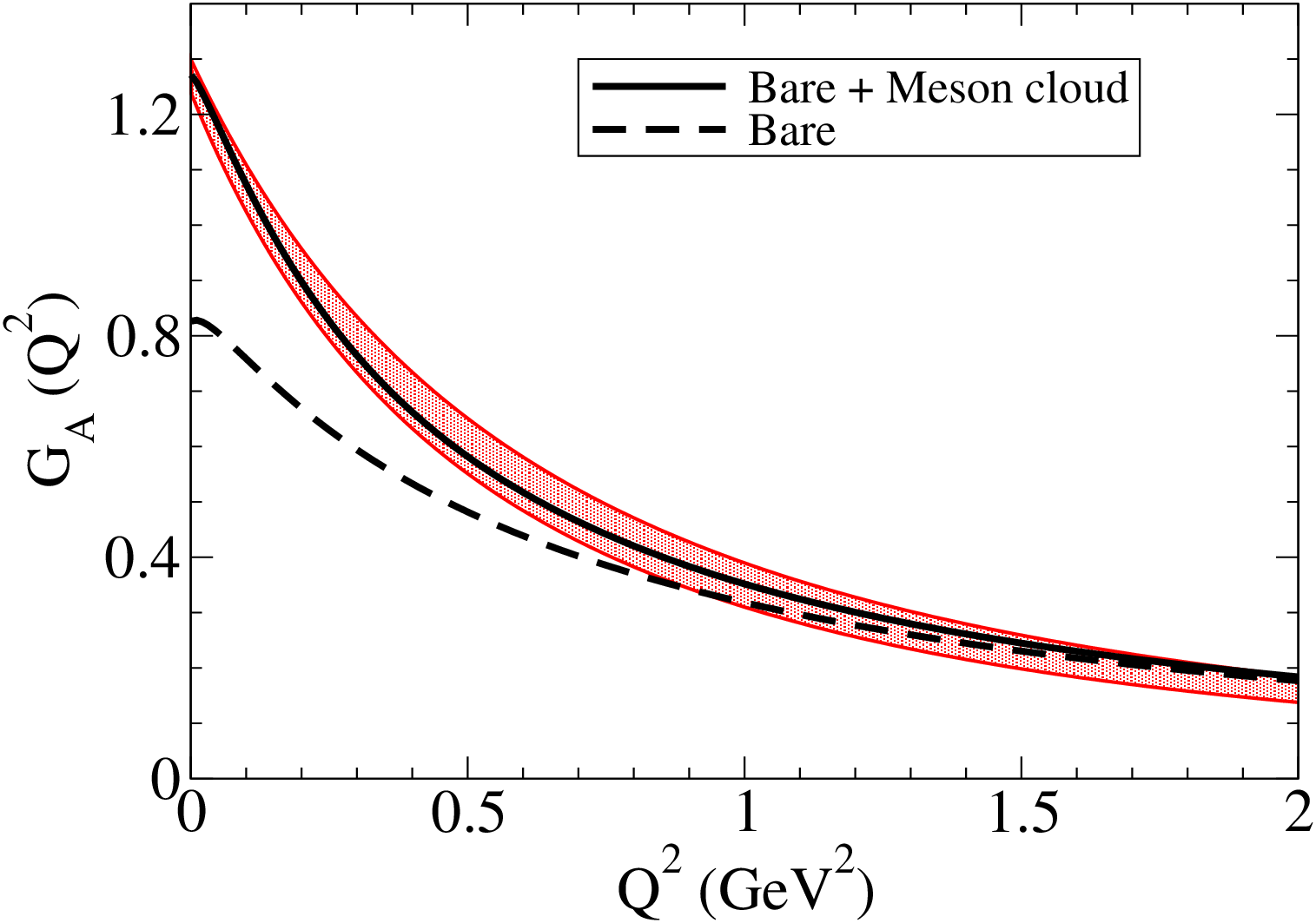}
}}
\caption{\footnotesize{
Result for $G_A(Q^2)$ in the physical limit.
The dashed-line (Bare) is the result
from the extrapolation of the lattice data 
for the physical regime with $Z_N=0.7434$.
The solid-line (Bare + Meson cloud) is the sum 
of the bare contribution with $G_A^{MC}(Q^2)$ given by Eq.~(\ref{eqGAmc}),
with $\Lambda = 1.05$ GeV. 
The bands are a representation of the 
experimental data.
}}
\label{figGAphys}
\end{figure}

In Fig.~\ref{figGAphys}  we present 
the bare contribution for the form factor $G_A$ 
(dashed-line) determined by the value  $Z_N= 0.7343$.
The deviation from the empirical data $G_A^{{\rm exp}\pm} (Q^2)$,
represented by the red band, from the result 
$\tilde G_A^B(Q^2)$, can be interpreted 
as the result of the meson cloud effect.
Since it is expected that 
the meson cloud effects are suppressed 
by the factor $1/Q^4$ relative to that of 
the valence quark contributions 
for large $Q^2$ according to 
perturbative QCD arguments~\cite{Carlson}, 
we parametrize the meson cloud contribution as
\ba
G_A^{MC}(Q^2) = Z_N \frac{G_A^{MC0}}{\left( 
1 + \frac{Q^2}{\Lambda^2}\right)^4 },
\label{eqGAmc}
\ea
where $\Lambda$ is a cutoff parameter  
and $G_A^{MC0}$ is the relative magnitude of the 
meson cloud contribution for $G_A(0)$.
Note that, according  to the normalization 
of the nucleon wave function 
Eq.~(\ref{eqNucleonWF}),  
both the valence and the meson cloud 
components are multiplied by the
normalization factor $Z_N$.
Therefore, for convenience the 
normalization factor $Z_N$ is included 
in the definition of $G_A^{MC}$.

We have also tried some variations of the quadrupole 
expression~(\ref{eqGAmc}), e.g., such as a product 
of dipoles, however,  the quadrupole expression 
with \mbox{$\Lambda = \frac{1}{2}(M_{A+} + M_{A-})= 1.05$} GeV,
gives a description of the data with a 
quality equivalent to a product of dipoles
with two adjustable cutoff parameters.
We can interpret then Eq.~(\ref{eqGAmc})  
as one of the best parametrizations for the meson cloud, 
with the value of $\Lambda$ fixed by 
the average cutoff from the global parametrizations of $G_A$, 
given in Eq.~(\ref{eqGAexp}).
The best fit from Eq.~(\ref{eqGAmc})
for the data fixes 
$G_A^{MC0}=0.6077$, leading to $G_A^{MC}(0)= 0.4462$
(35\% of meson cloud for $G_A$ at $Q^2=0$).
The result of the combination 
for the bare and meson cloud contributions,  
is presented in Fig.~\ref{figGAphys} (solid-line).

\begin{figure}[t]
\vspace{.3cm}
%\vspace{2cm}
\centerline{
\mbox{
\includegraphics[width=3.1in]{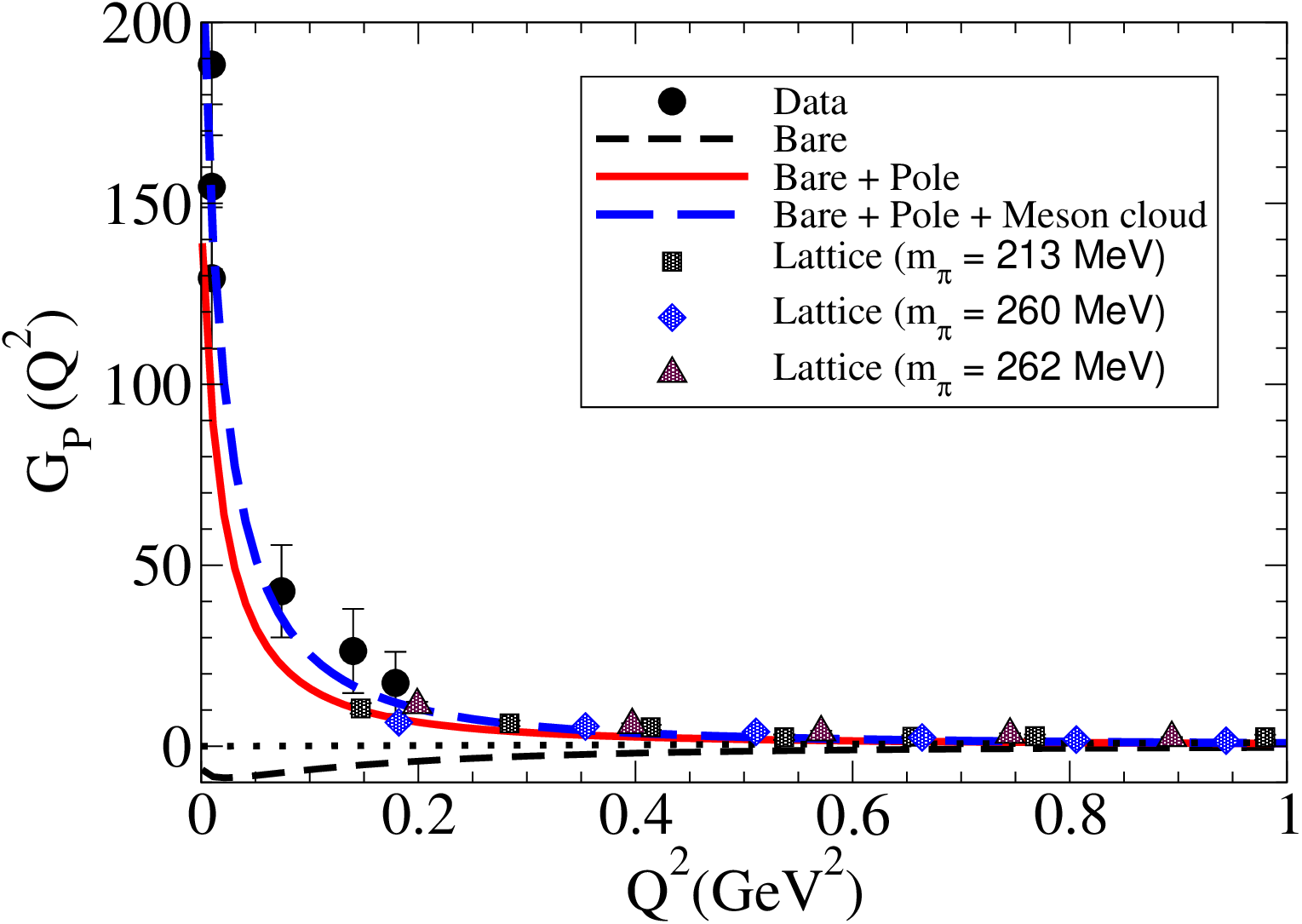}
}}
\caption{\footnotesize{
Physical result for $G_P(Q^2)$.
The data are from Refs.~\cite{Bernard02,Choi93}.
Lattice QCD data are from Ref.~\cite{Alexandrou13}.}}
\label{figGPphys}
\end{figure}

To finalize the study of the nucleon axial 
form factors in the physical regime,  
we represent in Fig.~\ref{figGPphys} 
the results for the form factor  $G_P$
in comparison with the available data ($Q^2 < 0.2$ GeV$^2$).
Since there are no physical data for $Q^2 > 0.2$ GeV$^2$,
we compare the model also 
with the lattice QCD data with 
small pion masses  $m_\pi =213, \, 260, \, 262$ MeV.
Lattice QCD data for $G_P$ can also be found in 
Refs.~\cite{Bratt10,Sasaki08,Sasaki09,Bali15}.

For the following discussion we 
recall that $G_P$ can be decomposed 
in the physical regime into
\ba
G_P (Q^2)= G_P^{\rm pole}(Q^2) + \tilde G_P^B(Q^2) + G_P^{MC} (Q^2),
\ea
where $G_P^{\rm pole}$ is the  
contribution from the pion pole defined by Eq.~(\ref{eqGPpole})
in terms of bare axial-vector form factor,  $\tilde G_A^B$,
in the physical limit, and  $G_P^{MC}$ is a possible contribution 
from the meson cloud.
For convenience we redefine 
the bare contribution $G_P^B$ as $\tilde G_P^B = Z_N G_P^B$ 
in the physical limit.

In Fig.~\ref{figGPphys} we represent the bare contribution 
by the short-dashed line.
The magnitude is small and negative. 
The sum of the pion pole term and the bare 
contribution is indicated by the solid line.
We can see in Fig.~\ref{figGPphys} that the line ``bare + pole'' 
(solid line) is very close to physical data.
In addition we show the full result, 
labeled as ``bare + pole + meson cloud'' 
(long-dashed line),
where we define the meson cloud contribution 
as $G_P^{MC}= \frac{4M^2}{m_\pi^2 + Q^2} G_A^{MC}$.
This procedure is equivalent 
to redefine the contribution from the pole contribution  
by the replacement $\tilde G_A^B \to G_A$,
when no $G_P^{MC}$ term contribution is included.
In the figure we can also see
 that the sum of all terms 
``bare + pole + meson cloud'' 
has a better agreement with the data than 
 ``bare + pole''.
Note also that final result (sum of all terms) agrees with 
both the physical data ($Q^2 < 0.2$ GeV$^2$) and 
the lattice QCD data for larger $Q^2$.

Overall, we conclude that the covariant 
spectator quark model, once fitted to 
the lattice QCD data and the experimental nucleon data for $G_A$,
gives a consistent description of 
the lattice and physical data for the nucleon.
The fit for the $G_A(Q^2)$ data  
fixes the amount of the meson cloud contribution 
for the physical nucleon  state 
as 27\% ($Z_N= 0.7324$), resulting   
the meson cloud  contribution for the axial-vector form factor  
at $Q^2=0$ as 0.4462  (35\% of the total), 
and the falloff of the meson cloud component 
as a quadrupole with a cutoff $\Lambda =1.05$ GeV.

\section{Results for the octet baryon axial form factors }
\label{secResults-Octet}

\begin{table}[t]
\begin{center}
\begin{tabular}{c r r r}
\hline
\hline
 & Model  & Ref.~\cite{Lin09} & Ref.\cite{Erkol10a}\\
\hline
$N$       & 1.125     & 1.210(05)    & 1.314(24) \\
$\Sigma$  & 0.900     & 0.900(30)    & 0.970(21) \\
$\Xi$     & $-0.225$  & $-0.270(10)$ & $-0.300(10)$ \\
\hline
\hline
\end{tabular}
\end{center}
\caption{Comparison of the results 
for the neutral current 
$G_A^B(0)$ in  the covariant spectator quark model 
and the lattice QCD results 
with $m_\pi \approx 500$ MeV 
from Refs.~\cite{Lin09,Erkol10a}. }
\label{tabGB0}
\end{table}

\begin{table}[t]
\begin{center}
\begin{tabular}{l r r}
\hline
\hline
 & Model &  Ref.\cite{Erkol10a}\\
\hline
$n \to p$               & 1.125    & 1.314(24) \\
$\Sigma^+ \to \Lambda$   & 0.551   & 0.655(14) \\
$\Sigma^- \to \Sigma^0$  & 0.636    & 0.686(15) \\
$\Xi^- \to \Xi^0$        &  $-0.225$  & $-0.300(10)$ \\
\hline
$\Lambda \to p$      &  $-0.827$  &   $-0.632(14)$\\
$\Sigma^- \to n$      &   0.225    &   0.339(12)\\
$\Xi^- \to \Lambda$  &  $-0.276$  &  $- 0.274(08)$\\
$\Xi^- \to \Sigma^0$ &   0.795    &   0.908(19) \\
$\Xi^0 \to \Sigma^+$ &   1.125     &   1.284(28) \\
\hline
\end{tabular}
\end{center}
\caption{Comparison  of the results 
for $G_A^B(0)$ in the covariant spectator quark model 
and the lattice QCD results 
with $m_\pi \approx 500$ MeV from Ref.~\cite{Erkol10a}. }
\label{tabGB0-2}
\end{table}

We present now the results for the octet baryon 
axial form factors.
First, we discuss the results for the valence 
quark contributions, and compare 
the results with those of the lattice QCD.
Next, we combine the valence quark contributions 
with the meson cloud contribution estimated based on 
an $SU(6)$ baryon-meson model, 
and an $SU(3)$ baryon-meson model defined by a fit to the data.
Combining the two contributions 
we obtain our final predictions for 
the octet baryon axial-vector form factors.
We finish with our predictions 
for the octet baryon $G_P$ form factors,
based on these two models.

A note of caution is in order 
concerning the following results.
Since the predictions of the model for 
the octet baryon axial form factors $G_A$ and $G_P$
are based on the calibration of the 
radial wave functions developed in Ref.~\cite{Medium},
for the study of the octet electromagnetic form factors,
the quality of the results is also limited 
by the numerical results in that study.
Therefore, we expect the results for 
the reactions with the nucleon, $\Lambda$
and $\Sigma$ to be more reliable than that with $\Xi$.

\begin{figure}[t]
\vspace{.5cm}
\centerline{
\mbox{
\includegraphics[width=3.0in]{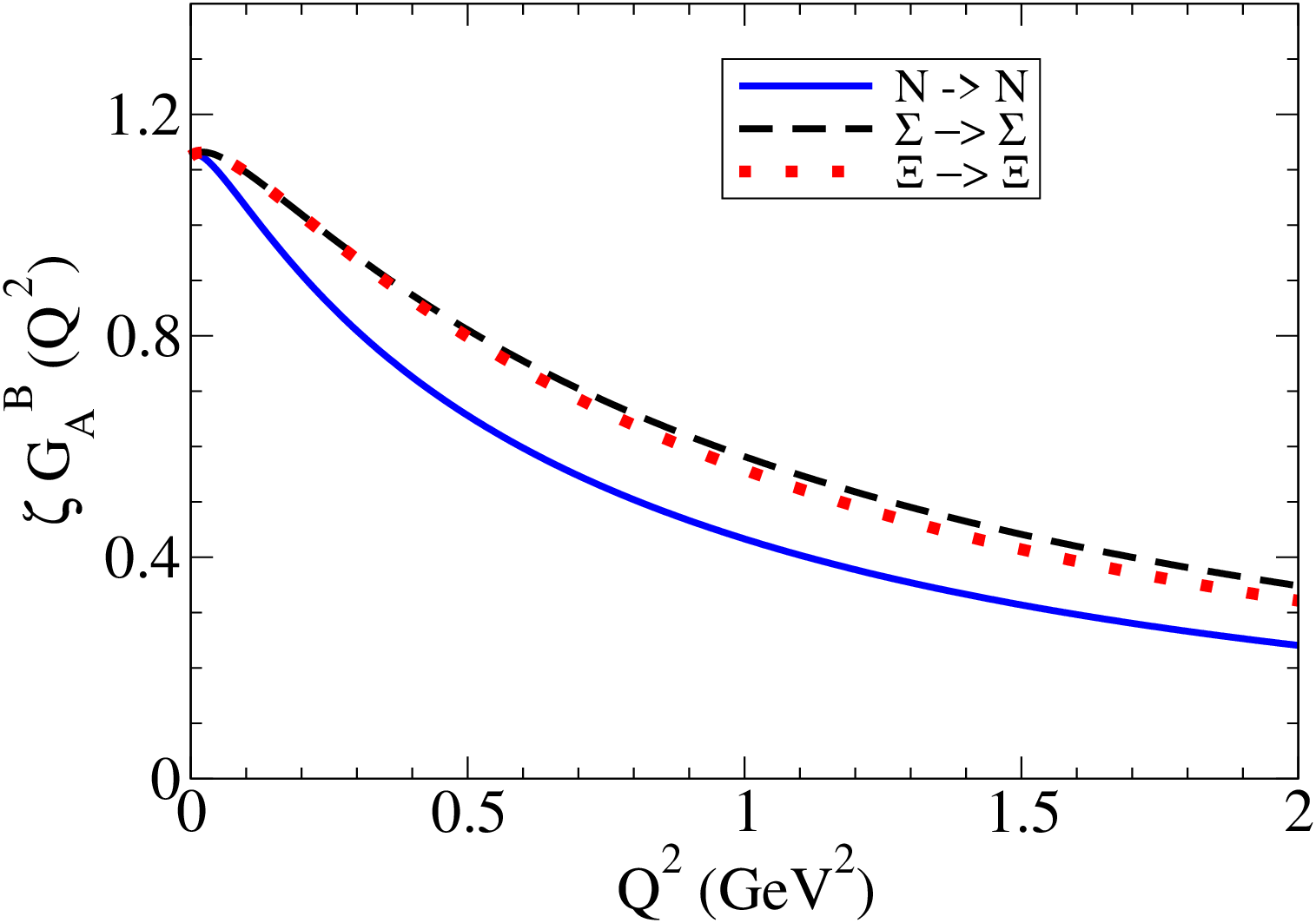}
}}
\caption{\footnotesize{
Result for $G_A^B(Q^2)$ for the octet baryon neutral current transitions 
normalized to the result of the nucleon at $Q^2=0$.
The factor $\zeta$ is defined as $\zeta= G_{A,N}^B(0)/G_A^B(0)$.
Then, one has $\zeta= 5/4$ for $\Sigma$
and $\zeta= -5$ for $\Xi$.
}}
\label{figGA-Octet1}
\end{figure}

\begin{figure}[t]
\vspace{.5cm}
%\vspace{2cm}
\centerline{\mbox{\includegraphics[width=3.0in]{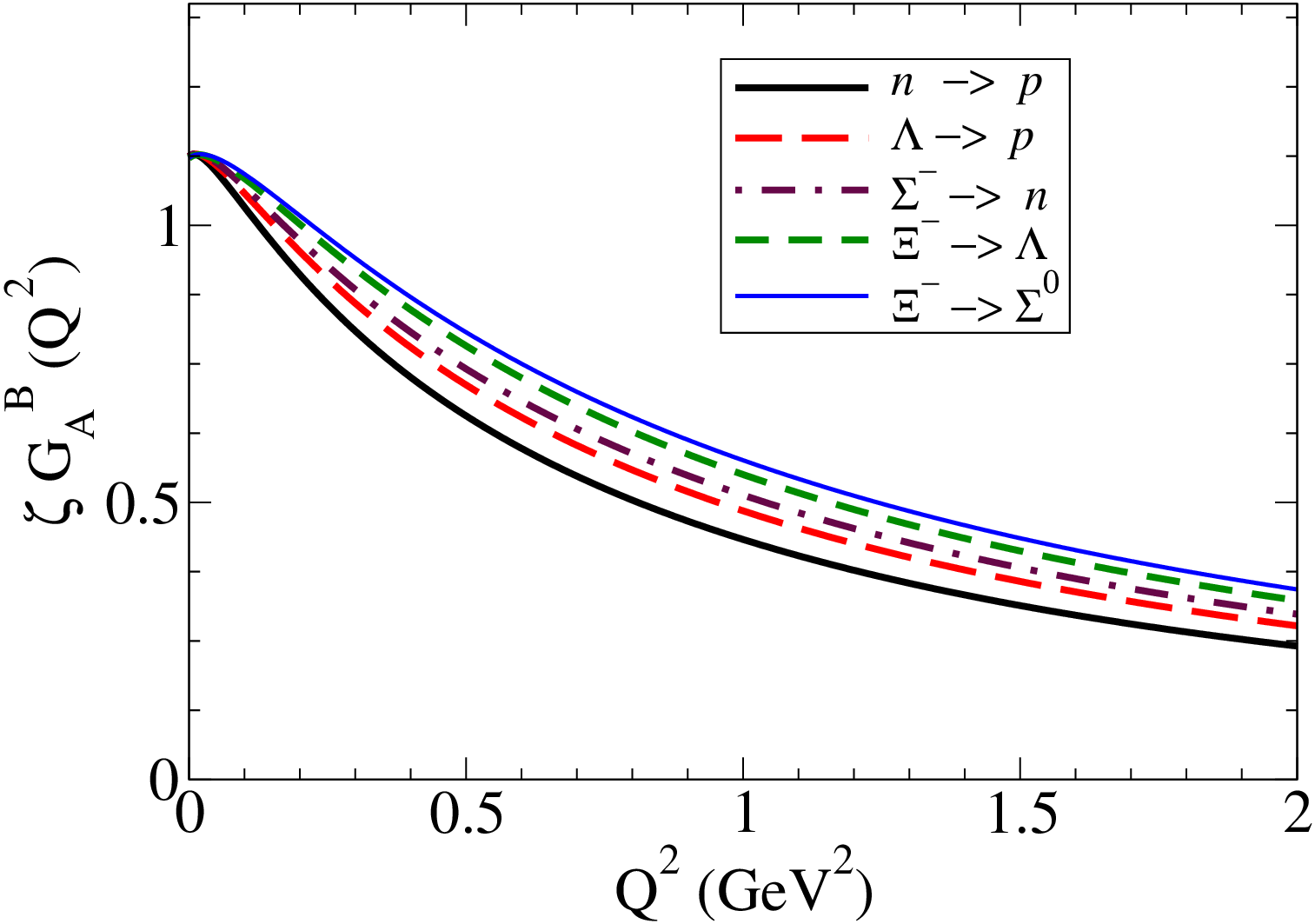}}}
\caption{\footnotesize{
Results for $G_A^B(Q^2)$ for the octet baryon  transitions
with  $\Delta S=1$ normalized to the result for the nucleon 
at $Q^2=0$.
}}
\label{figGA-Octet2}
\end{figure}

\subsection{Contribution of valence quarks for $G_A$}

We start the presentation of our results for the 
octet baryons discussing the effect of the valence quark 
contributions for the form factor $G_A$.  
Since we have not included the meson cloud contribution,  
it can be interesting to compare the 
results with the lattice QCD simulations first.

The comparison of our results 
with the lattice QCD simulations~\cite{Lin09,Erkol10a}
for the neutral current transitions ($N, \, \Sigma, \,\Xi$) 
are presented in Table~\ref{tabGB0}. 
The results for the $\Delta I=1$ and $\Delta S=1$ 
transitions compared with Ref.~\cite{Erkol10a} 
are presented in Table~\ref{tabGB0-2}. 
To avoid any contamination from meson cloud effects,  
we use lattice QCD simulations 
with $m_\pi \approx 500$ MeV from Refs.~\cite{Lin09,Erkol10a} 
for the  comparison.
From Tables~\ref{tabGB0} and~\ref{tabGB0-2}, 
one can see that the results 
of our model are close to the estimates of the lattice QCD.

The results of our model presented 
in Tables~\ref{tabGB0} and~\ref{tabGB0-2}  
are not calculated in the lattice QCD regime, 
as in the case of the nucleon 
[see Secs.~\ref{secNucleonGA} and \ref{secNucleonGP}].
However, this is not an approximation, 
since, due to the constraints of the model,
the transition form factors $G_A$ are independent 
of the masses for $Q^2=0$
(but for $G_P$ we have a correction).
This interesting propriety 
is a consequence of the definition of the
quark current at $Q^2=0$, independent of the 
hadron masses ($m_\rho$ and $m_N$),
and also a consequence of the fact that 
the normalization of the radial wave functions 
is independent of the masses of the baryons 
(normalization defined by the wave functions at the rest frame).

We can consider a more sophisticated 
model where $g_A^q(0)$ and $g_P^q(0)$ 
depend on the constituent quark mass as in Ref.~\cite{Lattice}  
at the expenses of an extra parameter. 
In that case we expect however only a small correction,
as in the case of the  electromagnetic transitions~\cite{Lattice}.
For the purpose  of the present study,  
the approximation that 
$g_A^q(0)$ and $g_P^q(0)$ are independent
of the constituent quark  
mass is sufficient.

In the calculation we use the octet baryon physical 
masses, $M_N= 0.939$ GeV,
$M_\Lambda= 1.116$ GeV, $M_\Sigma= 1.192$ GeV
and $M_\Xi=1.318$ GeV.
The values of $M_{BB'}$ are determined 
using these values.

Back to the discussion of the results in Table~\ref{tabGB0-2},
our calculations are  compatible with 
the lattice results   
within a $\pm\, 20\%$ deviation, 
with two main exceptions: 
the $\Sigma^- \to n$ and $\Xi^- \to \Xi^0$ transitions.
In the  $\Sigma^- \to n$ transition, the 
lattice value deviates also from the estimate of  
$SU(3)$ baryon-meson model 
for the other transitions.
In the case of the $\Xi^- \to \Xi^0$  transition,
the model and the lattice QCD result 
are both small in comparison with the
other transitions.
Since we compare the core effects  
with lattice simulations with $m_\pi \approx 500$ MeV,
one can regard the agreement as reasonable.
Looking in more detail  
for the result of the nucleon,
we note that the model underestimates the lattice result.
However, the lattice results from Ref.~\cite{Erkol10a}
are larger compared with 
similar lattice QCD simulations,
like for instance, the results of Ref.~\cite{Lin09} 
presented in Table~\ref{tabGB0}. 
When the pion mass decreases, the value 
of $G_A(0)$ becomes almost constant 
and close to the physical value,
as far as the lattice volume is 
not too small~\cite{Yamazaki08,Yamazaki09}.
If the lattice volume becomes smaller, 
the lattice result for $G_A(0)$ starts to deviate from 
the continuous limit,  
and strongly underestimates the physical result~\cite{Yamazaki08}.

The results for the neutral transitions and 
$\Delta S=1$ transitions
are presented respectively in Figs.~\ref{figGA-Octet1} and 
\ref{figGA-Octet2}. 
The results are normalized by the value of $G_A^B(0)$ 
for the nucleon in order to better 
observe the differences of falloffs.
The factor $\zeta$ that multiplied to $G_A^B(Q^2)$,  
is defined by $\zeta= G_{A,N}^B(0)/G_A^B(0)$.
We  do  not present a figure for the 
$\Delta I=1$ transitions, 
since those, with the exception 
of the $\Sigma^\pm  \to \Lambda$ case,
are proportional to the results for the neutral transitions
(see Tables~\ref{tabGB0} and~\ref{tabGB0-2}).
The result for $\Sigma^+ \to \Lambda$  
is very close to the $\Xi \to \Xi$
and $\Sigma \to \Sigma$.
The result for $\Sigma^- \to \Lambda$ 
has the opposite sign to that of $\Sigma^+ \to \Lambda$.

In Fig.~\ref{figGA-Octet1} one can see 
that the results for $\Sigma$ and $\Xi$ 
are very similar. 
In  Fig.~\ref{figGA-Octet2}
for simplicity we do not not include the line 
associated with $\Sigma^0 \to p$,
because it is almost on the top of the line 
for $\Lambda \to p$,
since the mass difference 
between $\Sigma$ and $\Lambda$ is small (about 80 MeV).

It is clear from Figs.~\ref{figGA-Octet1} and~\ref{figGA-Octet2} that 
the reactions with heavier baryons 
have slower falloffs with increasing $Q^2$, compared to the nucleon case.
That is a consequence of using  
the physical masses of the baryons in the calculation 
instead of using 
the one common value of the  octet baryon mass 
suggested by the exact $SU(3)$ symmetry,
as well as a consequence of 
the difference in parametrizations of the 
radial wave functions [see Sec.~\ref{secValenceOctet}].

It is also interesting to note 
that the form factors associated with heavy baryons 
(excluding the nucleon)
have very similar falloffs for large $Q^2$, 
although differ in the behavior 
in the range $Q^2= 0, ..., 0.5$ GeV$^2$

\subsection{Results for the axial-vector form factor ($G_A$) }

\begin{figure}[t]
\vspace{.5cm}
%\vspace{2cm}
\centerline{\mbox{\includegraphics[width=3.0in]{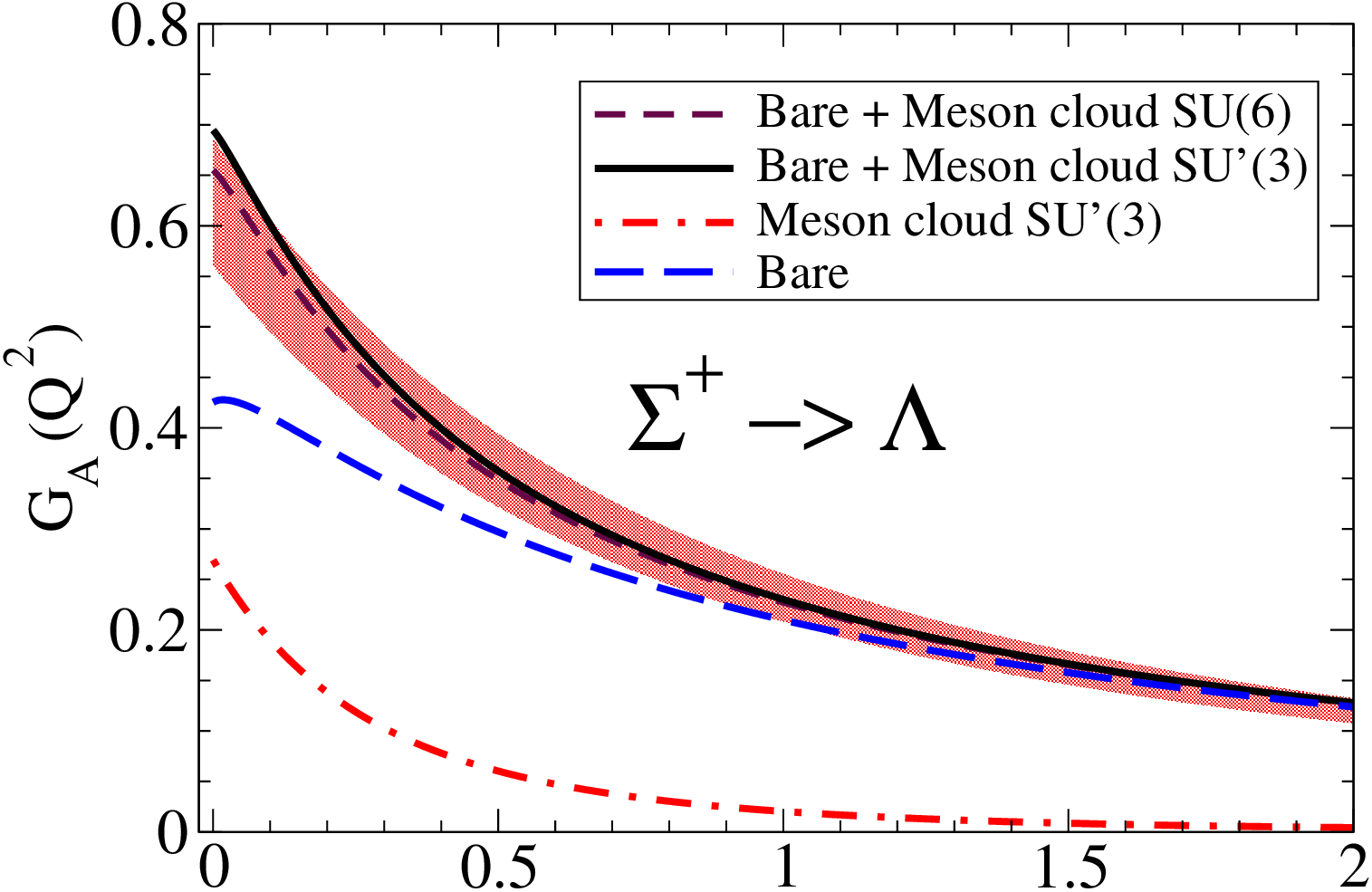}}}
\vspace{.9cm}
\centerline{\mbox{\includegraphics[width=3.0in]{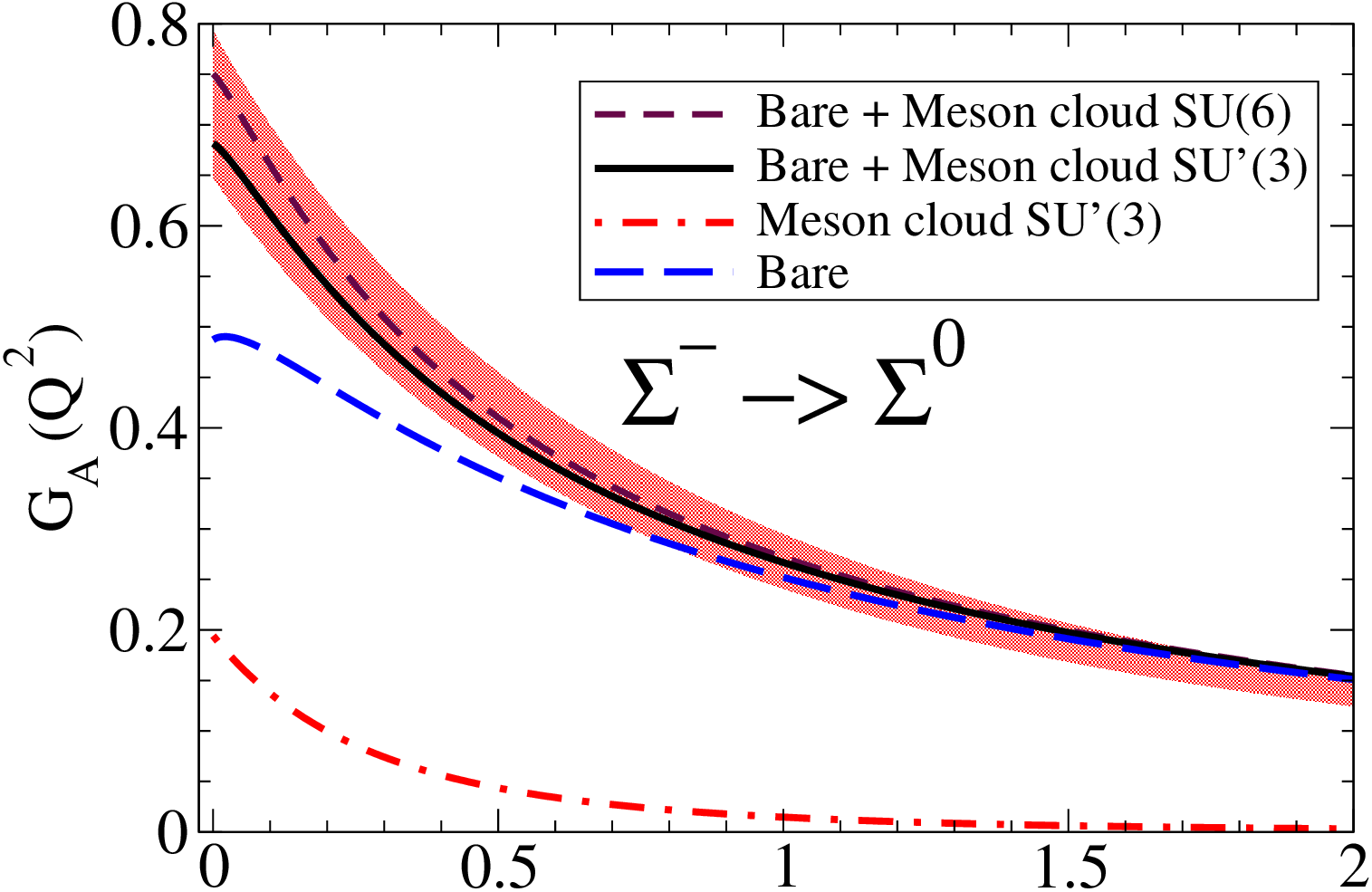}}}
\vspace{.9cm}
\centerline{\mbox{\includegraphics[width=3.0in]{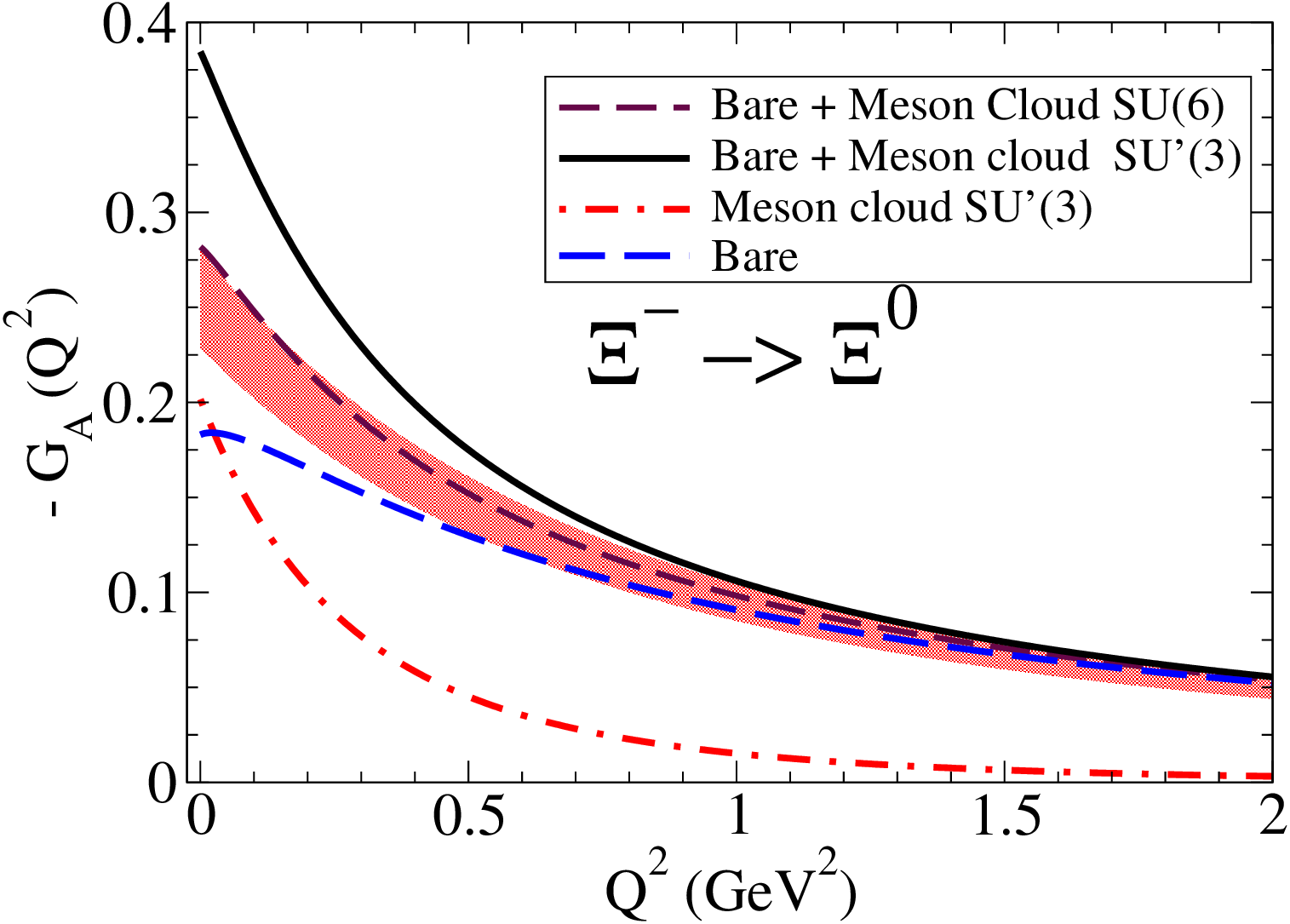}}}
\caption{\footnotesize{
Result of $G_A(Q^2)$ for the octet baryons  with  $\Delta I=1$.
For convenience we changed the sign of the function $G_A$ 
for $\Xi^- \to \Xi^0$.
}}
\label{figGA-DeltaI}
\end{figure}

We present now our predictions for the 
octet baryon $G_A$ form factors 
as functions of $Q^2$
based on the model calibrated in the previous sections 
by the nucleon data.

For the later discussions, it is important to mention  
that the results obtained up to now 
within the covariant spectator quark model
(valence quark contribution) for $G_A^B(0)$ 
correspond to  an $SU(6)$ model with parameters $F=0.675$ and $D=0.450$.
The model fails to describe the data 
because $F+D$ is too small, 
and also because the model breaks the  $SU(6)$ symmetry 
for finite $Q^2$, as already discussed.

As in the case of the nucleon, one has to correct 
the result from the valence quark contribution 
by the normalization factor due to the meson cloud.
We use then
\ba
\tilde G_A^B (Q^2) = \sqrt{Z_{B'}} \sqrt{Z_{B}} G_A^B(Q^2), 
\label{eqGAB-Octet}
\ea
where $Z_{B'}$, $Z_{B}$ are the normalization factors 
associated with the initial and final baryons.

As for the meson cloud contribution,
we consider two possible parametrizations 
that we label as $SU(6)$ and $SU'(3)$  hereafter.

We first explain the $SU(6)$ model for the meson cloud.
In the $SU(6)$ model we assume 
that $SU(6)$ symmetry holds for  
the valence quark component of the form factors 
as well as the meson cloud contribution.
In this case we can write the meson cloud contribution  in the form 
\ba
G_A^{MC}(Q^2)= \eta_{BB'}
\frac{
\sqrt{Z_{B'} Z_{B}}}{Z_N} 
 G_{A,N}^{MC}(Q^2),
\label{eqGMCoctet}
\ea 
where $G_{A,N}^{MC}$ represents 
the parametrization of the nucleon meson cloud contribution 
given by Eq.~(\ref{eqGAmc}) with $\Lambda = 1.05$ GeV.
The coefficient $\eta_{BB'}$ is the factor  
associated with the $SU(6)$ symmetry, the 
coefficient that multiplied to $G$
in the last column of Table~\ref{tabGA}. 
Note that as for the nucleon, 
we include the normalization factors
associated with the octet baryon wave functions.
The factor $1/Z_N$ is introduced to remove 
the dependence on the nucleon's normalization 
in the definition of $G_{A,N}^{MC}$.

% figure 9 and 10 here

\begin{figure}[t]
\vspace{.5cm}
%\vspace{2cm}
\centerline{\mbox{\includegraphics[width=3.0in]{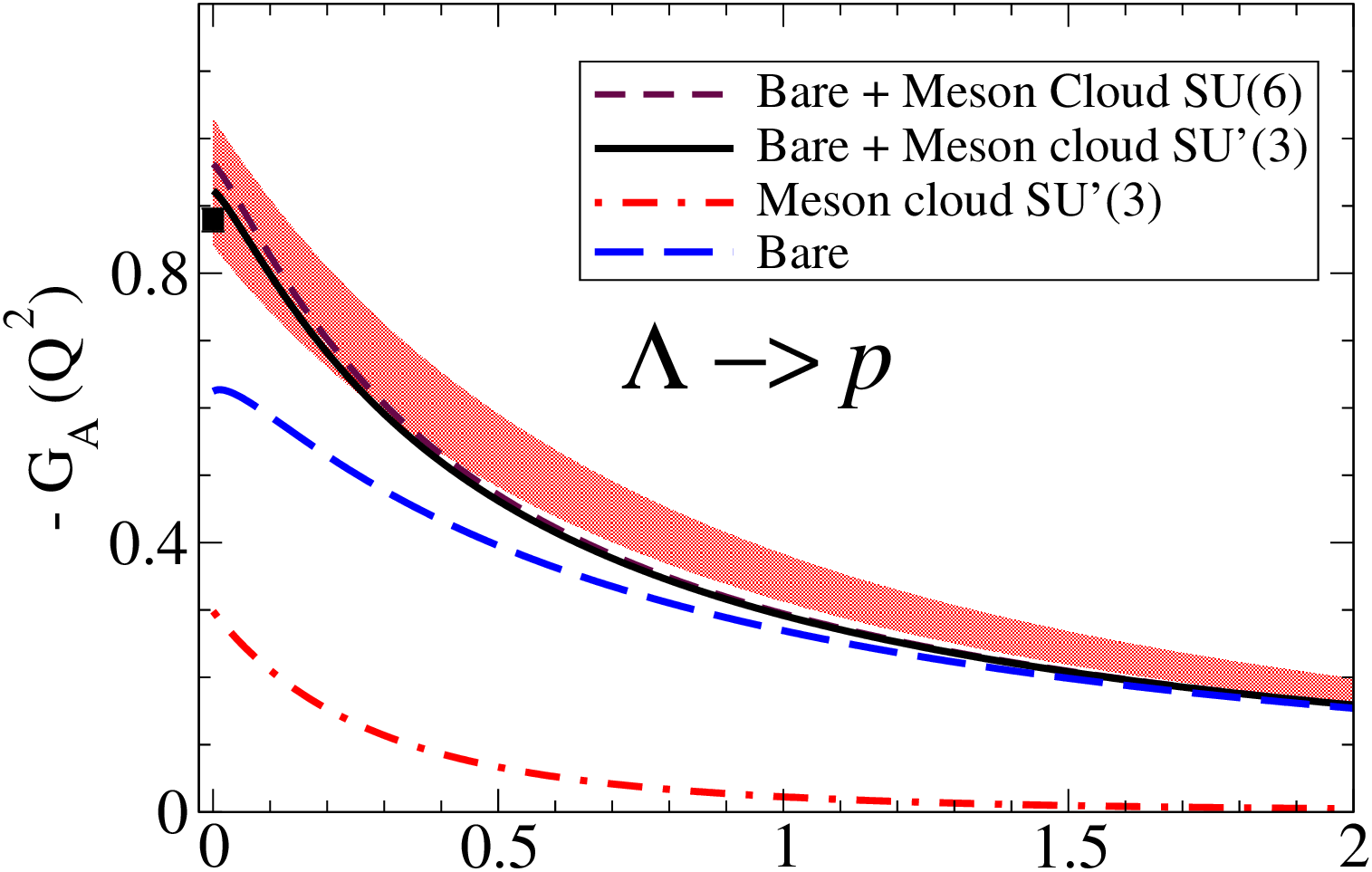}}}
\vspace{.9cm}
\centerline{\mbox{\includegraphics[width=3.0in]{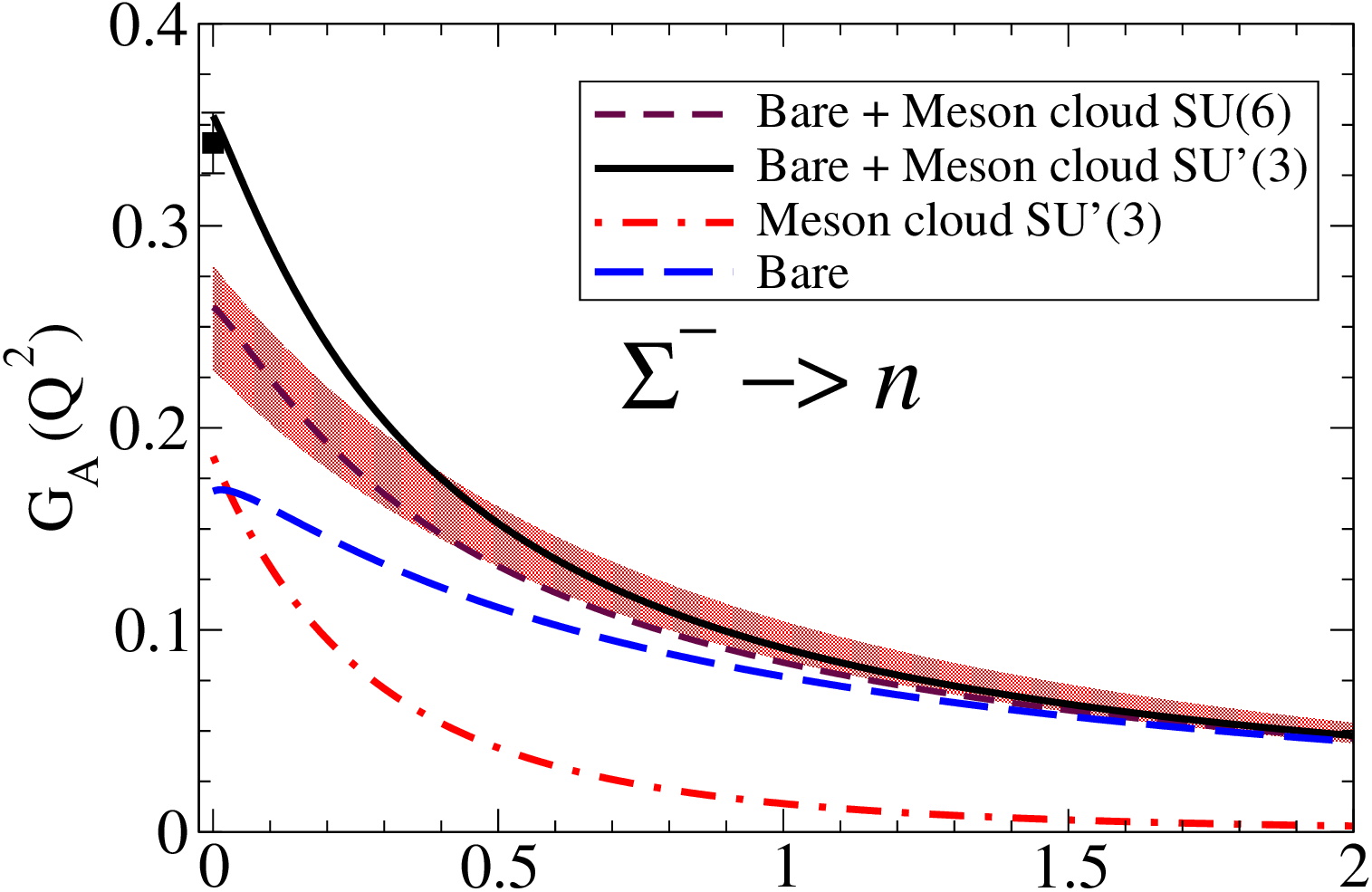}}}
\vspace{.9cm}
\centerline{\mbox{\includegraphics[width=3.0in]{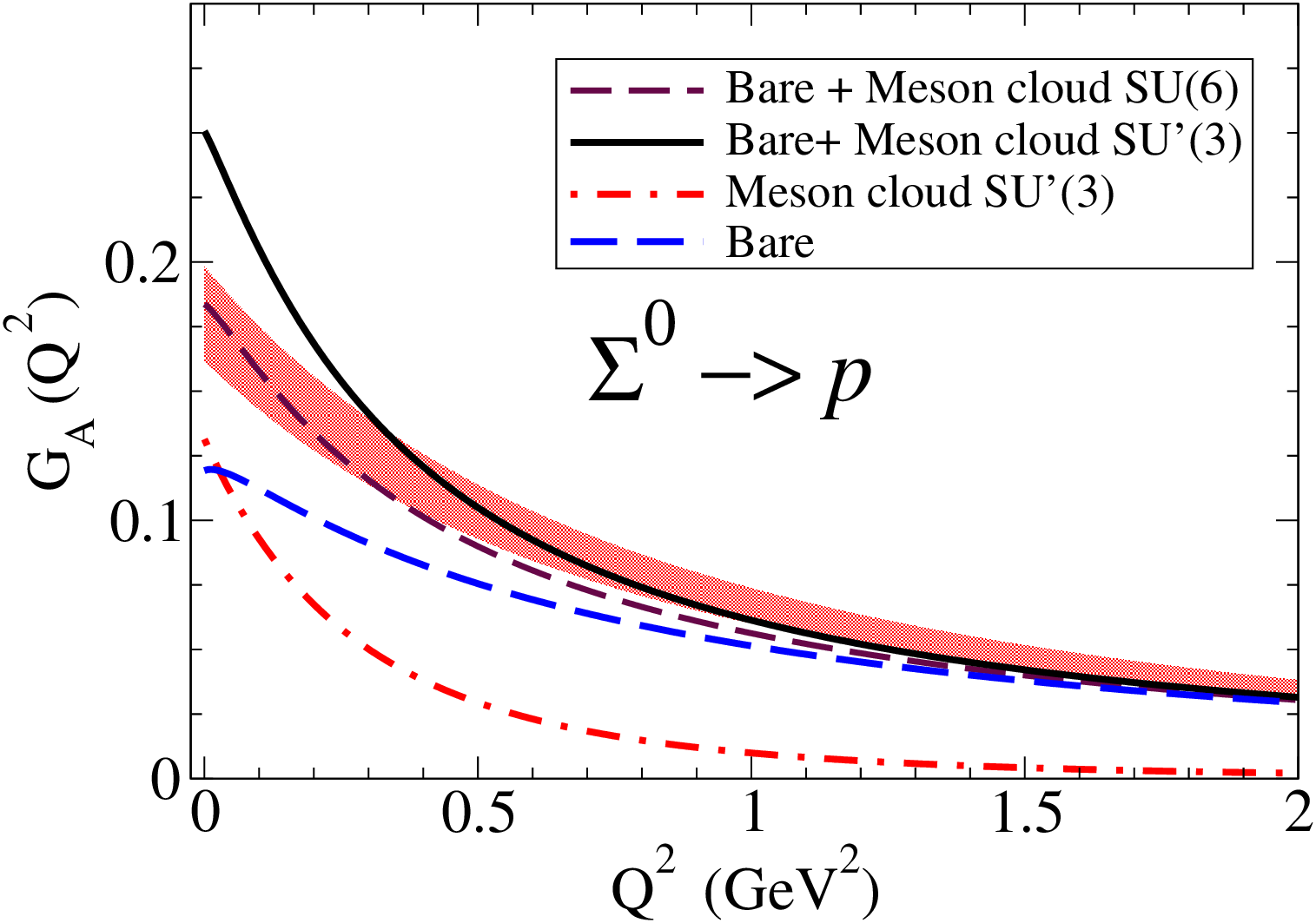}}}
\caption{\footnotesize{
Result of $G_A(Q^2)$ for the octet baryons with
$\Delta S=1$ (part 1).
For convenience we changed the sign of the function $G_A$ 
for $\Lambda  \to p$.
}}
\label{figGA-DeltaS}
\end{figure}
\begin{figure}[t]
\vspace{.36cm}
%\vspace{2cm}
\centerline{\mbox{\includegraphics[width=3.0in]{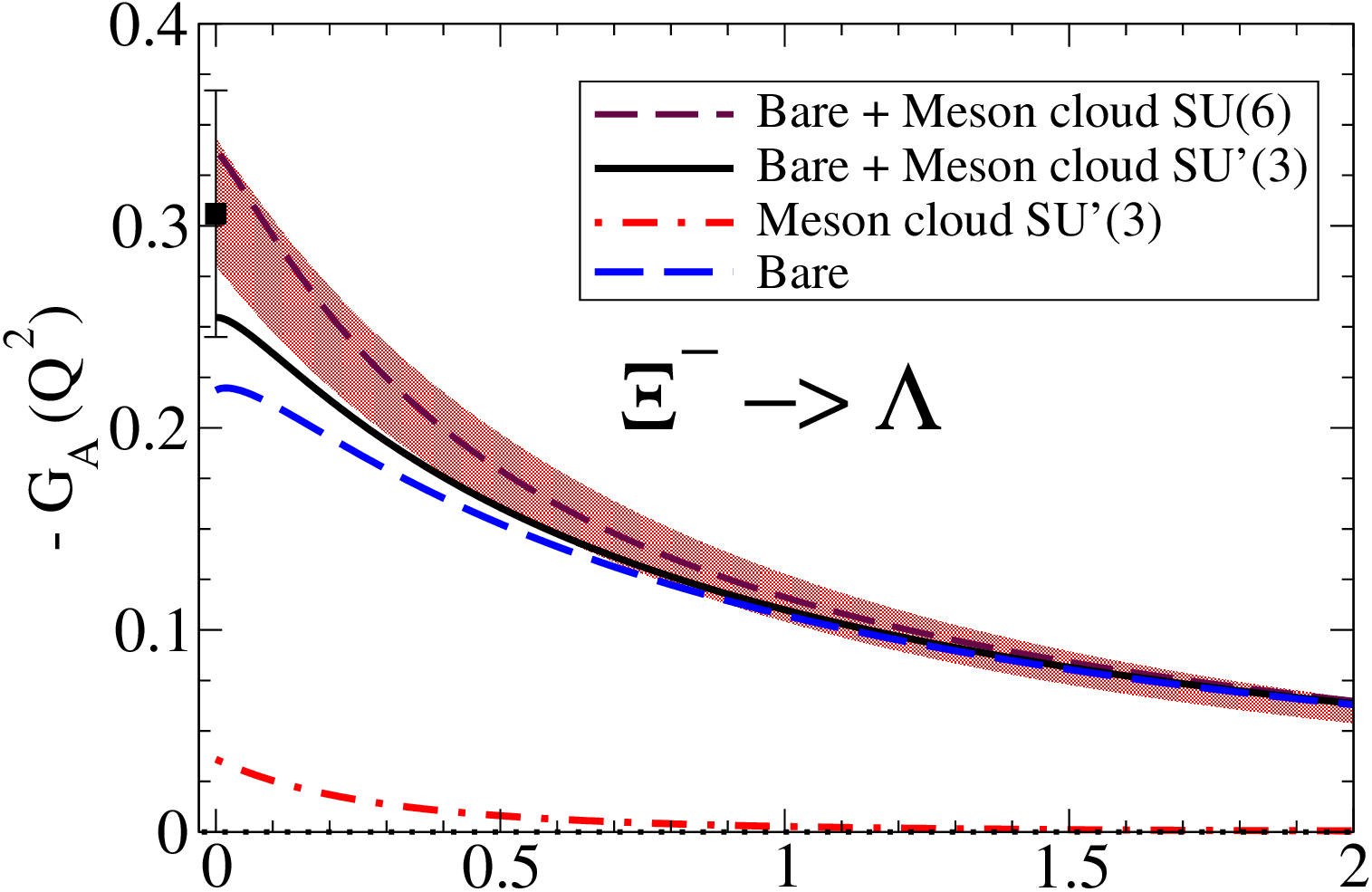}}}
\vspace{.9cm}
\centerline{\mbox{\includegraphics[width=3.0in]{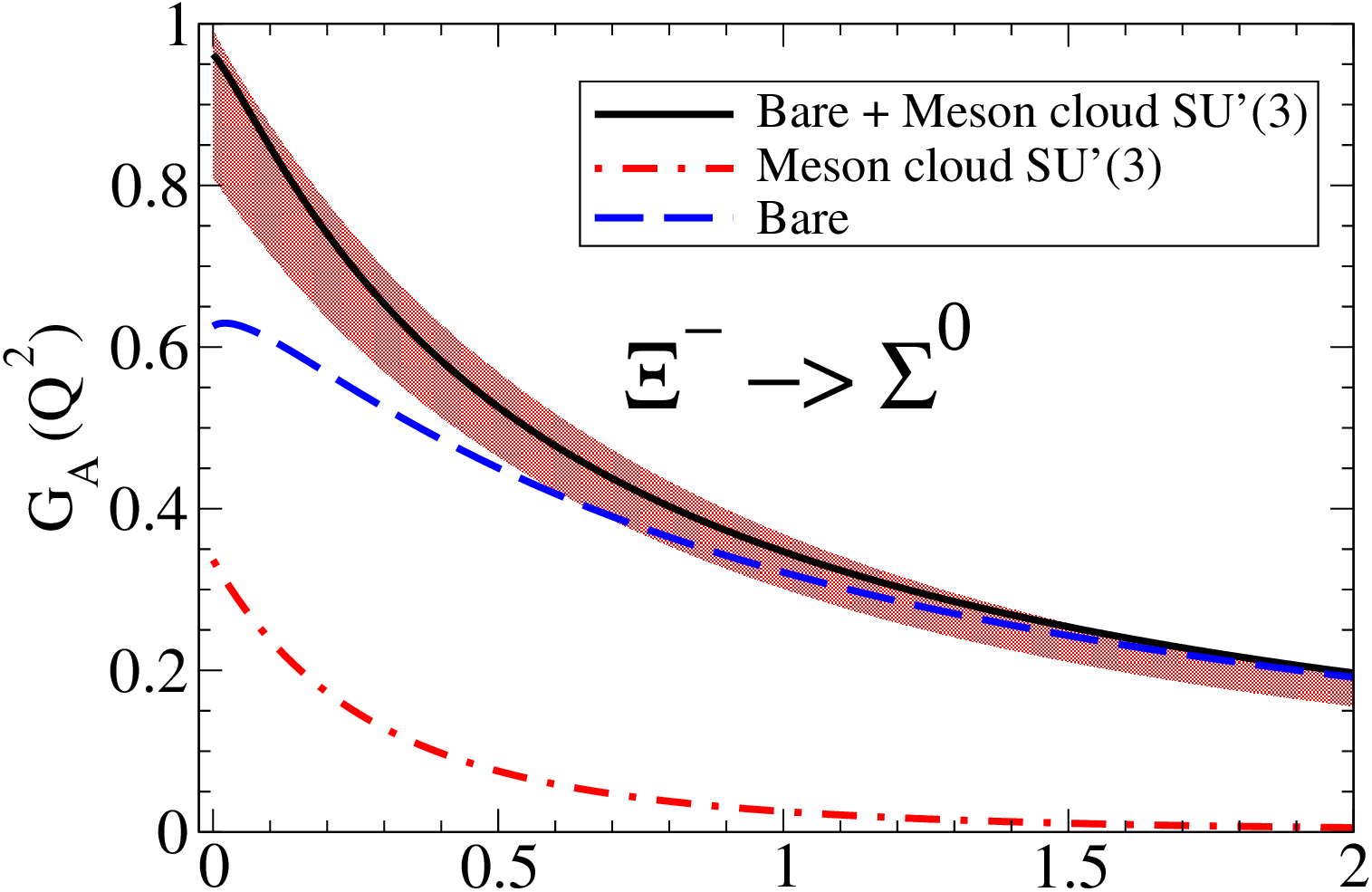}}}
\vspace{.9cm}
\centerline{\mbox{\includegraphics[width=3.0in]{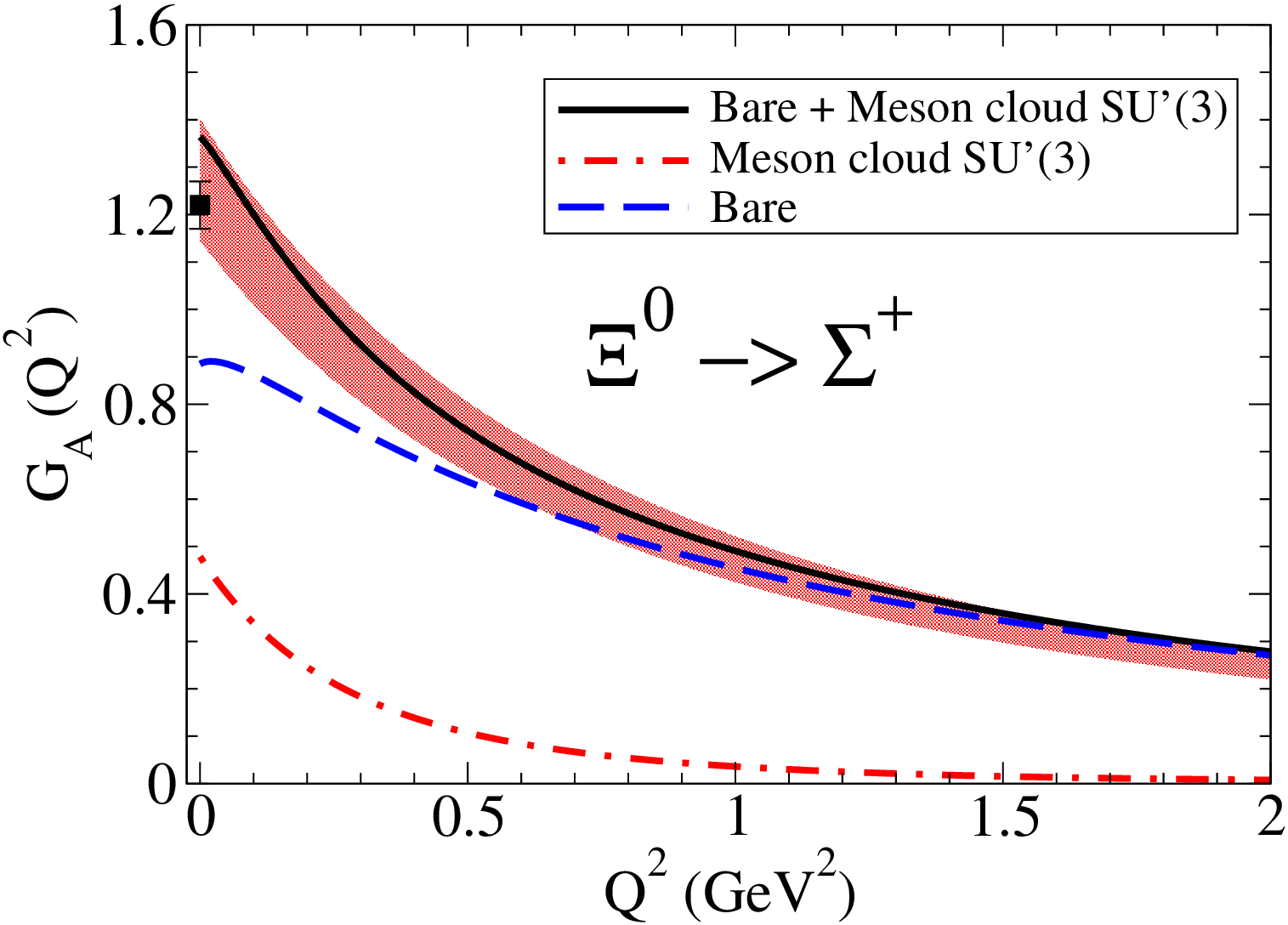}}}
\caption{\footnotesize{
Result of $G_A(Q^2)$ for the octet baryons with 
$\Delta S=1$ (part 2).
For convenience we changed the sign of the function $G_A$ 
for $\Xi^-  \to \Lambda$.
}}
\label{figGA-DeltaS2}
\end{figure}

Contrarily to the contribution 
from the valence quarks~(\ref{eqGAB-Octet}), 
the meson cloud contribution is independent 
of the baryons masses in the $SU(6)$ model.
We ignore here minor differences  due to the 
normalization constants $Z_B$ and $Z_{B'}$,
since the corrections are of about 2--10\%.
This is a consequence of the $SU(6)$ assumption, 
that the octet baryons have all the same mass.

Since $SU(6)$ symmetry
is not expected to work  well in general
for the meson cloud component of the form factors 
as discussed in Sec.~\ref{secSU6},
we consider the possibility of improving the
meson cloud model given by Eq.~(\ref{eqGMCoctet})
by a direct fit to the data at $Q^2=0$.
To achieve this goal, we use an alternative parametrization 
for the meson cloud, where the octet baryon data for $G_A(0)$
are fitted by an $SU(3)$ model for the meson cloud component.
In this model the $SU(6)$ structure 
for the quark model is preserved, but 
the meson cloud contribution is 
determined by an $SU(3)$ parametrization 
in terms of the two coefficients denoted by 
$F'$ and $D'$, that replace the coefficients $F$ and $D$
in the $SU(3)$ baryon-meson model  
discussed in Sec.~\ref{secSU6}.
We break then the $SU(6)$ symmetry in the meson cloud component.
To avoid misinterpretations, we label 
this model as the $SU'(3)$ model for the meson cloud.

In the $SU'(3)$ model we write the  meson cloud contribution as
\ba
G_A^{MC}(Q^2)= \eta_{BB'}^\prime
\frac{
\sqrt{Z_{B'} Z_{B}}}{Z_N} 
 G_{A,N}^{MC}(Q^2),
\label{eqGMCoctet2}
\ea 
where $\eta_{BB'}^\prime$ represents 
the expressions in the column labeled as $SU(3)$ in 
Table~\ref{tabGA}, 
with the replacements $F \to F'$,  $D \to D'$,
normalized by $F' +D'$.
From the fit to the data 
we obtain 
$F'=0.1784$ and $D'= 0.4273$.
The results obtained for the 
$n \to p$, $\Xi^- \to \Sigma^0$ 
and $\Xi^- \to \Sigma^0$ transitions, 
where the meson cloud contributions 
is proportional to $F' + D'$,  
are indistinguishable from the $SU(6)$ model.
This happens because the fit is strongly constrained 
by the result of $G_A(0)$ for the 
$n \to p$ transition due to the high accuracy of the datapoint.
Since in practice $F' + D'$ is fixed,  
we fit only the relative size of the coefficients $F'$ and $D'$.
Therefore the $SU'(3)$ model is the result of 
a fit with one parameter only.

It is worth to mention 
that models based on the $SU(3)$ symmetry 
fits to subsets of the data were used already in the past 
in attempts to interpret the impact of  
the $SU(3)$ symmetry breaking effect~\cite{Gaillard84,Kubodera88,Serrano14}.

The results for the octet baryon axial-vector form factors 
for $\Delta I=1$,
with the exception of the nucleon discussed earlier,
are presented in Fig.~\ref{figGA-DeltaI}.
The results for $\Delta S=1$
are presented in Figs.~\ref{figGA-DeltaS} 
and \ref{figGA-DeltaS2}.

In Figs.~\ref{figGA-DeltaI}, \ref{figGA-DeltaS} 
and~\ref{figGA-DeltaS2} we include the contribution from the 
valence quark component (long-dashed line),
the contribution from the $SU'(3)$ model 
for the meson cloud (doted-dashed line),  
and the final result for the 
model with the $SU'(3)$ meson cloud (solid line).
In addition, we present the final result 
(bare + meson cloud) for the  
$SU(6)$ meson cloud  model (short-dashed line).
The meson cloud contribution of the $SU(6)$ meson cloud
model is not presented to avoid the superposition of lines.
The difference between the meson cloud components
between the $SU'(3)$ 
and the $SU(6)$ models 
can however be estimated by 
the difference between the result 
``bare + meson cloud'' in the two models.
For the transitions 
$\Xi^- \to \Sigma^0$ 
and $\Xi^0 \to \Sigma^+$, we omit the indication 
of the $SU(6)$ model since it  
is equivalent to the $SU'(3)$ model, 
as discussed already.

In order to compare our results 
with the estimates of the 
$SU(6)$ baryon-meson model discussed in Sec.~\ref{secSU6},
which are independent of the baryon  masses,  
we present a band  (at red)
given by the parametrization inspired
by the fit to the nucleon data, 
\ba
G_A^{SU(6)}(Q^2) = \eta_{BB'}
\frac{G_A^{0}}{\left( 1 + \frac{Q^2}{M_A^2}\right)^2},
\label{eqTotalSU6}
\ea
where $M_A$ is cutoff parameter.
The band indicates a  $\pm \, 10$\% variation
from Eq.~(\ref{eqTotalSU6}).

It was suggested by Gaillard and Sauvage~\cite{Gaillard84} 
that $M_A= 1.05$ GeV for $\Delta I=1$
and  $M_A= 1.25$ GeV for $\Delta S=1$. 
Using the two different parametrizations 
for $\Delta I=1$ and  $\Delta S=1$
we take into account in an effective way 
the modification due to the octet baryon mass difference 
in the $SU(6)$ baryon-meson model.
We realize however, that our model
cannot be compared with the results of 
$M_A= 1.05$ GeV 
for both cases, $\Delta S=1$ and $\Delta I=1$, 
except for  the case of the 
nucleon, discussed in Sec.~\ref{secNucleonGA}.
Therefore, we compare all ours estimates with $M_A= 1.25$ GeV.
At $Q^2=0$, we compare also the results 
with the data from Particle Data Group (PDG)~\cite{PDG2014}.

We recall that, although we present different 
parametrizations for the meson cloud 
that differ at low $Q^2$,  
our results may be considered true predictions 
in the high $Q^2$ region, 
since the result is extrapolated 
from the model calibrated by the lattice QCD data 
as well as the high $Q^2$ data for the nucleon.
In the large $Q^2$ region  the  meson cloud contributions 
are very small and the valence quark effects dominate.

The results of both models are close 
to the data, but the model $SU'(3)$ 
gives a better description of the 
\mbox{$\Sigma^- \to n$} data.
Larger difference between the 
$SU(6)$ and the $SU'(3)$ parametrizations 
is also observed for the reactions 
with small magnitude for $G_A(0)$
($\Xi^- \to \Xi^0$, $\Sigma^- \to n$  and $\Sigma^0 \to p$).
This is  a consequence of 
the large meson cloud contributions 
compared to those of the valence quarks.
In any case, we should not expect an excellent agreement 
with the $G_A(0)$ data, since the 
$SU(3)$ symmetry at the quark level is already broken
(based on the octet baryon masses 
the violation is about 20\%).

In the comparison with the data at $Q^2=0$
the deviation is less than 
five standard deviations, and better than 24\% 
for the model $SU(6)$. 
As for the model $SU'(3)$ the deviations
is less than three standard deviations 
and better than 17\%.

It is also interesting to note 
that the estimate of the meson cloud effects 
based on the $SU(6)$ parametrization 
is in general larger than the 
estimate of the $SU(6)$ baryon-meson model 
(given by the central value of the red band for $Q^2=0$),
particularly for the transitions involving $\Xi$  
[see  Fig.~\ref{figGA-DeltaS2}]. 
This happens because our model 
corrects the estimate made for the nucleon 
with the normalizations of the octet baryon 
wave functions according to Table~\ref{tabZB}.
Therefore, the contributions of the 
valence quark core and the meson cloud 
are enhanced by the factor 
$\sqrt{\frac{Z_{B'}}{Z_N}}\sqrt{\frac{Z_{B}}{Z_N}}>1$.

As mentioned above, the falloffs of the 
form factors from Figs.~\ref{figGA-DeltaI}, 
\ref{figGA-DeltaS}  and \ref{figGA-DeltaS2} 
are slower than the falloff estimated for the nucleon 
($M_A \simeq 1.05$ GeV).
The estimate of the falloff 
based on a dipole form near $Q^2=1$ GeV$^2$ 
gives values of $M_A$ in the range of 1.2--1.4 GeV.
We conclude that the falloffs 
for the octet baryon $G_A$ form factors, 
except for the nucleon, are consistent with 
the conjecture made by  Gaillard and Sauvage ($M_A= 1.25$ GeV)
for the $\Delta S=1$ transitions~\cite{Gaillard84}.

\subsection{Results for the induced pseudoscalar 
\mbox{form factor ($G_P$)}}

\begin{table*}[t]
\begin{center}
\begin{tabular}{|l | r r |r r | r r |}
\hline
\hline
       &  &  & $SU(6)$ & &   $SU'(3)$ &  \\
\hline 
 & $G_P^B(0)$  & $G_P^{\rm pole}(0)$ & $G_P^{\rm pole *}(0)$ & $G_P(0)$ 
& $G_P^{\rm pole *}(0)$ & $G_P(0)$  \\
\hline
 $n \to p$ & $-5.532$ & 152.160 & 234.339 & 228.806  
& 234.339 & 228.806 \\
$\Sigma^+ \to \Lambda$
  & $-5.186$   &  118.365 &  182.292 &   177.107
               &  193.283 &   188.098 \\
$\Sigma^- \to \Sigma^0$
  & $-6.444$   &  144.635  &  118.365 &   216.308     
                           &  202.289 &   195.845 \\
$\Xi^- \to \Xi^0$
 &   2.946 & $-66.420$    &   $-102.294$ & $ -99.346$  
                         & $  -139.590$ & $ -136.643$ \\
\hline
$\Lambda \to p$ & 5.567   & $-10.771$   &  $  -16.591$    & $-11.023$  
                          & $ -10.771$ & $-5.203$ \\
$\Sigma^- \to n$ & $-1.675$  &  4.816 &  3.141 
                          & 3.126 &  6.572 &  4.897 \\
$\Sigma^0 \to p $ &  $ -1.185$ &   2.211  &  4.816  &  3.140  
                               &   4.647  &  3.462 \\
$\Xi^- \to \Lambda$ & 2.964 & $-5.282$  & $  -8.144$    & $ -5.180$    
                                & $-6.153$ & $-3.189$ \\   
$\Xi^- \to \Sigma^0$ & $-9.158$ & 16.084 &  24.783 &  15.625 
                                         &  24.783 &  15.625 \\
$\Xi^0 \to \Sigma^+$ & $-12.951$ &   22.747 &  35.049 &  22.097 
                                            &  35.049 &  22.097 \\
\hline
\hline
\end{tabular}
\end{center}
\caption{Contributions for $G_P$ at $Q^2=0$ 
for the meson cloud models $SU(6)$ and $SU'(3)$.
$G_P^{\rm pole *}(0)$ represent the contribution 
of the meson pole term with the replacement $G_A^B \to G_A$.}
\label{tabGP}
\end{table*}

We discuss now the results for the 
induced pseudoscalar 
form factors of the octet baryons.
The case of the nucleon has already been discussed in  
Sec.~\ref{secNucleonGP}.
We recall that $G_P$  has a contribution from 
a pseudoscalar meson pole (pion or kaon)
that subsequently decays 
into a lepton-neutrino pair.

%Thus,  
The reaction associated
with $\Delta I=1$ transitions have 
a contribution of the pion pole
(\ref{eqGPpole}), 
related with the $u \leftrightarrow d$ transitions.
Then, similarly to the case for the nucleon, 
recalling that  $M_{B'} + M_B= 2 M_{BB'}$,
we use
\ba
G_P^{\rm pole}(Q^2) = \frac{(M_{B'} + M_B)^2}{m_\pi^2 + Q^2} G_A^B(Q^2).
\label{eqGPoctet}
\ea 
In the above $G_P$ and $G_A^B$ represent now the 
form factors associated with the $B \to B'$ transition.
The pole term dominates in general the  $\Delta I=1$ transition 
as we will show.
In addition to the pole term, 
there is also the contribution from the quark core  
due the non-zero values of the 
quark form factor $g_A^q$ and $g_P^q$. 
As in the case of the nucleon, we take into account 
the meson cloud effect for $G_P$ 
replacing the contribution from $G_A^B$ 
(core contribution) by the dressed $G_A$,
given by the replacement  $G_A^B \to \tilde G_A^B + G_A^{MC}$.
The expressions for the contributions
$\tilde G_A^B$  and  $G_A^{MC}$
have already been discussed in the previous section.
Also the expressions derived for $G_P^B$,  
presented in Sec.~\ref{secValenceOctet},  
have to be corrected 
in the physical limit by the factor $\sqrt{Z_{B'} Z_B}$.

% atempt 1- figs 11 here
\begin{figure}[t]
\vspace{.5cm}
%\vspace{2cm}
\centerline{\mbox{\includegraphics[width=3.0in]{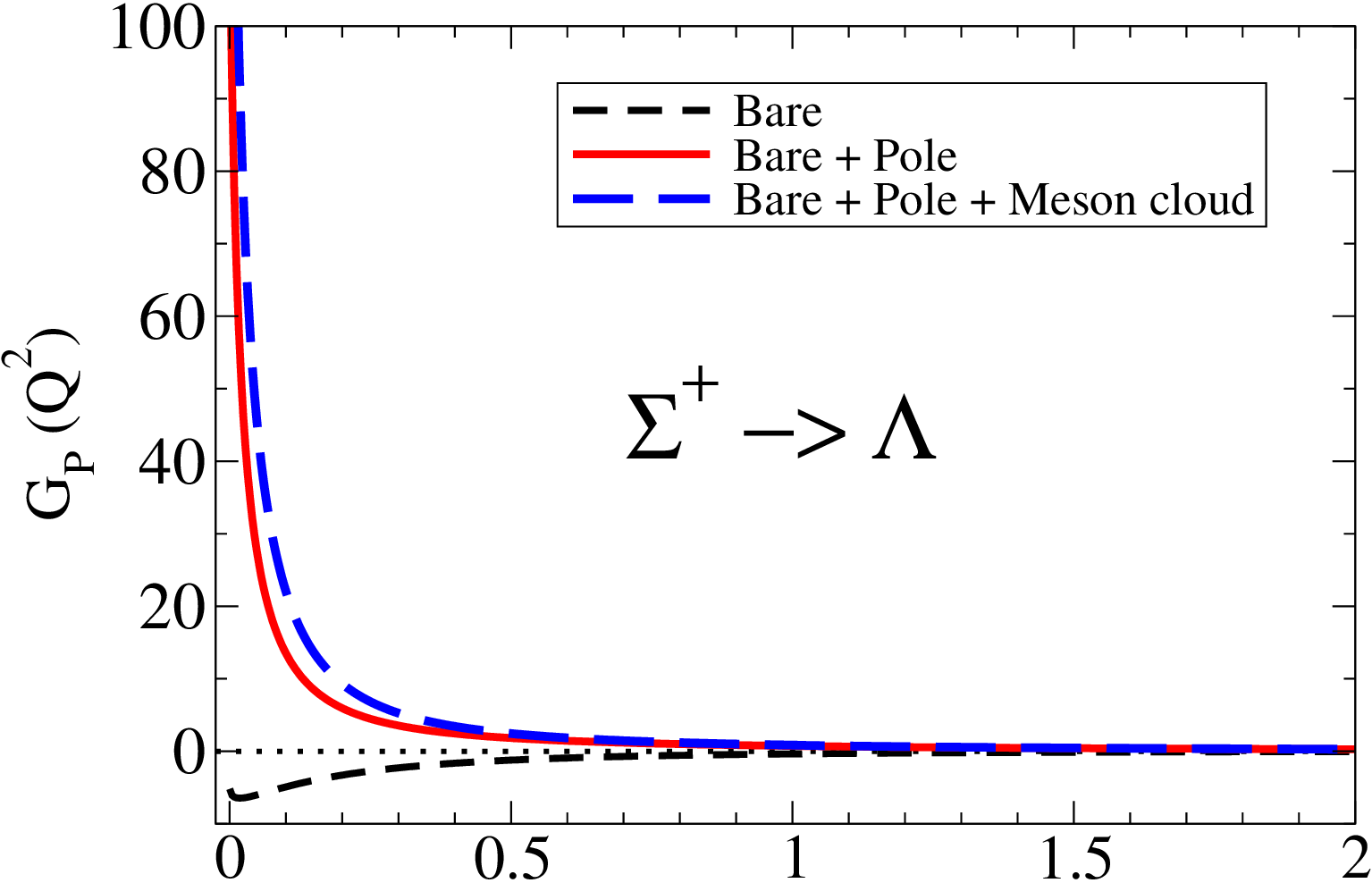}}}
\vspace{.9cm}
\centerline{\mbox{\includegraphics[width=3.0in]{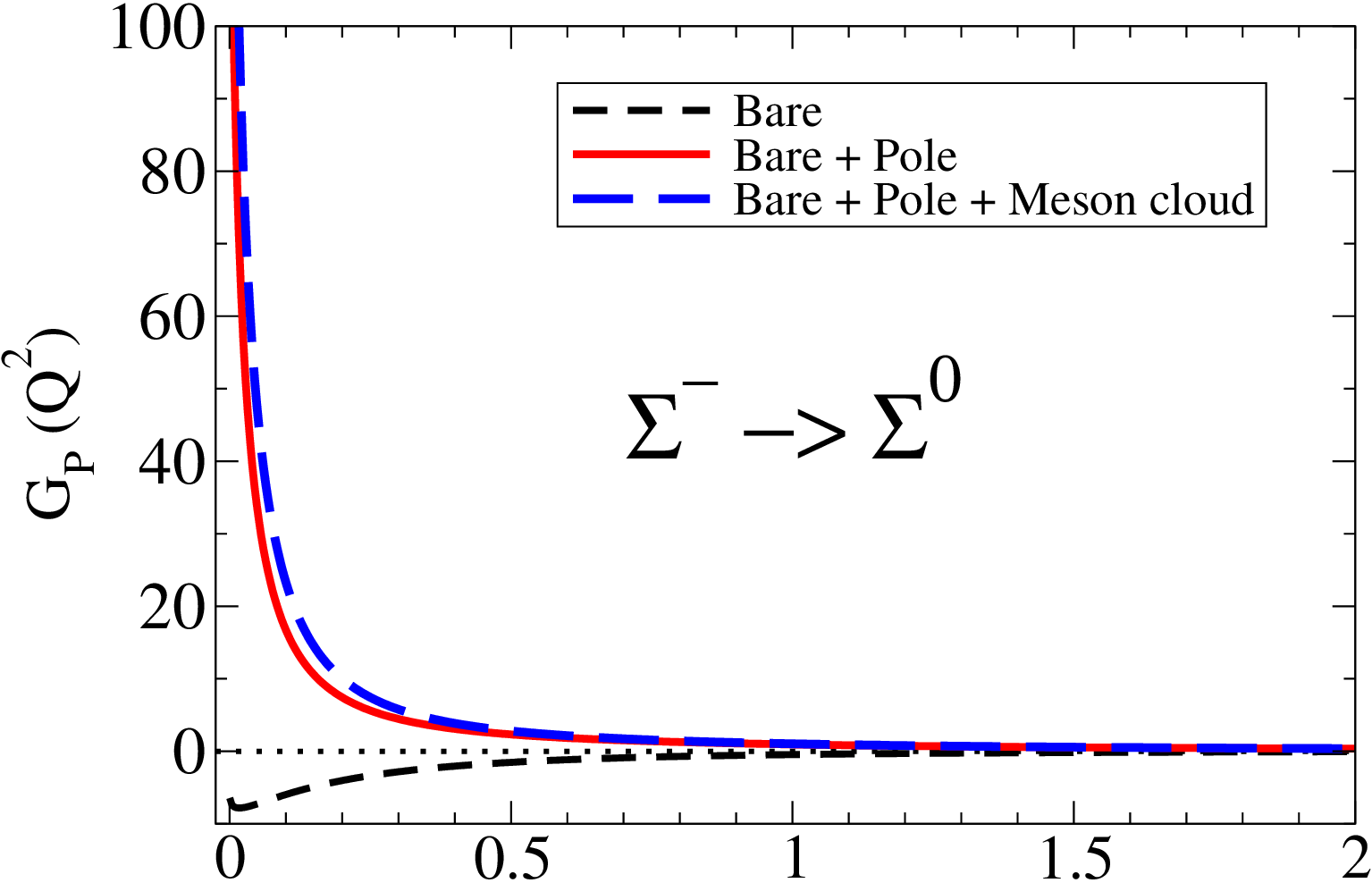}}}
\vspace{.9cm}
\centerline{\mbox{\includegraphics[width=3.0in]{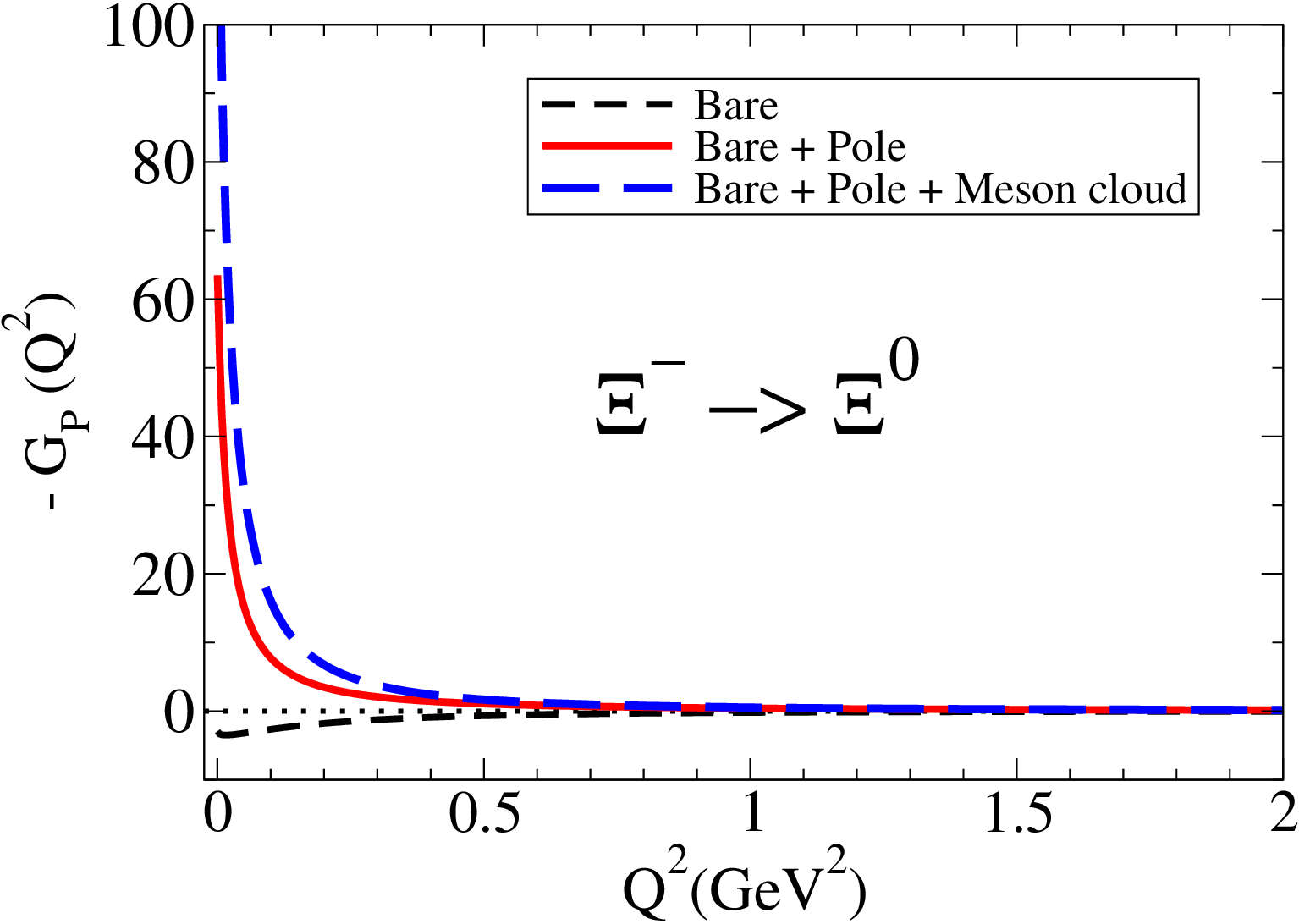}}}
\caption{\footnotesize{
Result of $G_P(Q^2)$ for the octet baryons for 
\mbox{$\Delta I=1$.}
For convenience we changed the sign of the function $G_P$ 
for $\Xi^- \to \Xi^0$.
}}
\label{figGP-DeltaI}
\end{figure}

The  $\Delta S=1$ transitions,
include the transitions between  
the quarks $u$ and $s$, 
are associated with a kaon pole.
Therefore in the $\Delta S=1$ case, 
we replace the pion pole 
on the $\Delta I=1$ case  by the kaon pole.
Replacing $m_\pi^2 \to m_K^2$ in  Eq.~(\ref{eqGPoctet}),  we obtain
\ba
G_P^{\rm pole}(Q^2) = \frac{(M_{B'} + M_B)^2}{m_K^2 + Q^2} G_A^B(Q^2).
\label{eqGPoctet2}
\ea 
Although it may be questionable  
that the pole contribution for $G_P$ obtained for 
the $\Delta I=1$ transition (and for the nucleon)
using PCAC in the chiral limit ($m_\pi$ negligible),
may be generalized for the $\Delta S=1$ transition,
as suggested in Ref.~\cite{Kubodera88},
there are arguments that support this generalization.
The first argument is based on the fact that 
lattice QCD simulations for the octet axial-vector couplings 
follow the generalization 
of the Goldberger-Treiman relation~\cite{Erkol10a},
which is related with Eq.~(\ref{eqGPoctet2}) near $Q^2=0$.
The second argument is that the overall 
description given by our model 
for the $G_A$ and $G_P$ 
lattice data in a wide range  of the
pion masses ($m_\pi =350,...,500$ MeV),  
motivate also the use of Eq.~(\ref{eqGPoctet2}) 
for $\Delta S=1$.

To obtain the ``bare'', ``bare + pole'' 
and total result "bare + pole + meson cloud''
for $\Delta S=1$,
we use the same procedure already discussed  
for $\Delta I=1$.

The results for the case $Q^2=0$ are 
presented  in Table~\ref{tabGP}
for the models $SU(6)$ 
and  $SU'(3)$.
The results for the nucleon, discussed in Sec.~\ref{secNucleonGP}, 
are also included for the sake of the discussion.
The differences between the  two models are 
the consequence of the difference in  
the meson cloud for $G_A(0)$.
Note the difference of results 
for $G_P^{\rm pole}(0)$ and consequently $G_P(0)$ 
in the cases $\Xi^- \to \Xi^0$ and $\Lambda \to p$.
The effect of the kaon pole for the  $\Lambda \to p$ transition 
($\Delta S=1$) is much smaller 
than the pion pole for the $\Xi^- \to \Xi^0$ transition ($\Delta I=1$).
The physical pion is closer to the chiral limit than the kaon.

\begin{figure}[t]
\vspace{.45cm}
%\vspace{2cm}
\centerline{\mbox{\includegraphics[width=3.0in]{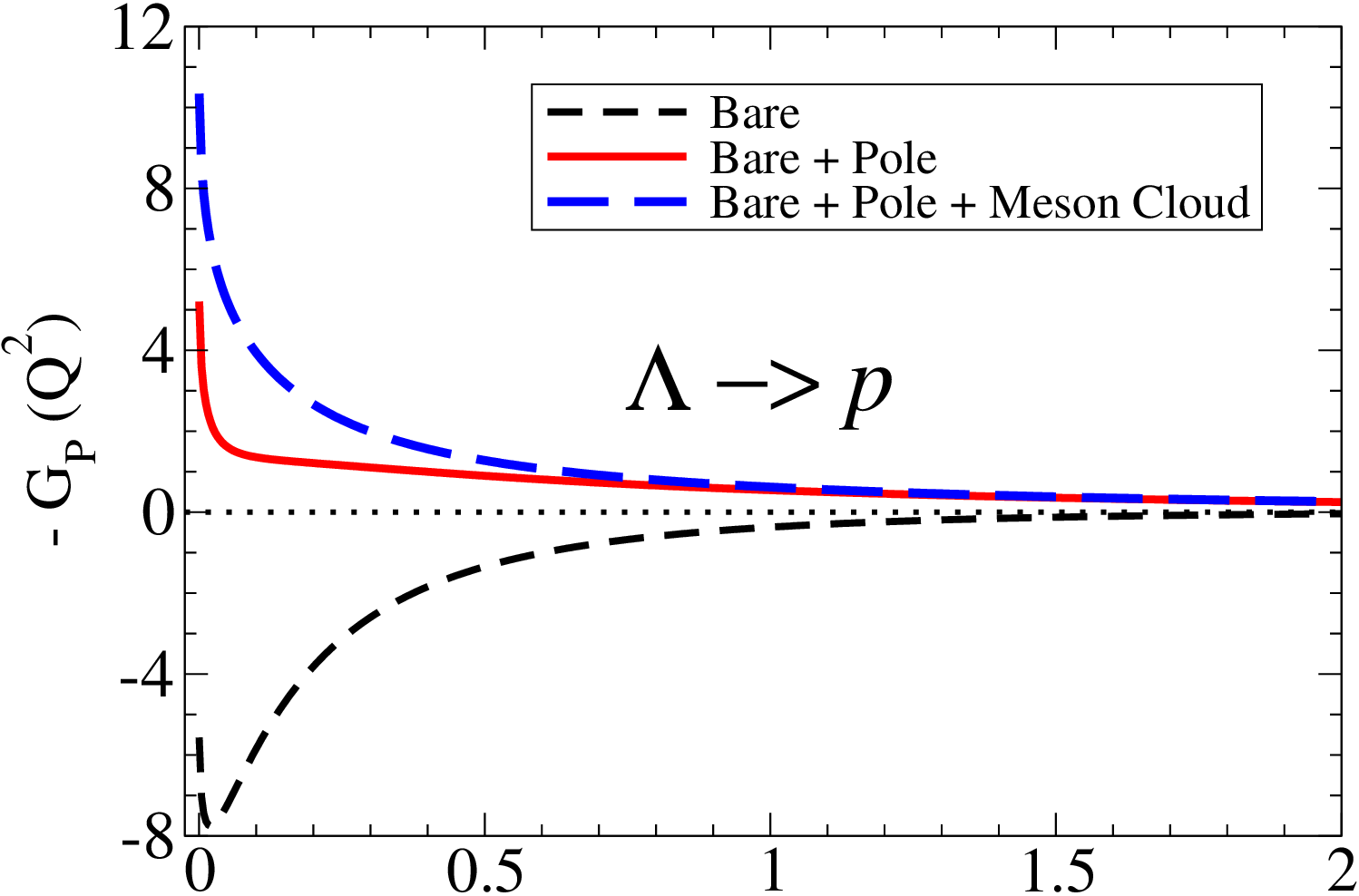}}}
\vspace{.95cm}
\centerline{\mbox{\includegraphics[width=3.0in]{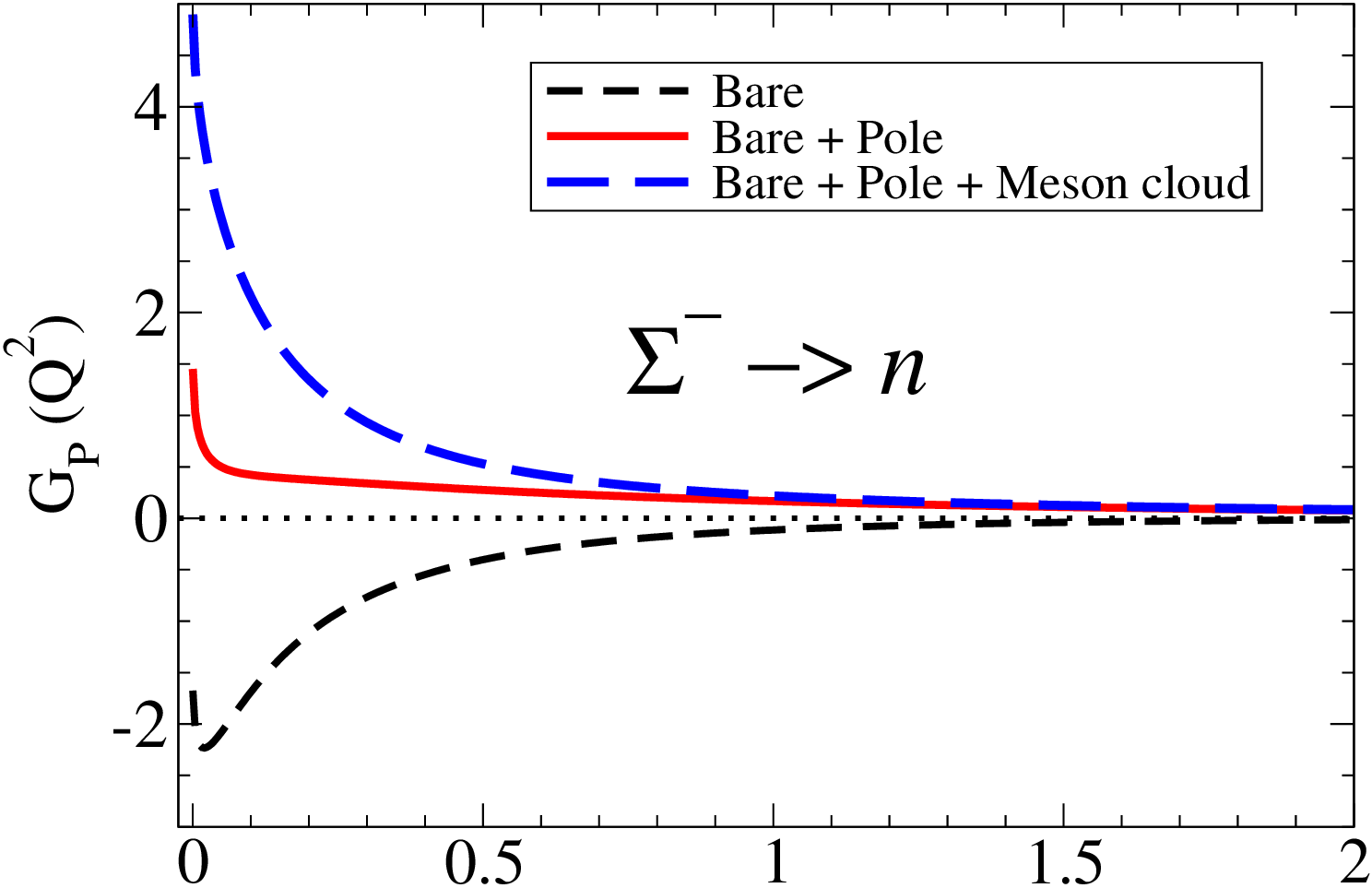}}}
\vspace{.85cm}
\centerline{\mbox{\includegraphics[width=3.0in]{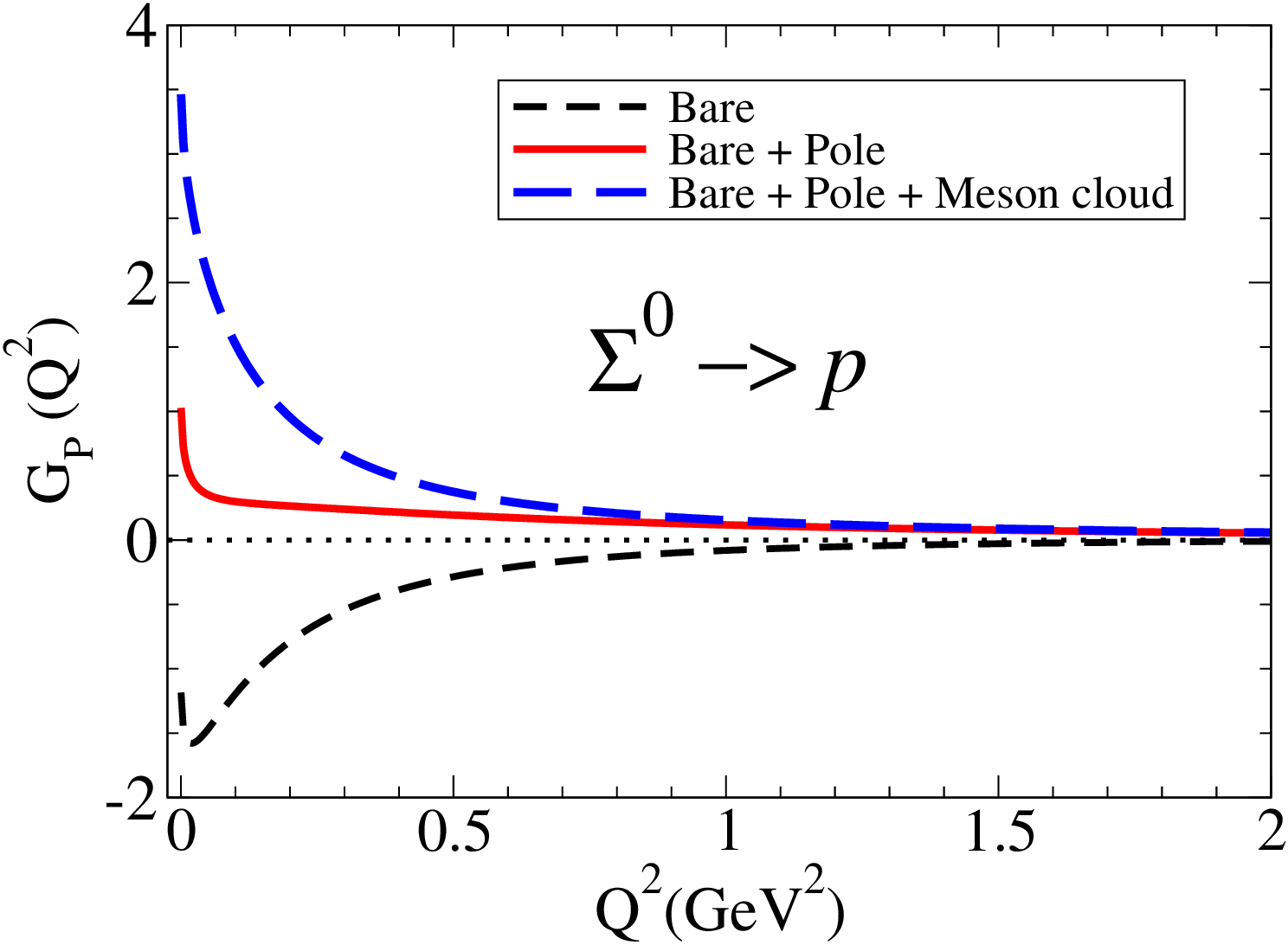}}}
\caption{\footnotesize{
Results of $G_P(Q^2)$ for the octet baryons for 
$\Delta S=1$ (part 1).
For convenience we changed the sign of the function $G_P$ 
for $\Lambda  \to p$.
}}
\label{figGP-DeltaS}
\end{figure}

\begin{figure}[t]
\vspace{.5cm}
%\vspace{2cm}
\centerline{\mbox{\includegraphics[width=3.0in]{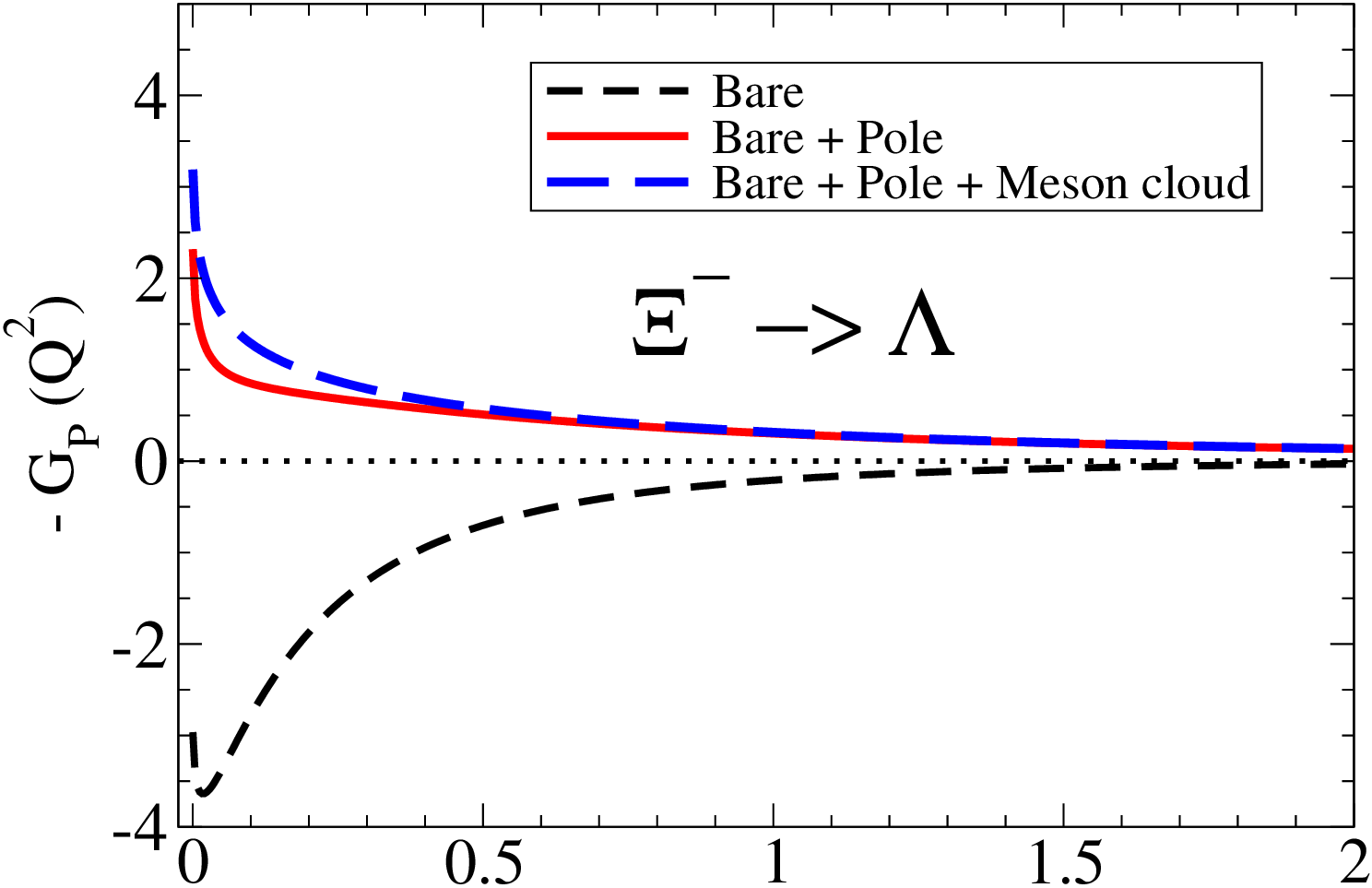}}}
\vspace{.9cm}
\centerline{\mbox{\includegraphics[width=3.0in]{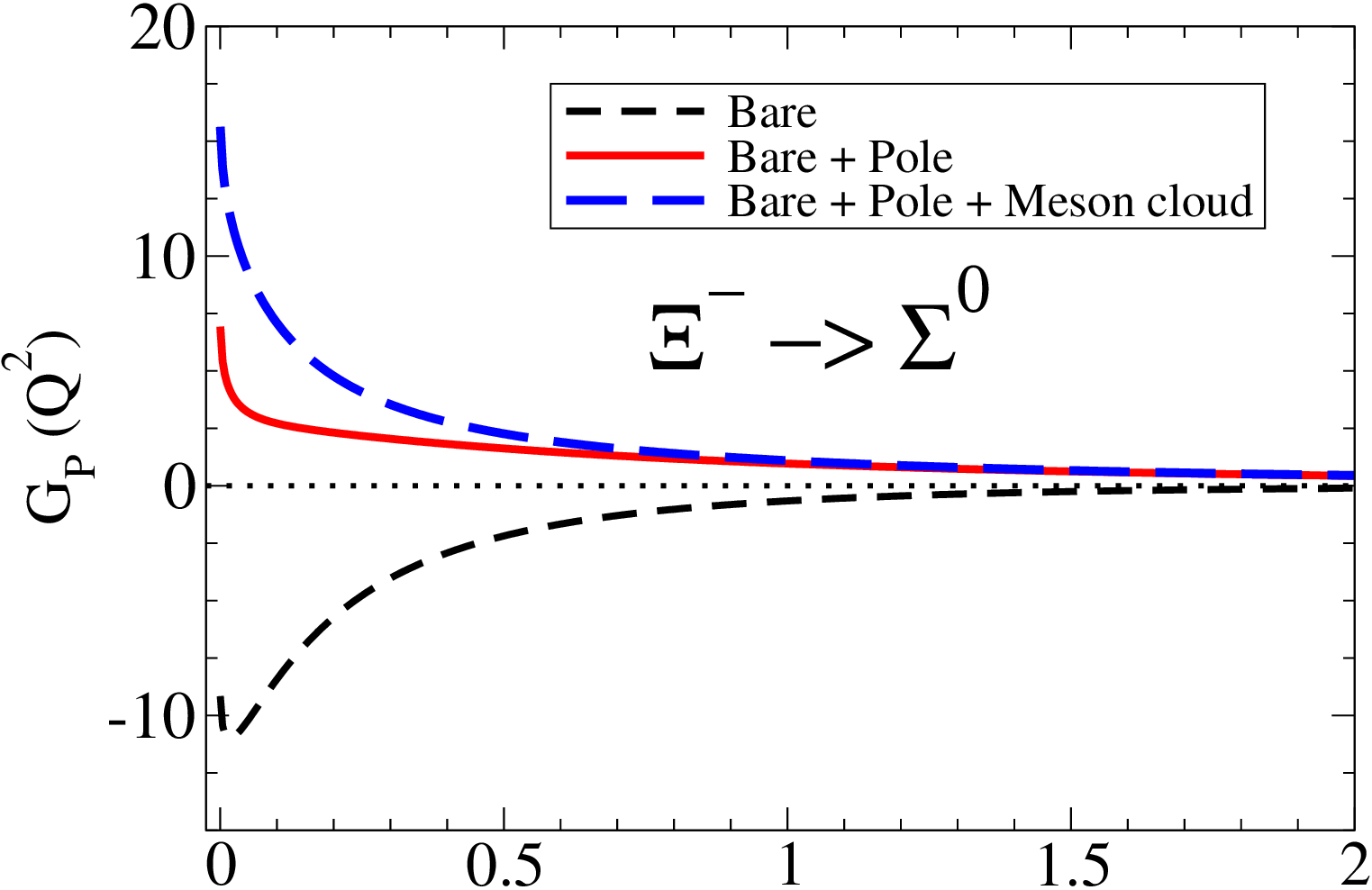}}}
\vspace{.9cm}
\centerline{\mbox{\includegraphics[width=3.0in]{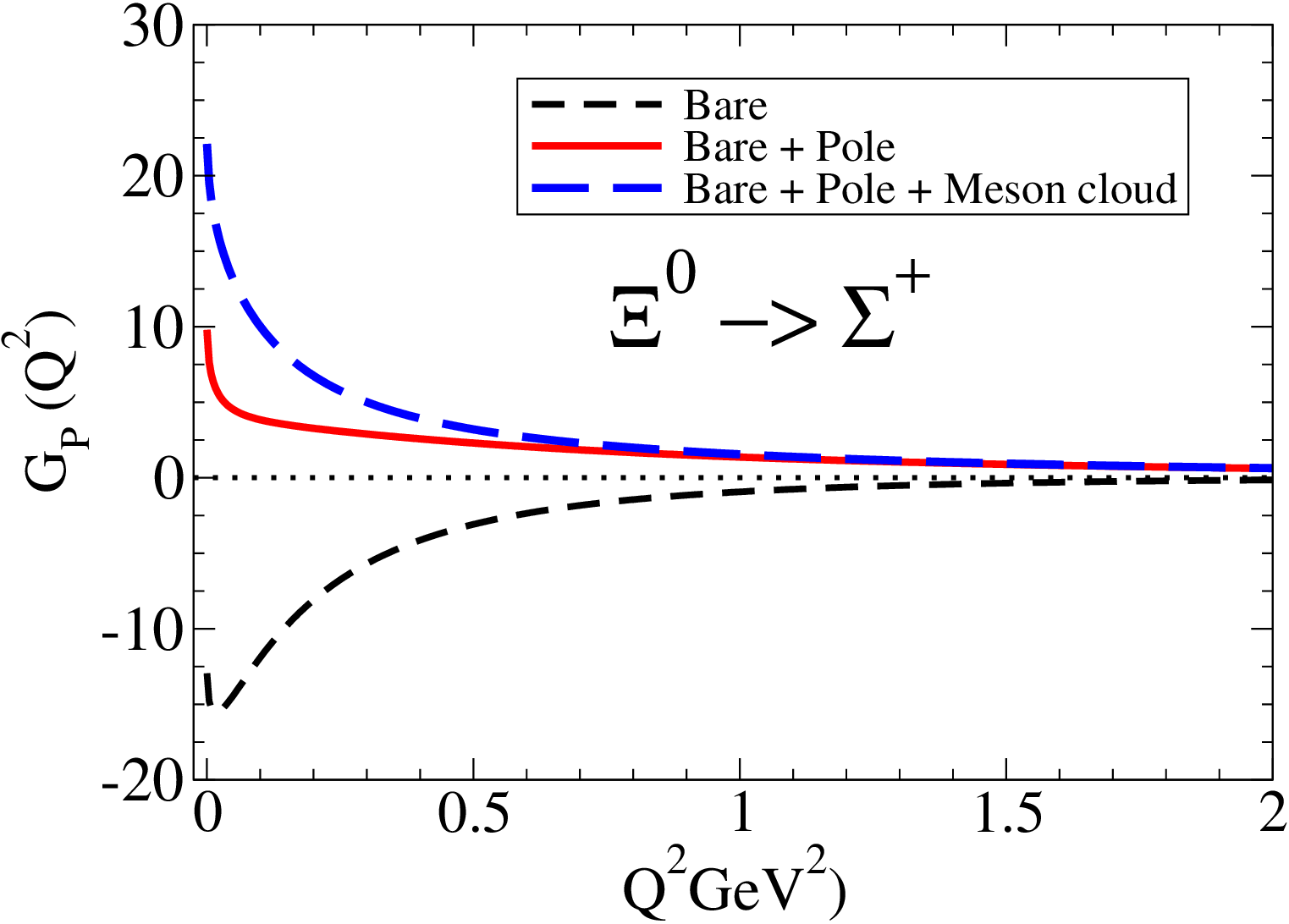}}}
\caption{\footnotesize{
Results of $G_P(Q^2)$ for the octet baryons for  
$\Delta S=1$ (part 2).
For convenience we changed the sign of the function $G_P$ 
for $\Xi^-  \to \Lambda$.
}}
\label{figGP-DeltaS2}
\end{figure}

Results for $G_P$ by the  
$SU'(3)$ meson cloud model 
related with $\Delta I=1$ transitions 
are presented in Fig.~\ref{figGP-DeltaI}, 
while those related with $\Delta S=1$ transitions
are presented in Figs.~\ref{figGP-DeltaS} and \ref{figGP-DeltaS2}.
The results associated with the $SU(6)$ meson cloud 
model are similar in shape, but differ in values for small $Q^2$.

In Fig.~\ref{figGP-DeltaI}  we can observe 
the dominance of the pole term for the $\Delta I=1$ transitions, 
since the values of "bare + pole`` are much larger 
in absolute value than the values of "bare``. 
All the results are very similar  
although the magnitude of the pole contribution
is small for the $\Xi^- \to  \Xi^0$ case,
because the magnitude of $G_A$ 
is also small in this transition.
The similarity is a consequence of the 
approximated $SU(6)$ structure of $G_A^B$ and $G_P^B$
that results from the quark model and the 
small $SU(6)$ violation from the meson cloud component.

In Figs.~\ref{figGP-DeltaS} and~\ref{figGP-DeltaS2}
we can notice that the magnitude of $G_P$
for the transitions $\Delta S=1$ 
is smaller than that for the case of the $\Delta I=1$ transitions.
In this case the contributions from 
the pole are reduced by about an order of magnitude,
due to the difference in the meson masses,
that contributes with a reduction 
of about $\frac{m_K^2}{m_\pi^2} \simeq 12.9$, 
corrected by the factors $M_{BB'}^2$ 
depending on the transition
[see Eqs.~(\ref{eqGPoctet}) and~(\ref{eqGPoctet2})].

Thus, in the $\Delta S=1$ transitions the 
bare and pole contributions are comparable in magnitude,
and the dominance of the pole term does not happen 
as in the case $\Delta I=1$.
In Fig.~\ref{figGP-DeltaS} for the transitions 
$\Lambda \to p$, $\Sigma^- \to n$ and $\Sigma^0 \to p$, 
we can observe a significant cancellation between 
the bare and pole contributions.
The evidence of the cancellation can be observed 
due to the small values of ``bare + pole'' 
when compared  with ``bare'' in absolute values.
The calculation is more significant for 
the  $\Sigma^0  \to p$ transition.
One can see however, that when $Q^2$ increases 
the pole contribution dominates (positive values for $G_P$ 
or $-G_P$ according to the transition).
This happens because although the bare 
contributions goes with $1/Q^4$   
as discussed in Sec.~\ref{secValenceOctet}, 
while the pole goes with $1/Q^6$ for very large $Q^2$,
the factor $1/(m_K^2 + Q^2)$ in the pole term 
dominates over the other terms 
in the region observed ($Q^2 < 2$ GeV$^2$).
Since $m_K^2 \simeq 0.25$ GeV$^2$,  
the factor $1/(m_K^2 + Q^2)$ has a strong impact
for the small and the intermediate $Q^2$ regions.
For much larger $Q^2$ the bare contribution will dominate.

The results for the transitions involving $\Xi$ presented 
in Fig.~\ref{figGP-DeltaS2} 
are very similar, apart the scale.
In the case of the $\Xi^- \to \Sigma^0$
and $\Xi^0 \to \Sigma^+$ transitions  
the similarity is a consequence of  
$SU(3)$ symmetry for $G_A$, 
since they differ only by the factor $1/\sqrt{2}$,  
and the masses are the same in both transitions.
As for $\Xi^- \to \Lambda$, this is  
a consequence of the approximated $SU(6)$ structure 
which implies a reduction of the pole contribution 
of the factor $\frac{5}{\sqrt{3}} \simeq 3$
compared to  $\Xi^- \to \Sigma^0$,
neglecting the effect of the (small) mass 
difference between  the $\Sigma$ and $\Lambda$.


Our result for the $G_P$ (full result)
associated with $\Xi^0 \to \Sigma^+$ transition  
has a magnitude similar to that of the 
lattice QCD results in Ref.~\cite{Sasaki08}.

To finalize, it is interesting to discuss the 
breaking of $SU(3)$ symmetry due to the 
difference in the transitions 
between $\Delta I=1$ and $\Delta S=1$.
This can be observed  by comparing the transitions  
$n \to p$ and $\Xi^0 \to \Sigma^+$,
which, according to the $SU(3)$ baryon-meson 
model, have the same results for 
the axial-vector form factor at $Q^2=0$,  
$G_A(0)= F+D$.
However, since the value of $G_P(0)$
for the pole is determined by the pion mass (squared)
for the reaction with $\Delta I=1$
and by the kaon mass (squared) for the reaction with $\Delta S=1$,
the magnitude of the $G_P^{\rm pole} (0)$ 
changes drastically 
from the $\Delta I=1$ to $\Delta S=1$ cases.
From Table~\ref{tabGP} we can conclude 
that the ratio between the pole terms 
for the nucleon and $\Xi^0 \to \Sigma^+$ 
is about 6.7, which is essentially 
the result of the combination    
of the baryon and meson masses, 
$\frac{M^2}{M_{BB'}^2}\frac{m_\pi^2}{m_K^2} \simeq 7.2$,
where $M_{BB'}$ is the average mass 
of the initial and final baryons 
in the $\Xi$ transitions.
The value 7.2 has to be 
corrected by a factor 7\% due to the 
difference in the values of $G_A(0)$
which correspond to the deviation from the 
$SU(6)$ baryon-meson model.
We can conclude then, 
that the pole term (pion or kaon) 
breaks  the $SU(3)$ symmetry for $G_P$, 
but that the magnitude 
of the breaking is mainly a consequence 
of the ratio $\frac{m_\pi^2}{m_K^2}$. 

% figure 12-13 here

\section{Summary and conclusions}
\label{secConclusions}

Weak interaction axial form factors 
of the octet baryons have been studied extensively 
based on a large numbers of theoretical frameworks.
However, such studies have mostly been restricted 
to the axial charges and proprieties at $Q^2=0$.
In this work we use the covariant spectator quark 
model to probe the weak interaction axial structure of the octet baryons  
and take advantage of the covariance 
to make predictions on  the $Q^2$ dependence of 
the axial form factors $G_A(Q^2)$ and $G_P(Q^2)$, 
for all the octet baryon weak interaction axial transitions.

In the covariant spectator quark model 
the quarks have their own structure   
characterized by the electromagnetic and axial form factors,
that can be used to calculate electromagnetic 
and weak interaction transition form factors between baryons.
The model has been successfully used  in the past for the studies 
of several electromagnetic transitions 
between baryons, including in particular 
the electromagnetic structure of the 
nucleon, the octet and decuplet baryons, and other reactions.
In the future the present approach can be applied 
for the weak interaction vector transition form factors, 
replacing  the quark axial current $j_{Aq}^\mu$ by 
the quark electromagnetic-vector current.
Similar calculations were already performed for the 
octet baryon electromagnetic form factors
using an $S$-state model~\cite{OctetFF,OctetMM,Medium,LambdaSigma},
except that it is necessary now 
to consider also charged currents.

To study the weak interaction axial structure of the octet baryons 
the covariant spectator quark model is first calibrated by the 
lattice QCD and the physical data for nucleon.
After that, the model
is extended for the octet baryons  
using the $SU_F(3)$ (flavor) symmetry. 
The $SU_F(3)$ symmetry breaking effects are  taken into account by 
the  octet baryon masses and the shape 
of the radial wave functions, determined 
in previous works by the study of the electromagnetic properties
in the context of the covariant spectator quark model. 
For simplicity we neglected the difference 
in the masses of the initial and the final baryons. 
The axial form factors 
are then calculated in the relativistic impulse approximation 
in terms of the covariant wave functions of the octet baryons 
and the  quark axial current, defined by the quark 
axial form factors  $g_{A}^q(Q^2)$ and $g_{P}^q(Q^2)$.
The wave functions of the octet baryons are 
determined by a dominant $S$-state component
defined in previous works, and   
a $P$-state  is introduced in this work, 
in order to better describe 
the  axial-vector form factors of the octet baryons.
The addition of the extra  $P$-state 
was suggested by  some studies based 
on the quark degrees of freedom.

The calibration of the  present model is done as follows:
the quark form factor $g_{A}^q(Q^2)$
is assumed to have the same form  
as that of the quark electromagnetic isovector form factor 
$f_{1-}(Q^2)$,  and the quark form factor $g_{P}^q(Q^2)$ 
has a form analogous to the 
Pauli form factors of the quarks,
motivated by vector meson dominance
with  two adjustable parameters.
The unknown parameters of the model are, 
the $P$-state mixture coefficient,  
and the parameters of the $g_{P}^q(Q^2)$ function,  
and they are determined by a fit to the nucleon axial form factor 
data in the lattice QCD regime.
In this regime the contamination 
of the form factors due to the meson cloud 
is significantly suppressed and the 
physics associated with the valence quarks 
can be estimated more accurately. 

The results obtained for the octet baryon 
$G_A$ form factors are consistent 
with the nonrelativistic $SU(6)$ quark models in the 
equal mass case ($M_B= M$)  
when the $P$-state component is dropped ($n_P=0$).
In addition, the model at $Q^2=0$ 
has the same structure of an $SU(6)$ quark model 
or an $SU(6)$ baryon-meson model, 
even when the $P$-state is included.

We conclude that the  axial form factors  
of the nucleon, both $G_A$ and $G_P$, can be very well explained 
in the lattice regime of our constituent quark model  
with a $P$-state mixture of about 26\%.
Once the parameters of the model are fixed,
the results can be extrapolated to the physical regime
and used to calculate the contributions 
of the valence quarks for the nucleon axial form factors.
As in previous works on the nucleon 
axial-vector form factor,
we conclude also that only the effects of the valence quarks
underestimate the $G_A$ data in the physical regime.
Under the assumption that the missing part 
is due to the meson cloud component 
in the physical nucleon state, we used the model, 
which is well calibrated  in the high $Q^2$ region, 
to estimate the size of the meson cloud contribution 
in the physical nucleon state.
The results obtained for the nucleon 
are in agreement with the experimental 
data for $G_A$ and $G_P$, 
when the fraction of the  meson cloud 
in the nucleon wave function is about 27\%.
With the use of the 
lattice QCD and physical data 
for the axial form factors,
we obtain in principle a better 
constraint for the magnitude of the meson cloud than 
when we use only the constraint
of the nucleon electromagnetic form factor data.

Using the $SU_F(3)$ symmetry at the quark level 
we generalize the model for the octet baryons, 
and predict all the axial form factors of 
the octet baryons.
It is expected that the present estimates is accurate 
for  $Q^2 >$ 1 GeV$^2$ (large $Q^2$ regime)
except for a  small correction due to the 
normalization factor of baryon the wave functions,  
that result from the meson cloud component.
The corrections are estimated based on the relation 
for the meson cloud in the nucleon wave function  
and the other members of the octet baryon wave functions.
As for the low $Q^2$ regime ($Q^2 < 1$ GeV$^2$), 
the estimates based exclusively 
on the valence quark degrees of freedom are expected to fail.
We provide however, effective descriptions 
for  the region $Q^2 < 1$ GeV$^2$,
based on simple parametrizations for the meson cloud contributions
constrained by $SU(3)$ and/or $SU(6)$ symmetries.
Those parametrizations can be useful 
in the future for the studies associated with the properties of 
octet baryons.
The meson cloud model labeled as $SU'(3)$, 
gives the  best description of the data 
within an deviation of three standard deviations.

We conclude in general, that the naive $SU(6)$ baryon-meson model  
is expected to fail at large $Q^2$ for the $\Delta I=1$ transitions.
In this case the falloff observed for the nucleon 
given by the  cutoff parameter $M_A \simeq 1.05$ GeV in 
the dipole parametrization, should be replace by a value near 1.25 GeV.
As for the transitions $\Delta S=1$, although dependent 
on the  transitions, are consistent with the estimate    
$M_A \simeq 1.25$ GeV, proposed long time ago~\cite{Gaillard84}.
Predictions for the induced pseudoscalar 
form factors are also presented in this work 
based on the contribution of the meson pole 
(pion for $\Delta I=1 $ and kaon for $\Delta S=1$ transitions),
complemented by a contribution from the bare core.
The bare contribution used in this work 
is derived from the quark substructure of the baryons, 
and it is calibrated using lattice QCD data.
As far as the authors are aware, this is the first time 
that the $G_P$ form factors with finite $Q^2$ are estimated 
using the results from lattice QCD as input.

To summarize, we present the   
result of a phenomenological fit with some constraints based on a
constituent quark model combined with a parametrization 
of meson cloud effects 
for the octet baryon  weak interaction 
form factors $G_A(Q^2)$ and $G_P(Q^2)$. 
The model presented is covariant and can therefore be used 
for the studies of the reactions at large $Q^2$.
The predictions of the model, in particular 
the falloff of the $G_A(Q^2)$,
can be tested in the near future by lattice QCD simulations,
or hopefully  by upcoming experiments.

\begin{acknowledgments}
The authors thank P.~Guichon for sharing the data for $G_P$.
GR thanks Alireza Tavanfar for his comments and suggestions.
GR was supported by the Brazilian Ministry of Science,
Technology and Innovation (MCTI-Brazil).
KT was also supported by FAPESP, 2014/26892-8. 
GR and KT were also supported by Conselho Nacional de Desenvolvimento
Cientif{\'{i}}co e Tecn\'ologico (CNPq) 400826/2014-3.
\end{acknowledgments}

\appendix

\section{Calculation of the transition currents}

We discuss here how we calculate the factors
associated with radial wave functions
using the symmetries of the functions.

First, we present identities that 
can be used in the calculation 
of transitions between the states of the same mass. 
Next, we discuss the angular integration that can be
used to simplify the calculations for the nucleon,
including the terms associated 
with the same state ($S$ or $P$)
as well as the terms associated with the 
$S$ and $P$ mixture (different states).
Finally, we explain how the procedure 
can be extended for transitions between 
the different baryon states  which requires also
different parametrizations between 
the initial and the final states.

\subsection{Integral identities}

The following relations
are valid under the integration symbol 
when the radial wave functions 
of the initial and final state 
are defined for the states with the same mass:
\ba
k^\alpha &=& \frac{P' \cdot k}{(P')^2} (P^\prime)^\alpha \\
k^\alpha k^\beta &=&
S_1 g^{\alpha \beta} + \frac{m_D^2 + S_2}{(P')^2} (P')^\alpha (P')^\beta
\nonumber \\
&& +
\frac{S_3}{Q^2} q^\alpha q^\beta \\
(k \cdot q) k^\alpha &=&   - c_2 q^\alpha,
\ea
where $\sqrt{P^{\prime 2}}= M(1+ \tau)$ and
$S_1 = \frac{1}{2}(c_2 -c_1)$, $S_2 = - \frac{1}{2}(c_2 -3 c_1)$
and $S_3 = \frac{1}{2}(3 c_2 -c_1) $.

\subsection{Angular integration -- nucleon case}

The expression for the transition currents
depends on a few covariant integrals.
The integrals are by definition
frame independent, however, the symmetries
of the radial wave functions are better
understood by fixing a frame.
We consider in particular the Breit frame.
In the Breit frame $P'= (M\sqrt{1 + \tau},0,0,0)$
and $q=(0,0,0,Q)$.
In this reference frame we can represent
the initial radial wave function $\psi_i(P_-,k)$
as a function of $\omega_- = \frac{P_+ \cdot k}{M}= a + b k_z$
and the final radial wave function $\psi_f(P_+,k)$
as a function of  $\omega_+ = \frac{P_+ \cdot k}{M}= a - b k_z$.
Here $a$ and $b$ are functions of ${\bf k}$
and independent of the angles.\footnote{The explicit
expressions are
\ba
a= \frac{P' \cdot k}{M}, \hspace{.3cm}
b k_z= - \frac{q \cdot k}{M}.
\nonumber
\ea
In the Breit frame $a=  \sqrt{1 + \tau}E_D $
and $b= \frac{Q}{2M}$.}
When the initial and the final states are
the same ($i=f$) as in the case
of the transition between $S$ states or $P$ states,
we can conclude right away that the product
of the wave functions becomes 
\ba
\psi_i(P_+,k) \psi_i(P_-,k)= F(z),
\ea
where $F$ is an implicit function of $|{\bf k}|$, 
and $z= k_z/|{\bf k}|$ represents
$\cos \theta$ ($\theta$, the angle between ${\bf k}$
and the $z$-axis).
Since the integration range in $z$ is bounded 
by the $-1$ and $+1$, one can conclude that
\ba
\int_k k_z \psi_i(P_+,k) \psi_i(P_-,k) = 0.
\ea
Therefore, the terms proportional to $k_z$
vanishes in the integration.
The same happens trivially for the terms
proportional to $k_x$ or $k_y$.

In the case of the transition
between the S and the P states
one can have terms in
$\psi_P(P_+,k) \psi_S(P_-,k)= F(z)$,
and $\psi_S(P_+,k) \psi_P(P_-,k)= F(-z)$,
where the argument $z$ changes sign
from the $S \to P$ to the $P \to S$ cases.
In this case we have to combine the
contributions from the  both processes
which can take the form
$t_1 F(z) + t_2 F(-z)$,
where $t_1,t_2$ are independent of $z$.
We note that one can change $-z \to z$
in the second term, 
under the integration symbol,
since
\ba
\int_{-1}^1 F(z) dz = \int_{-1}^1 F(-z) dz.
\ea
Using the same argument one can conclude
also that
\ba
\int_{-1}^1 z \, F(z) dz = 0.
\ea
Therefore, the terms in $k_z$ vanishes also
in the integral appearing in the $S-P$ transition, 
\ba
\int_k k_z \psi_f(P_+,k) \psi_i(P_-,k) = 0,
\ea
and the same holds for the change of the initial and 
final state interchange, $i \leftrightarrow f$.

\subsection{Angular integration -- octet case}

The calculation of integrals associated with the transitions 
between the different states, which involves different 
parametrizations of the radial wave functions,
can be reduced to the case discussed in the previous section 
for the nucleon, provided that the radial wave functions are 
associated with the same mass.

In the equal mass case the discussion 
associated with the $S$- and $P$-states
can be generalized for radial 
wave functions with different parametrizations 
for the initial and final states.
The key point again is that 
we can re-write the factors associated 
with the radial wave functions 
in terms of the $\omega_+$
and $\omega_-$ as already discussed.

\clearpage

\end{document}